\documentclass[11pt]{article}
\pdfoutput=1

\usepackage[lmargin=50pt,rmargin=60pt,tmargin=60pt,bmargin=65pt]{geometry}
\usepackage{authblk} 
\usepackage{epsfig} 
\usepackage{amstext,amsmath,amssymb,amsfonts,amsmath,amsthm}
\numberwithin{equation}{section}
\usepackage[table]{xcolor} 
\usepackage{tikz}
\usepackage{makecell}
\usepackage{braket} 
\usepackage{dsfont} 
\usepackage{graphicx} 
\usepackage[export]{adjustbox} 
\usepackage{float}  
\usepackage{subfig}
\usepackage{caption} 
\usepackage{cite}
\usepackage{color}
\usepackage[colorlinks, linkcolor=blue, citecolor=blue, urlcolor=blue]{hyperref}

\theoremstyle{definition}

\theoremstyle{remark}

\numberwithin{equation}{section}

\definecolor{cardinal}{rgb}{0.6,0,0}
\definecolor{darkgreen}{rgb}{0,0.5,0}
\definecolor{golden}{rgb}{0.92, 0.7, 0}
\definecolor{midnight}{rgb}{0, 0, 0.5}
\definecolor{darkblue}{rgb}{0.2, 0, 0.8}

\newcommand{\sqbracket}[1]{\left[#1\right]}
\newcommand{\abs}[1]{\left| #1 \right|}

\newcommand{\be}{\begin{equation}}
\newcommand{\ee}{\end{equation}} 
\newcommand{\ba}{\begin{equation}\begin{aligned}}
\newcommand{\ea}{\end{aligned}\end{equation}} 
\newenvironment{systeme}{\left\{ \begin{array}{l}}{\end{array}\right.}

\newcommand{\f}{\frac}
\newcommand{\p}{\partial}
\newcommand{\la}{\langle}
\newcommand{\ra}{\rangle}

\DeclareMathOperator{\Tr}{Tr}

\newcommand{\vphi}{\varphi}
\def\tend{\rightarrow}

\renewcommand{\(}{\left(}
\renewcommand{\)}{\right)}
\def\half{\frac{1}{2}}

\def \kdk {k\partial_k}
\def \const{\ \text{const}}
\newcommand{\bigO}[1]{{O}\left(#1\right)}
\newcommand{\partd}[2]{\frac{\partial#1}{\partial#2}}

\def \dd  {{\rm d}}

\newcommand{\intpi}[2]{\frac{\dd^#2 #1}{\( 2\pi \) ^#2}} 

\renewcommand{\a}{\alpha} 
\renewcommand{\b}{\beta}  
\newcommand{\g}{\gamma} 
\renewcommand{\d}{\delta}
\newcommand{\eps}{\epsilon}

\renewcommand{\l}{\lambda}
\newcommand{\m}{\mu}

\newcommand{\vph}{\varphi}
 
\newcommand{\D}{\Delta}

\newcommand{\cM}{\mathcal{M}}

\newcommand{\cP}{\mathcal{P}}

\newcommand{\cT}{\mathcal{T}}

\newcommand{\Z}{\mathbb{Z}}
\newcommand{\R}{\mathbb{R}}

\graphicspath{{fig/}}

\begin{document}

\title{\bf Critical and multicritical Lee-Yang fixed points\\ in the local potential approximation}

\author[1]{Dario Benedetti}
\author[1]{Fanny Eustachon}
\author[2]{Omar Zanusso}

\affil[1]{\normalsize \it 
 CPHT, CNRS, \'Ecole polytechnique, Institut Polytechnique de Paris, 91120 Palaiseau, France
  \authorcr \hfill}

\affil[2]{\normalsize \it
 Universit\`a di Pisa and INFN - Sezione di Pisa, Largo Bruno Pontecorvo 3, 56127 Pisa, Italy
  \authorcr \hfill}

\date{}
\maketitle

\hrule\bigskip

\begin{abstract}
The multicritical generalizations of the Lee-Yang universality class arise as renormalization-group fixed points of scalar field theories with complex  $i\varphi^{2n+1}$ interaction, $n\in\mathbb{N}$, just below their upper critical dimension. 
It has been recently conjectured that their continuation to two dimensions corresponds to the non-unitary conformal minimal models $\mathcal{M}(2,2n+3)$.
Motivated by that, we revisit the functional renormalization group approach to complex $\mathcal{P}\mathcal{T}$-symmetric scalar field theories in the Local Potential Approximation, without or with wavefunction renormalization (LPA and LPA' respectively), aiming to explore the fate of the $i\varphi^{2n+1}$ theories from their upper critical dimension to two dimensions.
The $i\varphi^{2n+1}$ fixed points are identified using a perturbative expansion of the functional fixed-point equation near their upper critical dimensions, and they are followed to lower dimensions by numerical integration of the full equation.
A peculiar feature of the complex $\mathcal{P}\mathcal{T}$-symmetric potentials is that the fixed points are characterized by real but negative anomalous dimensions $\eta$, and in low dimension $d$, this can lead to a change of sign of the scaling dimensions $\Delta=(d-2+\eta)/2$, thus requiring a novel analysis of the analytical properties of the functional fixed-point equations.
We are able to follow the Lee-Yang universality class ($n=1$) down to two dimensions, and  numerically determine the scaling dimension of the fundamental field as a function of $d$. 
On the other hand, within the LPA', multicritical Lee-Yang fixed points with $n>1$ cannot be continued to $d=2$ due to the existence of unexpected non-perturbative fixed points that annihilate with the $i\varphi^{2n+1}$ fixed points.
\end{abstract}

\hrule\bigskip

\tableofcontents

\section{Introduction and summary}
\label{sec:introduction}

Unitarity is a fundamental axiom of quantum field theory (QFT), but it plays a lesser role in applications to critical phenomena, for which the relevant QFTs are defined on Euclidean space rather than Minkowski spacetime. 
In such setting, unitarity is equivalent to reflection positivity, but in general there is no fundamental reason to impose the latter on a classical statistical system, other than reducing the complexity of the task of classifying universality classes, or for invoking useful results that rely on unitarity. 
In fact, many interesting universality classes are described by nonunitary conformal field theories (CFTs). For example, percolation, self-avoiding walks, and critical systems with quenched disorder in two dimensions are associated to logarithmic CFTs with vanishing central charge \cite{Cardy_2003,Gurarie:2004ce}.
Another famous example is the universality class of the Lee-Yang edge singularity, that is associated to the infrared fixed point of a scalar field theory with  $i\vph^3$ interaction  \cite{Fisher:1978pf,Cardy:1985yy,Cardy:2023lha}: although the spectrum of scaling dimensions stays real thanks to $\cP\cT$ symmetry,\footnote{\label{foot:PT}By $\cP\cT$ symmetry we mean an antilinear version of $\mathbb{Z}_2$ symmetry, i.e.\ the invariance under a transformation that acts as $i\to -i$, $\vph\to-\vph$. It is well known (e.g.\ \cite{Bender:1998gh}) that a diagonalizable operator that is invariant under an antilinear symmetry, and whose eigenvectors are also invariant, has real eigenvalues, hence applying this argument to the dilation operator of a $\cP\cT$ symmetric theory it follows that scaling dimensions are real, as long as the scaling operators are $\cP\cT$ invariant.  We refer to \cite{Bender:2018pbv,Castro-Alvaredo:2017udm} and references therein for more on $\cP\cT$ symmetry in the field theoretic context.} 
the imaginary coupling leads to violations of unitarity, in the form of complex OPE coefficients and of scaling dimensions below the unitarity bounds.

Once we allow nonunitarity, the study and classification of universality classes faces a number of additional challenges. First of all, unitarity is a very constraining condition and lifting it leads to a proliferation of possible models, only partially tamed if we demand $\cP\cT$ symmetry. 
The enlarged class of models leads to questions about their interpretation and possible relevance to physical models. 
Moreover, nonunitarity introduces also a number of technical challenges. For example, the numerical bootstrap \cite{Poland:2018epd} relies crucially on the unitarity assumption, and bootstrapping nonunitary models remains an open problem.
Monte Carlo simulations also face difficulties in the case of models with complex couplings. Therefore, we are left with fewer options for studying such models, and it is important to develop further any approach that works.

A case in point for the above challenges is the Lee-Yang universality class.
In dimension $d=6-\epsilon$, for $\epsilon\ll 1$, the $i\vph^3$ theory has a weakly-coupled infrared fixed point, with coupling of order $\sqrt{\epsilon}$, allowing to establish the connection to the Lee-Yang edge singularity via semiclassical and perturbative arguments \cite{Fisher:1978pf}.
However, as the dimension is lowered, the fixed-point theory becomes strongly coupled, and nonperturbative methods become necessary. 
In two dimensions, the fixed-point theory is believed to coincide with the conformal minimal model $\cM(2,5)$ \cite{Cardy:1985yy}, although a full proof is missing. Such a missing proof is a further motivation for developing nonperturbative methods able to interpolate between $d=6$ and $d=2$.

A similar story holds for the multicritical generalization of the Lee-Yang universality class, associated to the nontrivial fixed point for the $\cP\cT$-symmetric scalar field theory with  $i\vph^{2n+1}$ interaction, with $n>1$, arising below the upper critical dimension $d_{2n+1} = \frac{4n+2}{2n-1}$.
As for the $n=1$ case, such fixed point is weakly coupled at $d=d_{2n+1} - \epsilon$, for $\epsilon\ll 1$, but it becomes strongly coupled as the dimension is decreased further towards $d=2$.
Precisely at $d=2$, the status of these multicritical Lee-Yang fixed points is more open than in the $n=1$ case.
In particular, assuming that indeed they can be continued to $d=2$, two different conjectures have been put forward for their corresponding CFTs: the first \cite{Becker:1991nr,Becker:1991nq,vonGehlen:1994rp,Zambelli:2016cbw,Lencses:2022ira,Katsevich:2025ojk} posits that such models correspond to the conformal minimal models $\cM(2,2n+3)$, while the second  \cite{Amoruso,Lencses:2024wib} associates them to  $\cM(2,4n+1)$.\footnote{It should be noticed that in some early forms of these conjectures, the kinetic term of the $i\vph^{2n+1}$ theory was unconventional, in particular when trying to generalize Zamolodchikov's OPE-based argument for the unitary multicritical models \cite{Zamolodchikov:1986db} to the nonunitary case (see for instance appendix A of \cite{Lencses:2022ira}).
This problem is overcome in the proposal of \cite{Katsevich:2025ojk}, according to which some of the relevant composite operators of the $i\vph^{2n+1}$ theory should be identified with quasi-primary operators of  $\cM(2,2n+3)$, i.e.\ operators that behave as primaries only under global conformal transformations.
}
Settling this question also requires the development of nonperturbative methods able to interpolate between $d=d_{2n+1}$ and $d=2$.

 Previous studies have approached the nonperturbative regime of the Lee-Yang universality class by different methods, in particular: resummation of perturbation theory \cite{deAlcantaraBonfim:1980pe,deAlcantaraBonfim:1981sy,Gracey:2015tta,Borinsky:2021jdb,Gracey:2025rnz}, high-temperature expansion \cite{Butera:2012tq}, Gliozzi's bootstrap method \cite{Gliozzi:2014jsa,Hikami:2017hwv}, functional renormalization group (FRG) \cite{An:2016lni,Zambelli:2016cbw,Rennecke:2022ohx}, and most recently
the fuzzy sphere approach  \cite{ArguelloCruz:2025zuq,Fan:2025bhc,EliasMiro:2025msj}.
Each method has its own limitations, some being affected by uncertainties due to uncontrolled truncations (Gliozzi's method, FRG) and others being restricted to integer dimensions (lattice regularizations of spheres).
Moreover, much less results are available for the multicritical theories (essentially only perturbative ones \cite{Codello:2017epp,Gracey:2017okb,Katsevich:2025ojk}).
Therefore, it is important to further develop these or other methods and try to apply them to the multicritical Lee-Yang universalities.

The purpose of this paper is to pursue the FRG approach to multicritical Lee-Yang theories, pushing further the type of analysis that was initiated in  \cite{Zambelli:2016cbw}.
In practice, this means that we will work with the Wetterich version of the FRG \cite{Wetterich:1992yh,Morris:1993qb} (see \cite{Delamotte:2007pf,Dupuis:2020fhh} for reviews), using two different realizations of the so-called local potential approximation (LPA and LPA'), with a complex but $\cP\cT$-symmetric potential for a single scalar field, and without or with a running wave function renormalization, respectively.
One advantage of this approach is  that it is easily continued in $d$, therefore allowing to continuously connect to the perturbative domain near the upper critical dimension of a given universality class, and to go further from it while staying at non-integer dimension, as needed for the higher multicritical models, for which $2<d_{2n+1}<3$ (if $n\geq 3$).\footnote{One might wonder if at noninteger dimension, evanescent operators could appear in the spectrum of the theory and lead to a loss of $\cP\cT$ invariance, in analogy to the loss of unitarity discussed in \cite{Hogervorst:2015akt}. In practice, one could contemplate the possibility that operator pairs with complex conjugate dimensions could appear in noninteger dimensions \cite{Gorbenko:2018ncu}. Given the different status of unitarity and $\cP\cT$ symmetry, the latter actually being a symmetry of the action, it is not clear to us how this could be compatible with the common understanding that the fixed point sits at the boundary between $\cP\cT$-symmetric and $\cP\cT$-broken phases. In any case, the derivative expansion of the FRG is probably blind to such evanescent operators, and a different type of study would be needed to address this question.}

The main results of our numerical analysis are the following:
\begin{itemize}
\item We are able to identify and continue all the way to $d=2$ the fixed point solution corresponding to the Lee-Yang universality class.
Despite the severe limits of the approximation, the quantitative results for the scaling dimension $\Delta_\vph$ of the fundamental field (and thus the anomalous dimension exponent $\eta$) are quite good, with a relative error  between $2.6\%$ and $7\%$ compared to the exact value in $d=2$.
The results for the scaling dimension of the composite operator $\vph^3$ (and thus the correction-to-scaling exponent $\omega$) appear instead to become unreliable for $d\lesssim 4$.
\item We are able to identify the fixed point solutions corresponding to the multicritical Lee-Yang universality classes below their respective upper critical dimensions $d_{2n+1}$.
The result for $\Delta_\vph$ in $d=3$ and $n=2$ is close, with a relative error of less than 4\% to that from the two-sided Pad\'e extrapolations of \cite{Katsevich:2025ojk} (based on the perturbative results of \cite{Codello:2017epp,Gracey:2017okb} and the exact result from $\cM(2,7)$), but the results for other scaling dimensions are inconclusive due to convergence issues.
\item Within the LPA' we find that the multicritical Lee-Yang fixed points cannot be continued to $d=2$:
below $d\simeq 2.72$ new nonperturbative fixed points branch off from the critical Lee-Yang, and they annihilate one by one with the multicritical fixed points at some lower dimensions, above $d=2$.
Further study would be needed, in particular involving the next order in the derivative expansion of the effective action, to be able to conclude whether this is an artifact of the approximation or a genuine new nonperturbative effect uncovered by the FRG. If the latter turns out to be the case, it would imply that the conjectures for a Ginzburg-Landau description of the minimal models  $\cM(2,2n+3)$ or $\cM(2,4n+1)$ should be revisited: at best, they would involve a version of the $\vph^{2n+1}$ theories not invariant under $\cP\cT$.
\end{itemize}

On the technical level, one main point of our analysis is the fact that in the LPA' the scaling dimension $\Delta_\vph$ for the Lee-Yang universality class becomes zero at some dimension $d_0>2$. This is not surprising, given that near $d=6$ we have $\Delta_\vph>0$ and at $d=2$ we know that $\Delta_\vph<0$. However, for the functional fixed-point equation this implies a drastic change of behaviour around $d_0$ and introduces a number of novel features in our analysis with respect to previous works using the LPA.
Moreover, we have that $d_0\simeq 2.72$ is precisely the dimension below which the new non-perturbative partners of the multicritical fixed points branch off from the critical one. It remains to be understood whether this is a technical problem of the LPA' or a new nonperturbative feature of this class of models.

\paragraph{Outline of the paper.}
In section \ref{sec:flow}, we begin with a concise review of the FRG setting and of the local potential approximation that we will be employing.
In section \ref{sec:perturbation}, we discuss the perturbative expansion at $d=d_{2n+1} - \epsilon$, within the FRG framework.
This will provide a first understanding of the properties of the functional fixed-point equation, and it will later provide a justification for identifying numerical solutions with the corresponding $i\vph^{2n+1}$ fixed points.
In section \ref{sec:analytic},  we analyse instead the nonperturbative properties of the functional fixed-point equation, such as singularities, large-field asymptotic behavior, and the special case with vanishing scaling dimension, which can be solved exactly, albeit in an implicit form. Such analytic study of the fixed-point equation serves as a basis for interpreting the numerical results. 
In section \ref{sec:numerical}, we present all our numerical results for the critical and multicritical Lee-Yang models.
In section \ref{sec:conclusions}, we summarize our findings, discuss their interpretation, and outline some possible future directions.
In appendix \ref{sec:osborn}, we present the perturbative expansion for the case of the Polchinski equation, in greater detail than what we do in section \ref{sec:perturbation}, in particular including a discussion of its extension beyond the LPA.
Lastly, in appendices \ref{app:asymptotics} and \ref{app:higher-truncation}, we provide some technical details for the large-field analysis of \ref{sec:analytic} and the redundancy of a quadratic truncation for the real part of the potential.

\section{Flow equations in the complex local potential approximation}
\label{sec:flow}

The idea behind Functional Renormalization Group (FRG) is to study the variation of the effective average action $\Gamma_k[\phi]$ with respect to the RG time $t=\ln k$. 
The effective action is defined by the Legendre transform of the logarithm of the partition function, which is modified by adding an infrared cutoff function $R_k(q)$. The latter implies that only the modes roughly between a UV cutoff $\Lambda$ and the IR scale $k$ are integrated out, thus generalizing  Wilson's momentum shell integration, which is recovered for sharp cutoffs and infinitesimal $\Lambda-k$. In the limit $k\tend 0$ all fluctuations below the UV cutoff are integrated, thus recovering the full effective action.
The resulting RG flow equation for $\Gamma_k[\phi]$ takes the form of an exact functional differential equation, called the Wetterich equation \cite{Wetterich:1992yh}
\be\label{eq:wetterich-flow}
\kdk \Gamma_k[\phi] = \frac{1}{2} \Tr \left[ \frac{\kdk R_k}{\Gamma_k^{(2)}[\phi] +R_k} \right] .
\ee
While this equation is not generally exactly solvable, several non-perturbative approximations allow to characterized the RG flow. These usually take the form of a truncated ansatz for the effective action, in which we fix some type of functional form for $\Gamma_k[\phi]$ and discard anything generated by the right-hand side that does not fit to the ansatz.
Among them, we will consider the derivative expansion. It consists in expanding the effective action in powers of derivatives, and truncate it to a given order, while keeping a functional dependence over the field.
The crudest approximation is known as local potential approximation (LPA)
\be\label{eq:lpa-ansatz}
\Gamma_k[\phi] = \int\dd^d x\, \sqbracket{\half Z_k (\partial\phi)^2+V_k(\phi)},
\ee
where we keep a full potential plus a minimal two-derivative term.
We refer to the above \eqref{eq:lpa-ansatz} as the LPA when $Z_k=1$, that is when no wave-function renormalization is taken into account and the anomalous dimension $\eta = -(\kdk Z_k) /Z_k$ identically vanishes;
and as the LPA' when $Z_k\neq 1$.

Under the ansatz \eqref{eq:lpa-ansatz}, the Wetterich equation can be projected on the potential part by evaluating the flow \eqref{eq:wetterich-flow} on a constant field configuration $\phi(x)=\phi$, which leads to the flow equation
\be\label{eq:wetterich-V}
\kdk V_k(\phi) = \frac{1}{2}\int\intpi{q}{d}\,G_k(q^2)\,\kdk R_k(q^2),
\ee
where $G_k(q^2)$ denotes the Green function in Euclidean momentum space
\be
G_k(q^2) = \(\Gamma^{(2)}_{k,q,-q}+R_k(q)\)^{-1} = \(Z_k q^2 + V_k''(\phi)+R_k(q^2)\vphantom{Gamma^{(2)}_{k,q,-q}}\)^{-1}.
\ee

The anomalous dimension is instead obtained by projecting on the $\bigO{p^2}$ part of the flow of the second derivative of the effective action \eqref{eq:lpa-ansatz}.\footnote{
We use
\[
\kdk \Gamma^{(2)}_{k,p,-p}[\phi_0] = \frac{1}{2} \int\intpi{q}{d}\, \left[G_k(q^2)^2 \, \left(V_k^{(3)}(\phi_0)^2 G_k((p+q)^2) - V_k^{(4)}(\phi_0) \right) \right]_{\phi=\phi_0}\kdk R_k(q^2) ,
\]
which is a function of $p^2$. However, the integrand depends on $p_\m$ via $(p+q)^2$, so we Taylor-expand the Green function
$$
 G_k((q+p)^2) = G_k(q^2)+p_\mu \frac{\partial}{\partial q_\mu}G_k(q^2)
 +\frac{1}{2}p_\mu p_\nu \frac{\partial^2}{\partial q_\mu\partial q_\nu}G_k(q^2) + O(p^3) \,,
$$
which, using the chain rule, becomes
$$
 G_k((q+p)^2) = G_k(q^2)+2p_\mu q^\mu G_k'(q^2)
 +p^2 G_k'(q^2) + 2 (p \cdot q)^2 G_k''(q^2) + O(p^3) \,,
$$
having denoted with a prime the derivatives with respect to the argument $q^2$. Then, inside the momentum integration we use rotational invariance to replace
$$
 G_k((q+p)^2) \to G_k(q^2) + p^2 \left(G_k'(q^2) + \frac{2}{d}q^2 G''_k(q^2)\right)+ O(p^3) \,.
$$
}
The result is
\ba\label{eq:etak}
    \eta &= - \frac{1}{Z_k}  \lim_{p^2\tend 0}\kdk \partd{}{p^2}\left.\Gamma^{(2)}_{k,p,-p}\right|_{\phi=\phi_0}\\
        &= -\frac{V_k^{(3)}(\phi_0)^2}{Z_k}\int\intpi{q}{d}\left.\,G_k^2(q^2)\(G'_k(q^2)+ \frac{2}{d}q^2\,G''_k(q^2)\)\,\kdk R_k(q^2)\right|_{\phi=\phi_0},
\ea
where now the prime on $G_k$ denotes derivative with respect to $q^2$.
Strict consistency with the ansatz requires taking $\phi_0=0$, but for $\Z_2$ even potentials this would lead to a trivial result, hence in that case $\phi_0$ is typically defined as the minimum of the effective potential $V_k$, such that it represents the order parameter. When the global $\Z_2$ symmetry is broken explicitly, $V_k^{(3)}(0)\neq 0$, hence we can take $\phi_0=0$.

A standard cutoff choice, greatly simplifying practical computations, is the ``optimized'' Litim cutoff \cite{Litim:2001up}
\be
   R_k(q) = Z_k\, \(k^2 - q^2\)\Theta\(k^2 - q^2\),
\ee
where $\Theta(z)$ is the Heaviside step function.\footnote{When choosing the Litim cutoff in \eqref{eq:etak}, the only nonvanishing contribution comes from a term proportional to $\d(k^2-q^2)$ in $G''_k(q^2)$. 
In order to deal with products of distributions, we assume a suitable regularization of $\Theta(z)$ and $\d(z)$, preserving the relation $\Theta'(z)=\d(z)$, from which one finds in particular that in the limit that the regularization is removed $\int {\rm d} z \, \d(z)\Theta(z) = 1/2$ (see \cite{Morris:1993qb}, equation 3.19).
}
Using this cutoff, the flow equations \eqref{eq:wetterich-V} and \eqref{eq:etak} can be computed analytically and cast into dimensionless variables, $\vphi = Z_k^{1/2} k^{\frac{2-d}{2}}\phi$ and $v(\vphi)= k^{-d}\,V_k(\phi)$. At the fixed point $\kdk v=0$ they reduce to 
\ba\label{eq:realode}
d&\, v(\vphi) - \Delta\, \vphi\, v'(\vphi) =  \(1-\frac{\eta}{d+2}\)\frac{c_d}{1+v''(\vphi)}, \qquad \Delta= \frac{d-2+\eta}{2},\\
&\eta = \begin{cases}
0 & \quad \text{(LPA)}\\
c_d \frac{\(v'''(\vphi_0)\)^2}{\(1+v''(\vphi_0)\)^4} & \quad \text{(LPA')}.
\end{cases}
\ea
The left hand side corresponds to the scaling part and only depends on the scaling dimension of the field $\Delta$, while the right hand side depends on the cutoff choice and corresponds to the nontrivial features of the RG flow.
The coefficient 
\be\label{eq:cd}
c_d = \frac{S_{d-1}}{(2\pi)^d}\frac{1}{d} = \frac{1}{(4\pi)^{d/2}\Gamma(\frac{d}{2}+1)}
\ee
is a strictly positive numerical coefficient depending monotonically on the dimension. In our range of interest it takes values between $c_2\simeq 7.96\times 10^{-2}$ and $c_6\simeq 8.40\times 10^{-5}$. Later we will absorb it in a rescaling of the field and potential (see section \ref{sec:analytic}).


The ordinary differential equation \eqref{eq:realode} is of second order, hence it has in principle a two-parameter family of solutions. 
However, most of the solutions either lead to an unstable potential, or they
develop a \emph{moveable singularity} at some finite value of $\vphi$, due to the nonlinearity of the equation. Because such singularity prevents from taking the limit $k\to 0$ (from the definition of dimensionless variables, small $k$ corresponds to large dimensionless $\vphi$), we should demand that valid fixed-point solutions correspond to global solutions of the fixed-point equation \cite{Hasenfratz:1985dm,Felder:1987,Morris:1994ki}, with real part bounded from below.
Previous analyses show
us that in general bounded global solutions form a discrete set of isolated solutions.


In the case of a complex potential
\be\label{eq:complex-split}
 v(\vphi)=u(\vphi)+ih(\vphi)
\ee
the fixed-point equation \eqref{eq:realode} splits into a coupled system for the real and imaginary parts (for the sake of readability, from here on we will generally omit the argument $\vphi$ in the differential equations)
\be\label{eq:coupled}
\begin{cases}
d\, u- \Delta\, \vphi\, u' &= c_d \(1-\frac{\eta}{d+2}\)\frac{1+u''}{\(1+u''\)^2+\(h''\)^2},\\
d\, h- \Delta\, \vphi\, h' &= - c_d \(1-\frac{\eta}{d+2}\)\frac{h''}{\(1+u''\)^2+\(h''\)^2}
\end{cases}
\ee
 
As anticipated in the introduction, we impose invariance of the complex potential  under an antilinear $\cP\cT$-symmetry, $v^*(-\vphi)=v(\vphi)$, which is expected to ensure that the spectrum and critical exponents are real (see footnote \ref{foot:PT}).
Thus, the real and imaginary parts must be taken to be even and odd functions of $\vph$, respectively: $u(\vphi)=u(-\vphi)$, $h(\vphi)=-h(-\vphi)$.
Under these additional assumptions, the coupled system \eqref{eq:coupled} is endowed with initial conditions parametrized by two real parameters $(\lambda_2,g)$ chosen to correspond to the $\vphi^2$ and $i\vphi^3$ couplings, respectively:
\be \label{eq:init-cond}
u'(0) = 0,\quad u''(0)=  \lambda_2 \in \R, \quad h(0)=0, \quad h'''(0)=g\in \R \,.
\ee
Notice that with these initial conditions, \eqref{eq:coupled} immediately fixes to zero all the odd derivatives of $u(\vphi)$ and all the even derivatives of $h(\vphi)$, at $\vphi=0$. 

As mentioned below equation \ \eqref{eq:etak}, in the case of the complex potential the anomalous dimension \eqref{eq:realode} can be evaluated at the origin, and thus it is fully determined by the initial conditions
\be\label{eq:eta-init}
\eta = -c_d\frac{g^2}{(1+\lambda_2)^4} \,.
\ee
Therefore, \eqref{eq:coupled} is solved in the LPA' scheme by plugging back \eqref{eq:eta-init} into the differential equations.
In the LPA scheme, we set instead $\eta=0$ in the fixed point equation, but we will still use \eqref{eq:eta-init} to give an estimate of the scaling dimension at the fixed point in the LPA. Reiterating such procedure -- i.e.\ plug back the fixed-point value of $\eta$ in the LPA equations, search again for fixed points, evaluate the new $\eta$, and so on -- leads to results equivalent to the LPA', when the iteration converges.
Notice that the LPA and LPA' differential equations only depend explicitly on $g$ through $g^2$, hence the solutions are invariant under the transformation $g\to -g$, $h\to -h$.

The resulting coupled differential equations are nonlinear in both the LPA and the LPA' schemes and in general do not have known analytical solutions.
Several strategies can be employed to circumvent the problem such as employing perturbative techniques, expanding the potentials in a polynomial truncation, solving analytically in the large-$\vphi$ limit or solving numerically while keeping the potential fully functional.
In the following sections, we will present a combination of different methods with the goal of better understanding $\cP\cT$-symmetric single scalar universality classes and clarifying the limitations of the LPA and LPA'.

\section{Perturbative treatment: $\eps$-expansion}
\label{sec:perturbation}

In this section our objective is to discuss the perturbative solutions of the local potential approximation of the Wetterich equation \eqref{eq:wetterich-flow}. We follow and generalize a procedure, illustrated previously in \cite{ODwyer:2007brp} for the Polchinski equation \cite{Polchinski:1983gv}, that allows to construct a consistent $\eps$-expansion of multicritical models below their upper critical dimensions. We discuss the procedure in much more detail in appendix \ref{sec:osborn} using Polchinski equation
because it is structurally simpler and therefore more transparent, while here we summarize the main steps and concentrate on the more peculiar aspects of the application to the Wetterich local potential and on the special case of the Lee-Yang model.\footnote{See \cite{Litim:2005us,Morris:2005ck} for a detailed comparison of the Polchinski and Wetterich equations in the LPA.}

We start with Eqs.\ \eqref{eq:realode}, that we write in the form
\ba\label{eq:realode-2}
 k\partial_k v &= -d \, v(\vphi) + \frac{d-2+\eta}{2}\, \vphi\, v'(\vphi) +  \(1- \frac{\eta}{d+2}\)\frac{c_d}{1+v''(\vphi)} \,,
 \\
 \eta &= c_d \frac{v'''(\vphi_0)^2}{\(1+v''(\vphi_0)\)^4} \,.
\ea
Notice that $k\partial_k v \neq 0$ for $v(\vphi)=0$, implying that
even the Gaussian solution has a nonzero constant contribution. We can however factor
it out by considering the constant shift $v(\vphi) = v_0 + \tilde{v}(\vphi)$
with
\ba
 v_0=\frac{1}{d}\left(1-\frac{\eta}{d+2}\right) c_d \,,
\ea
in which case the running for the shifted potential is
\ba\label{eq:realode-3}
 k\partial_k \tilde{v} &= -d \, \tilde{v}(\vphi) + \frac{d-2+\eta}{2}\, \vphi\, \tilde{v}'(\vphi) + c_d \(1- \frac{\eta}{d+2}\)\(\frac{1}{1+\tilde{v}''(\vphi)} -1\) ,
\ea
while the form of $\eta$ of eq.\  \eqref{eq:realode-2} is unchanged under the simple replacement of $v$ with $\tilde{v}$ as it depends only on derivatives of the potential. Notice that this operation does not coincide with factoring out the zero-point energy, which would correspond to the minimum of $v(\vphi)$ rather than the limit $v(\vphi)\to 0$, and, in fact, solutions $\tilde{v}$ will have nonvanishing minima.

The shift simply ensures that Gaussian solutions of $k\partial_k\tilde{v}=0$ are $\tilde{v}=0$.
Assuming that we are close to such Gaussian regime, we may expand the nonlinear parts of Eqs.\ \eqref{eq:realode-2} and \eqref{eq:realode-3}
\begin{eqnarray}\label{eq:realode-4}
 k\partial_k \tilde{v} & \approx & -d \, \tilde{v}(\vphi) + \frac{d-2+\eta}{2}\, \vphi\, \tilde{v}'(\vphi) - X_1 \tilde{v}''(\vphi)+X_2 \tilde{v}''(\vphi)^2 -X_3 \tilde{v}''(\vphi)^3+\cdots ,
 \\
 \eta & \approx & \left(Y_0 - Y_1 \tilde{v}''(\vphi_0) +\cdots \right) \tilde{v}'''(\vphi_0)^2 ,
\end{eqnarray}
and the dots contain higher powers of $\tilde{v}''(\vphi)$.
In a regime close to the Gaussian fixed point it is natural to choose $\vphi_0=0$.
The coefficients $X_l(d,\eta)$ and $Y_l(d,\eta)$ carry the explicit dependence on the cutoff and in general they may depend on both the dimension $d$ and the anomalous dimension $\eta$, so the latter is determined implicitly by the second equation.
For the cutoff chosen in this paper, the constants are
\begin{eqnarray}\label{eq:XY-coefficients}
    X_l(d,\eta) =  \(1- \frac{\eta}{d+2}\) c_d , \qquad  Y_l(d,\eta)=    \frac{(l+1)(l+2)(l+3)}{6} c_d \,.
\end{eqnarray}
For an optimized cutoff, the $X_l$ coefficients do not actually depend on $l$, and the $Y_l$ coefficients do not depend on $\eta$, although they could for a more complicated cutoff.
Most of the considerations of this section are independent of the choice of the cutoff,
as long as the first few coefficients in \eqref{eq:XY-coefficients} are positive, which, for our choice, is true as long as $\eta<d+2$. 
Other cutoffs that have been used in the literature can be shown to also verify the positivity of the first few coefficients. The solutions will of course be scheme dependent.

We now assume the more stringent requirement that $\eta > 2-d$, i.e.\ $2\Delta =d-2+\eta >0$.
This is certainly possible perturbatively and, given that for a  solution near the Gaussian one we expect $\eta\approx0$, we are in practice restricting to $d>2$.
Then we can perform the rescaling
\ba
 x = \frac{1}{2}\(\frac{d-2+\eta}{X_1}\)^{1/2}\vphi, \qquad\qquad w(x) = \tilde{v}(\vphi),
\ea
so that, up to a multiplicative constant, $k\partial_k \tilde{v}$ becomes
\begin{eqnarray}\label{eq:wetterich-rescaled}
  k\partial_k \tilde{v} & \propto &\frac{2d}{d-2+\eta} w(x) - x w'(x) + \frac{1}{2} w''(x) - F_2 w''(x)^2 +F_3 w''(x)^3 +\cdots ,
 \\
 \eta &=& \left(E_0 -E_1 w''(x_0)+\cdots\right)w'''(x_0)^2,
\end{eqnarray}
where the rescaled coefficients are defined
\ba
 F_n(d,\eta) = \frac{2\left(d-2+\eta\right)^{n-1}}{4^n (X_1)^n  }X_n \,, & \qquad\qquad & E_n(d,\eta)=\frac{\left(d-2+\eta\right)^{n+3}}{4^{n+3} (X_1)^{n+3} } Y_n \,,
\ea
and remain positive under the same assumptions.
It is straightforward to compute the terms contained in the dots by expanding \eqref{eq:realode-3} in $\tilde{v}''(\vphi)$ and redefining the coefficients appropriately.
As a result of the shift, eq.\  \eqref{eq:wetterich-rescaled}
admits the Gaussian solution $w(x)=0$ and we can try to determine under which conditions the potential is close to this Gaussian fixed point.
Equation \eqref{eq:wetterich-rescaled} should be compared with the Polchinski counterpart given in \eqref{eq:polchinski-rescaled} of appendix \ref{sect:osborn-polchinski}
that differs only in the nonlinear ``interaction'' part.

The Gaussian fixed point is $w(x)=\tilde{v}(\vphi) = 0$,
which implies that also $\eta = 0$, so it must be possible to find solutions of eq.\  \eqref{eq:wetterich-rescaled} that are close to the Gaussian one as corrections of the linearized equation
\be\label{eq:wetterich-linearized}
 \frac{2d}{d-2} w(x) - x w'(x) + \frac{1}{2} w''(x) \approx 0
 \,,
\ee
which now coincides with eq.\  \eqref{eq:polchinski-linearized} given in section \ref{sect:osborn-polchinski}
for the case of the Polchinski LPA.
As discussed in Ref.\ \cite{ODwyer:2007brp} and in appendix \ref{sec:osborn}, global solutions of the linearized equation that are polynomially bounded are possible only if the coefficient of the first term is a nonzero natural number. In such case, the solutions are proportional to the Hermite polynomials
\be
 w(x) \approx  \lambda \, H_n(x) \,, \qquad \qquad n = \frac{2d}{d-2} \in \mathbb{N} \,,
\ee
for some undetermined constant $\lambda$, given that the equation is linear.
The use of Hermite polynomials suggests us to introduce a natural inner product for polynomially bounded functions
\be\label{eq:w-norm}
 \left(f,g\right) \equiv \int {\rm d}x \, {\rm e}^{-x^2} f(x) \, g(x) \,.
\ee
We can then use inner product to project our equations on the natural basis of polynomials \cite{ODwyer:2007brp}, and we can use the associated norm $\lVert f \rVert \equiv  \left(f,f\right)$ to give meaning to the smallness condition.
The smallness requirement is satisfied only approaching critical dimensions such that $d \to d_n= \frac{2n}{n-2} $ for polynomial solutions. Recall also that $H_n(x) = (2x)^n +\cdots$ at large values of $x$, so the leading polynomial contribution to the potential is $v(\varphi) \sim \vphi^n$. Furthermore, the Gaussian fixed point is recovered for $\lambda \to 0$.

In this paper we are interested in odd solutions that are not bounded from below.
We perform the replacement $n\to 2n+1$, which reveals the critical dimensions of the odd models $d_{2n+1} = \frac{4n+2}{2n-1}=6, \frac{10}{3}, \cdots$.
The $\epsilon$-expansion is then defined below $d_{2n+1}$ as
$$
 d = d_{2n+1} -\epsilon \,.
$$
To set up the expansion properly, we must use the inner product \eqref{eq:w-norm}. A careful analysis reveals that $w(x)$ admits an expansion in semi-odd powers of $\eps$, while $\eta$ in powers of $\eps$ itself. We expound how the expansion comes about in greater detail in appendix \ref{sect:osborn-polchinski} where we also address the case of even models. The starting point for the Wetterich and Polchinski equations is the same, i.e., Eqs.\ \eqref{eq:wetterich-linearized} and \eqref{eq:polchinski-linearized} respectively, while they differ for the nonlinear ``interactions'' present in Eqs.\ \eqref{eq:wetterich-rescaled} and \eqref{eq:polchinski-rescaled}. Here, we mostly concentrate on the special case $n=1$, that is, the Lee-Yang model, which is the most important for this paper.

\paragraph{The Lee-Yang solution.}
To solve \eqref{eq:wetterich-linearized} in the case $n=1$, we first make the ansatz
\begin{eqnarray}\label{eq:w-ansatz-main}
 w(x) &=& \eps^{1/2} \lambda  H_3(x) + \eps \lambda ^2 \sum_{q=0}^\infty \alpha_{q} H_q(x) + O(\eps^{3/2}) ,
 \\
 \eta &=& \eps \lambda^2 \eta^{(1)} + O(\eps^{3/2}) \,,
\end{eqnarray}
in which we have rescaled the constant $\lambda$ by $\eps^{1/2}$. The product $\eps^{1/2} \lambda$ could be thought of as the fixed point coupling in the $\eps$-expansion (compare \eqref{eq:w-ansatz-main} with \eqref{eq:w-ansatz} for the case $n=1$).
By construction the ansatz solves eq.\  \eqref{eq:wetterich-linearized} for $n=1$, so it solves
\eqref{eq:wetterich-rescaled} with $d = 6 -\eps$
at the leading order in $\eps^{1/2}$, but $\lambda$ is still arbitrary.

We begin by projecting the equation that determines $\eta$ on $H_0(x)=1$, i.e., interpreting it as the beta function of a field dependent wavefunction renormalization $k\partial_k z =- \eta z(\varphi)$ (we elaborate more on this in section \ref{sect:polchinski-odd-beyond}). We compute $\left(H_0 , k\partial_k z\right)=\left(1 , k\partial_k z\right)=0$ using the inner product \eqref{eq:w-norm}.
The solution to this equation is
\be
 \eta^{(1)} = E_0 2^8 3^2
 \,.
\ee
Notice that we solve the equation for fixed coefficients $E_l$ and $F_l$, which are themselves functions of $d$ and $\eta$, so indirectly they are functions of $\eps$. This is not a problem because we can substitute their $\eps$ series at the end and re-expand the final result.

In order to solve the equation for $w(x)$ at order $\eps$, we first perform the projection
$\left(H_p, k\partial_k v\right)=0$ for arbitrary $p$.
The computation requires the integral of three Hermite polynomials, whose coefficients must sum to an even number and satisfy a triangular inequality. As a consequence,
the natural number $q$ that is summed over in eq.\  \eqref{eq:w-ansatz-main}
is bounded \emph{and} even. The solution is
\be
 \alpha_{2p} = 
 \frac{ F_2  3^2 2^{7-p} }{(3-2 p)
   p!^2  (1-p)!}
 \,,
\ee
and confirms the bound $0\leq p \leq 1$ using analytic continuation, while $\alpha_{2p+1}=0$.

The constant $\lambda$ is determined by projecting $\left(H_{3},k\partial_k v\right)=0$ and solving at order $\eps^{3/2}$ (recall that it already is a solution to order $\eps^{1/2}$). Inserting the previously determined $\eta^{(1)}$ and $\alpha_{2p}$ and summing over the range $p=0,1$
gives a cubic polynomial with two complex conjugate roots
\be
 12 \sqrt{\pi} \lambda (1 + 6912 (8 F_3 - E_0) \lambda^2)=0\,,
\ee
with nontrivial solutions
\be
 \lambda = \pm i \frac{1}{48 \sqrt{3} \sqrt{E_0-8 F_2}} \,.
\ee
For general $n$ the above procedure can be repeated, but gives considerably more complicate solutions.
The result is qualitatively similar to the Polchinski case given in Sects.\ \ref{sect:polchinski-odd} and \ref{sect:polchinski-odd-beyond}. In providing
the fixed point solution for $\lambda$, we have implicitly made the assumption that $E_0-8F_2>0$.
This can be argued in general, because in $d=6$ the constants $X_3(d=6,\eta=0)$ and $Y_0(d=6,\eta=0)$
are actually scheme-independent, while $X_2(d=6,\eta=0)>0$ for any reasonable cutoff function.\footnote{%
A simple way to see it is to notice that the integrals corresponding to $X_3$ and $Y_0$ scale as $k^{d-6}$, which means that they are those responsible for the one-loop logarithmic divergences when integrating the flow of the effective action. See for example appendix C of \cite{Codello:2017hhh}.
}
The final solution for the $n=1$ case in $d=6-\eps$ is
\ba\label{eq:perturbation}
 v(\vphi)&= 
 \frac{1}{2304 \pi ^3}
 +i \eps^{1/2}\(\frac{4\sqrt{2}}{3\sqrt{3}}\pi^{3/2}\vphi^3-\frac{\vphi}{96 \sqrt{6} \pi^{3/2}}\)
 -\eps\, \(\frac{\vphi^2}{18}-\frac{29-12\gamma+12\ln(4\pi)}{55296\pi^3}\)
 +O(\eps^{3/2})
 \\
 &\simeq 1.4\cdot 10^{-5}
 +i \eps^{1/2}  \left(  6.1 \, \vphi^3-7.6\cdot 10^{-4}\,\vphi\right)
 -\eps \left(5.6 \cdot 10^{-2} \,\vphi^2-3.1\cdot 10^{-5}\right)
 +O(\eps^{3/2})
\ea
with the anomalous dimension $\eta=-\frac{\eps}{9}$ which matches the universal result from standard perturbation theory.

Corrections to the complex conjugate pair of solutions can, in principle, be iterated to further orders by incorporating in the ansatz \eqref{eq:w-ansatz-main} more and more terms. The general structure reveals that semi-odd powers of $\eps$
multiply odd Hermite polynomials and have purely imaginary coefficients,
while the integer powers of $\eps$ multiply even Hermite polynomials
with real coefficients. This agrees with the expected symmetry under the  $\cP\cT$ transformation: $v(\varphi) \to v^*(-\varphi)$.

\paragraph{The first multicritical solution.}
Before discussing some general properties of the multicritical family, we give the $i\varphi^5$ multicritical solution in $d=\frac{10}{3}-\epsilon$ dimensions. The final expression can be displayed analytically in analogy with \eqref{eq:perturbation}, but the final form is very complicated and not particularly illuminating. For simplicity we give it numerically:
\begin{eqnarray}\label{eq:perturbation5}
  v(\vphi)&\simeq&
  2.9\cdot 10^{-3}
  + i \epsilon^{1/2}  \left( 1.2\cdot 10^{-1}\, \vphi ^5- 1.7\cdot 10^{-2} \, \vphi^3 +3.8\cdot 10^{-4} \, \vphi \right)
  \nonumber \\&&
   +\epsilon \left(8.1 \cdot 10^{-2}\, \vphi ^6-2.8 \cdot 10^{-2}\, \vphi
   ^4+1.6 \cdot 10^{-3}\, \vphi ^2+5.7 \cdot 10^{-3}\right)
   +O(\eps^{3/2}) \, ,
\end{eqnarray}
with anomalous dimension $\eta = -\frac{\eps}{4797}$.

\paragraph{Properties of the general multicritical cases.} There are also some crucial observations that relate to the $\eps$-expansion that can be formulated at this point. 
We can show that the solutions fail for some large $x$ at finite $\epsilon$ and we do it for the general case.
First we show the general structure of the solutions in $d=d_{2n+1}-\epsilon$ dimensions: equation \eqref{eq:w-ansatz-main} is replaced by
\begin{eqnarray}\label{eq:w-ansatz-main-general}
 w(x) &=& \eps^{1/2} \lambda_n  H_{2n+1}(x) + \eps \lambda_n ^2 \sum_{p=0}^{2n-1} \alpha_{2p} H_{2p}(x) + O(\eps^{3/2}) \,,
\end{eqnarray}
which is similar to the Polchinski's case given in eq.\  \eqref{eq:w-ansatz} of the appendix and in which we already inserted the fact that the subleading corrections are even and there is a bound on the index $p$ due to the integration of Hermite polynomials: $0\leq p \leq 2n-1 $.\footnote{%
Notice that the bound is more stringent for the Wetterich equation, because for the Polchinski equation it is $0\leq p \leq 2n $.}
To see where perturbation theory fails, recall that $H_{2n+1}(x) \sim x^{2n+1}$ for large $x$, in which case
we can compare the leading and subleading contributions to see when they are of the same order in terms of $\eps$ and $x$.
Using the odd part of the potential, the $\eps^{3/2}$ correction to eq.\  \eqref{eq:w-ansatz} can be shown to include the highest polynomial $H_{6n-1}(x)$, so the subleading corrections are of the same order as $H_{2n+1}(x)$
when $\eps^{3/2}H_{6n-1}(x) \sim \eps^{1/2} H_{2n+1}(x)$, resulting in 
$\eps \, x^{4n-2} \sim 1$. Thus for any finite value of $\eps$ any nonperturbative solution must be complemented by the appropriate large field analysis, which we touch upon in the next section.
We also notice that the polynomial at order $\eps$ grows as $x^{4n-2}$, i.e.\ with one power less than the  polynomial at order $\eps^{1/2}$. Therefore, when $x$ approaches the breaking point of the perturbative series, $x\sim (\f{1}{\eps})^{\f{1}{4n-2}}$, the odd part scales like $ (\f{1}{\eps})^{\f{2n+1}{4n-2}-\f12}= (\f{1}{\eps})^{\f{1}{2n-1}}$, while the even part scales like $ (\f{1}{\eps})^{\f{4n-2}{4n-2}-1}= 1$, and thus the latter is subleading.\footnote{
Whether this behavior is reflected order by order in perturbation theory it is not obvious, as the analysis requires bounds that come from the integrals of several Hermite polynomials. It would be interesting to verify it in more generality.
}
Hence later on this will provide further support to our claim that in the large-field the real part of the potential can be neglected.

Lastly, we notice that even if the critical dimensions $d_{2n+1}$ approach $d=2$ in the limit $n\to \infty$, the expansions do not converge (not even weakly) when $\eps$ is such that $d$ approaches two, i.e.\ $\eps=\eps_n \equiv d_{2n+1}-2 \sim \f{2}{n}$. Neglecting $\eta$ for a moment, the rescaling $x \sim (d-2)^{1/2} \varphi$ is not analytic when we approach $d=2$, and the entire construction cannot be appropriately set up. In fact, the radius of validity of the expansion $x\sim (\f{1}{\eps_n})^{\f{1}{4n-2}}$ becomes of order one as $n\to \infty$, meaning that the perturbative solutions breaks down very soon in $x$, and we inevitably require a nonperturbative framework.
This actually happens in general when the scaling dimension of $\varphi$, $\Delta = \frac{d-2+\eta}{2}$, approaches zero.
In the case of even models, this problem is mitigated by the fact that $\eta>0$
as long as $d>2$, so $\Delta>0$ above $d=2$. However, in the case of the odd models, we have $\eta <0$
and we expect that by lowering $d$ the value of $\Delta$ will approach zero and become negative already at some $d>2$. This problem will manifest and be addressed numerically in section \ref{sec:numerical}.

To close this section, let us mention that the method of Ref.\ \cite{ODwyer:2007brp} can also be used to compute the spectrum of critical solutions. In appendix \ref{sec:osborn} we give an example of such computation for both even and odd models using the Polchinski equation, which gives relatively more compact final results.


\section{Analytic properties of the fixed-point equations}
\label{sec:analytic}

Before moving beyond the perturbative domain and performing numerical integrations of the equations \eqref{eq:coupled}, it is useful to figure out as much as possible of their analytic properties, such as the structure of singularities and asymptotic expansions of the solutions, in order to have sufficient elements for interpreting the numerical results. 
The asymptotic analysis is also motivated by the results of section \ref{sec:perturbation}, where we saw that at finite $\eps$ the perturbative solution breaks down beyond some finite value of $\vphi$, and thus we need a separate analysis of the large-$\vphi$ behaviour of solutions.

We start by rewriting the equations in a simplified form that highlights their dependence on only one parameter.
Assuming $\eta\neq d+2$, that is $\Delta\neq d$, we divide \eqref{eq:coupled} by $d$ and do the following rescaling of field and potential:
\be
\vphi = \(\f{c_d \(1-\frac{\eta}{d+2}\)}{d}\)^{\f12}\, \bar{\vphi} , \qquad v(\vphi) = \f{c_d \(1-\frac{\eta}{d+2}\)}{d} \, \bar{v}(\bar{\vphi}) .
\ee

The equations simplify to 
\be\begin{systeme}\label{eq:coupled-rescaled}
\bar{u}- \gamma\,  \bar{\vphi}\, \bar{u}' = \, \frac{1+\bar{u}''}{\(1+\bar{u}''\)^2+\(\bar{h}''\)^2}\\
\bar{h}- \gamma\,  \bar{\vphi}\, \bar{h}'= - \, \frac{\bar{h}''}{\(1+\bar{u}''\)^2+\(\bar{h}''\)^2}
\end{systeme}
\ee
where we defined $\gamma = \Delta/d$. 
The initial conditions will be given in the same form as before, namely
\be \label{eq:init-cond-rescaled}
\bar{u}'(0) = 0,\quad \bar{u}''(0)= \lambda_2 \in \R, \quad \bar{h}(0)=0, \quad \bar{h}'''(0)=\bar{g}\in \R.
\ee
However, notice that while the second derivative of the potential is invariant under the rescaling, and hence $\lambda_2$ is the same as before, the third derivative is not invariant, hence $\bar{g}$ differs from the $g$ in \eqref{eq:init-cond} by a multiplicative factor:
\be
\bar{g} = \(\frac{c_d (1-\frac{\eta}{d+2})}{d}\)^{\f12} \, g .
\ee

In the rescaled variables, the only difference between LPA and LPA' is that in the former $\gamma$ is only a function of $d$, while in the latter it is a function of $d$ and $\bar{g}$.
One slight complication brought in by the rescaling is in the expression of $\eta$ in terms of the new initial conditions \eqref{eq:init-cond-rescaled}, because \eqref{eq:eta-init} becomes a second order equation for $\eta$, 
\be\label{eq:eta-rescaled-eq}
\eta = - \frac{d (d+2)}{d+2-\eta} \frac{\bar{g}^2}{(1+\lambda_2)^4} ,
\ee
whose solutions are
\be \label{eq:eta-rescaled}
\eta_{\pm} = \frac{d+2}{2} \(1\pm \sqrt{1+\frac{4 \, d\, \bar{g}^2}{(d+2) (1+\lambda_2)^4}} \) ,
\ee
and of which we should pick the one that goes to zero at $\bar{g}=0$, that is $\eta_-$, which is also the negative one.
However, in this section we will be mostly using $\gamma$ as a free parameter, thus making no distinction between LPA and LPA', and we will only use this expression for $\eta$ at the end for the $\Delta=0$ case.

In the numerical analysis of section \ref{sec:numerical} we will make the further approximation of discarding the real part.
Besides the obvious practical reason of simplifying the analytical and numerical study, we are motivated also by the observation made at the end of  section \ref{sec:perturbation} that for the multicritical Lee-Yang models the real part is subleading  at large  $\vph$. 
Therefore, we will be dealing with a single equation, that we write explicitly for convenience, dropping from now on the bar on the rescaled variables:
%
\be \label{eq:FP-pureIm}
 {h}- \gamma\,   {\vphi}\,  {h}'= - \, \frac{ {h}''}{1+( {h}'')^2} ,
\ee
with initial conditions $ {h}(0)=0$, $ {h}'''(0)= {g}\in \R$.
For simplicity, and for direct comparison to the next section, from now on we will consider only the single decoupled equation \eqref{eq:FP-pureIm}, and only occasionally comment about the full system \eqref{eq:coupled-rescaled}. 

Let us note some general facts that will be used later on. First, evaluating  \eqref{eq:FP-pureIm} at $ {\vphi}=0$ and using $ {h}(0)=0$, we find $ {h}''(0)=0$. This pattern in fact goes on and all the even derivatives of the potential vanish at the origin, as demanded by $\cP\cT$ symmetry.

Second, taking one derivative of \eqref{eq:FP-pureIm}, and solving for $  {h}'''( {\vphi})$, we find
\be \label{eq:h3}
  {h}''' = \frac{ ((1- \gamma)  {h}' -\g\,   {\vphi}\,  {h}'') (1+( {h}'')^2)^2}{( {h}'')^2-1} .
\ee
Evaluated at $ {\vphi}=0$, and using $ {h}(0)= {h}''(0)=0$, this fixes also $ {h}'(0)=- {g}/(1-\gamma)$.

The equation \eqref{eq:FP-pureIm}, or the system \eqref{eq:coupled-rescaled}, can be studied analytically in the neighbourhood of some given point $ {\vphi}\sim  {\vphi}_0$.
Near an ordinary point, such as the origin  $ {\vphi}_0=0$, the solution is smooth and we can develop the solution as a power series in $( {\vphi}- {\vphi}_0)$, essentially by evaluating higher derivatives from \eqref{eq:h3}.
The analysis of singular points, such as moveable singularities or the point at infinity, is more involved, and we discuss it below, together with the special case $\gamma=0$, for which an analytic (albeit implicit) solution of  \eqref{eq:FP-pureIm} is possible.

\subsection{Movable singularities}
\label{sec:sing}

As a general fact, nonlinear ODEs may lead to movable singularities, i.e.\ singularities whose location depends on the initial conditions. 
Let us check whether the LPA/LPA' equations have any such singularities, and of what type.

We first notice that for real $ {h}( {\vphi})$ the RHS of \eqref{eq:FP-pureIm} takes values in $[-1/2,1/2]$, and  hence so must do also the LHS.
The boundaries of such range are reached for $ {h}''=\pm1$, and we might expect that generic solutions develop singularities at these points.
We will confirm this below.

Assume that a singularity is located at $ {\vphi}= {\vphi}_s$, with $0< {\vphi}_s<\infty$ (by $\cP\cT$-symmetry the same singularity is found also at negative $ {\vphi}$), and that for $ {\vphi}\lesssim {\vphi}_s$ the potential behaves as
\be \label{eq:sing-ansatz}
 {h}( {\vphi}) \sim K ( {\vphi}_s -  {\vphi})^{\a} (1+O( {\vphi}- {\vphi}_s))+ \text{analytic} \;,
\ee
for some constant $K\in\mathbb{R}$, and an exponent $\a\in\mathbb{R}$ that is not a positive integer.
We distinguish two cases:
\begin{enumerate}
    \item $\a<2$, in which case $ {h}''( {\vphi})$ diverges for $ {\vphi}\to  {\vphi}_s$;
    \item $\a>2$, in which case $ {h}''( {\vphi}_s)<\infty$, but some higher derivative diverges.
\end{enumerate}

\paragraph{The $\a<2$ case.}
Let us consider first the $\D\neq 0$ case, i.e.\ $\g\neq 0$ in equation \eqref{eq:FP-pureIm}. 
Using the ansatz \eqref{eq:sing-ansatz} and matching LHS and RHS of \eqref{eq:FP-pureIm} at leading order in $| {\vphi} -  {\vphi}_s|$, imposes that
\be \label{eq:sing-h2}
\a = \f32 \;. 
\ee
Somewhat surprisingly, the leading singular behavior is independent of $\g$ as long as it is not zero.

If instead $\D=0$ ($\gamma=0$), the matching leads to $\a=1$, which is inconsistent with the assumption that $\a$ is not integer.
In this case, we find that the ansatz \eqref{eq:sing-ansatz} should be modified to include a logarithmic behaviour:
\be \label{eq:sing-ansatz-log}
 {h}( {\vphi}) \sim K ( {\vphi}_s -  {\vphi}) \ln| {\vphi}_s -  {\vphi}| + \ldots \;.
\ee

\paragraph{The $\a>2$ case.}
For $\a>2$, the second derivative is finite at the singular point, $ {h}''( {\vphi}_s)<\infty$, but some higher derivative diverges. 
Using \eqref{eq:h3}, we see that if $(1- \gamma)  {h}'( {\vphi}_s) -\g \,  {\vphi}_s\,  {h}''( {\vphi}_s)\neq 0$, then $ {h}'''( {\vphi}_s)$ diverges when $ {h}''( {\vphi}_s)=\pm 1$, that is the point at which the RHS of \eqref{eq:FP-pureIm} reaches the value $\pm1/2$, i.e.\ the boundary of its range for real $ {h}( {\vphi})$. 
In this case we must have $2<\a<3$. Plugging the ansatz $ {h}''( {\vphi}) \sim \pm1+ \a(\a-1)K ( {\vphi}_s -  {\vphi})^{\a-2} $ back into \eqref{eq:h3} and matching powers, we find
\be \label{eq:sing-h3}
\a=\f{5}{2}.
\ee
It is easy to see that no other singular solutions with $\a>3$ are possible: taking further derivatives of \eqref{eq:h3} we find that $ {h}^{(n)}$ is always a fraction involving products of $ {h}^{(n')}$ with $n'<n$ in the numerator and powers of $( {h}'')^2-1$ in the denominator, hence either $  {h}'''( {\vphi}_s)$ is singular or all of the $ {h}^{(n)}( {\vphi}_s)$ for $n\geq 0$ are regular.

\

The two cases above can also be understood by first solving \eqref{eq:FP-pureIm} for $ {h}''$. 
For real $ {h}$, the denominator $1+( {h}'')^2$ is strictly positive, hence \eqref{eq:FP-pureIm} reduces to a quadratic equation in $ {h}''$, and we thus have two solutions: 
\be \label{eq:normal_ode}
 {h}'' = \frac{-1 \pm \sqrt{1-4\, ( {h}- \gamma\,   {\vphi}\,  {h}')^2}}{2\, ( {h}- \gamma\,   {\vphi}\,  {h}')} \equiv F_{\pm}( {h}, {h}').
\ee
This can only be divergent ($\a<2$ case above) at points where $ {h}- \gamma\,   {\vphi}\,  {h}'=0$, if we choose $F_{-}( {h}, {h}')$. However, such choice of sign is incompatible with the initial conditions \eqref{eq:init-cond-rescaled}.
On the other hand, the solution for  $F_{+}( {h}, {h}')$, which is compatible with the initial conditions, can only become non-analytic at points where the argument of the square root (i.e.\ the discriminant) vanishes, and then we are in the $\a>2$ case above.
Therefore, we conclude that for the initial value problem given by  \eqref{eq:FP-pureIm} and \eqref{eq:init-cond-rescaled} the generic singular behaviour is the one with $\a>2$.
In particular, if
\be
1-4\, ( {h}- \gamma\,   {\vphi}\,  {h}')^2 \sim A ( {\vphi}_s- {\vphi}) , \qquad \text{for }  {\vphi}\to {\vphi}_s^- ,
\ee
then by taking one derivative of \eqref{eq:normal_ode} we find that $ {h}'''$ has a $( {\vphi}_s- {\vphi})^{-1/2}$ singularity, in agreement with \eqref{eq:sing-h3}.

Lastly, we notice that if the approach to zero of the discriminant is quadratic,
\be \label{eq:discrim-condition}
1-4\, ( {h}- \gamma\,   {\vphi}\,  {h}')^2 \sim A ( {\vphi}_s- {\vphi})^2 , \qquad \text{for }  {\vphi}\to {\vphi}_s ,
\ee
then $ {h}'''$ has a finite limit for $ {\vphi}\to {\vphi}_s$, but discontinuous. Indeed in this case $ {h}''$ develops a $| {\vphi}_s- {\vphi}|$ cusp.
However, the solution can be extended to a regular one if for $ {\vphi}> {\vphi}_s$ we switch to the solution for the  $F_{-}( {h}, {h}')$ branch. If and when such continuation leads to a global solution, rather than to a singularity for $ {h}''$ (case \eqref{eq:sing-h2}) at a larger value of $ {\vphi}$, is a hard question to answer analytically, but it will become clear in the following that if a global solution exists for $\gamma>0$, then it must cross the singularity at $ {h}''=\pm1$, and the one just described is the only possibility for that to happen.
%

\subsection{Large-field asymptotic behaviour at \texorpdfstring{$\Delta\neq 0$}{Delta not 0}}

We now turn to the asymptotic behaviour at large $| {\vphi}|$, for $\gamma\neq 0$.
Assume that a solution of  the equations  \eqref{eq:FP-pureIm}, with initial conditions at either the origin or some finite (possibly large) value of $ {\vphi}$, reaches infinite $ {\vphi}$.
We then try to solve \eqref{eq:FP-pureIm} in the limit of $ {\vphi}\to +\infty$ (the behavior at $ {\vphi}\to -\infty$ can be obtained by symmetry).
We leave the details of this analysis to Appendix~\ref{app:asymptotics}, where we also discuss the full system \eqref{eq:coupled-rescaled}, and we summarize here only the relevant behaviour for the numerical solutions studied in section \ref{sec:numerical}.

Excluding the case $\gamma\geq 1/2$, i.e. $\Delta\geq d/2$, which we will never meet, and unphysical solutions with real part of the potential unbounded from below, which are only found for initial conditions with $\lambda_2<-1$, the asymptotic behaviour of possible solutions is of the following form:.

\begin{itemize}
\item For $0<\gamma<\frac{1}{2} $, i.e.\ $0<\Delta<\frac{d}{2}$, we have either the isolated Gaussian solution
\be
 {h} =0 ,
\ee
or a nontrivial solution with asymptotic behaviour
\be \label{eq:asymp-posDelta}
 {h}( {\vphi}) \sim B  {\vphi}^\frac{1}{\gamma} - \frac{ \gamma^2}{2 B(1-\gamma)^2} {\vphi}^{2-\frac{1}{\gamma}}+\bigO{ {\vphi}^{4-\frac{2}{\gamma}}} ,
\ee
for some arbitrary parameter $B$. 
The fact that we have only one free parameter, despite the equation being second order, is a signal that any global solution will likely be unstable under small perturbations of the initial conditions at the origin, that is, global solutions are likely isolated.\footnote{See example 4 in section 4.3 of \cite{Bender:1999} for a clear example of this phenomenon.}

We also notice that since $1/\gamma>2$, then $ {h}''\sim  {\vphi}^{\frac{1}{\gamma}-2} \to \infty$ for $ {\vphi}\to\infty$. Given that initial conditions imply $ {h}''(0)=0$, any global solution in this range of $\gamma$ must cross the singularity at $ {h}''=\pm 1$, as anticipated above.


\item For $\gamma<0$, i.e.\ $\Delta<0$: 
\ba \label{eq:asymp-negDelta}
  {h}( {\vphi}) \sim b_1  {\vphi}^\frac{1}{\gamma} \(1 +\bigO{ {\vphi}^{-2}}\)
 + b_2\, e^{\f{\g}{2}  {\vphi}^2} \,  {\vphi}^{-\frac{1}{\gamma} -1}  \(1 +\bigO{ {\vphi}^{-2},e^{\g  {\vphi}^2} \,  {\vphi}^{-\frac{2}{\gamma} -2} }\) ,
\ea
for some arbitrary parameter $b_1$ and $b_2$.
Notice that in this case we have two free parameters, hence a continuum family of global solutions is to be expected, at least in some range of $ {g}$.

\end{itemize}

The case $\gamma=\Delta=0$ is special, and will be discussed below.

\subsection{Representation of the fixed-point equation as a dynamical system}
\label{sec:dyn-syst}

\begin{figure}[H]
\centering
\includegraphics[scale=.27,valign=t]{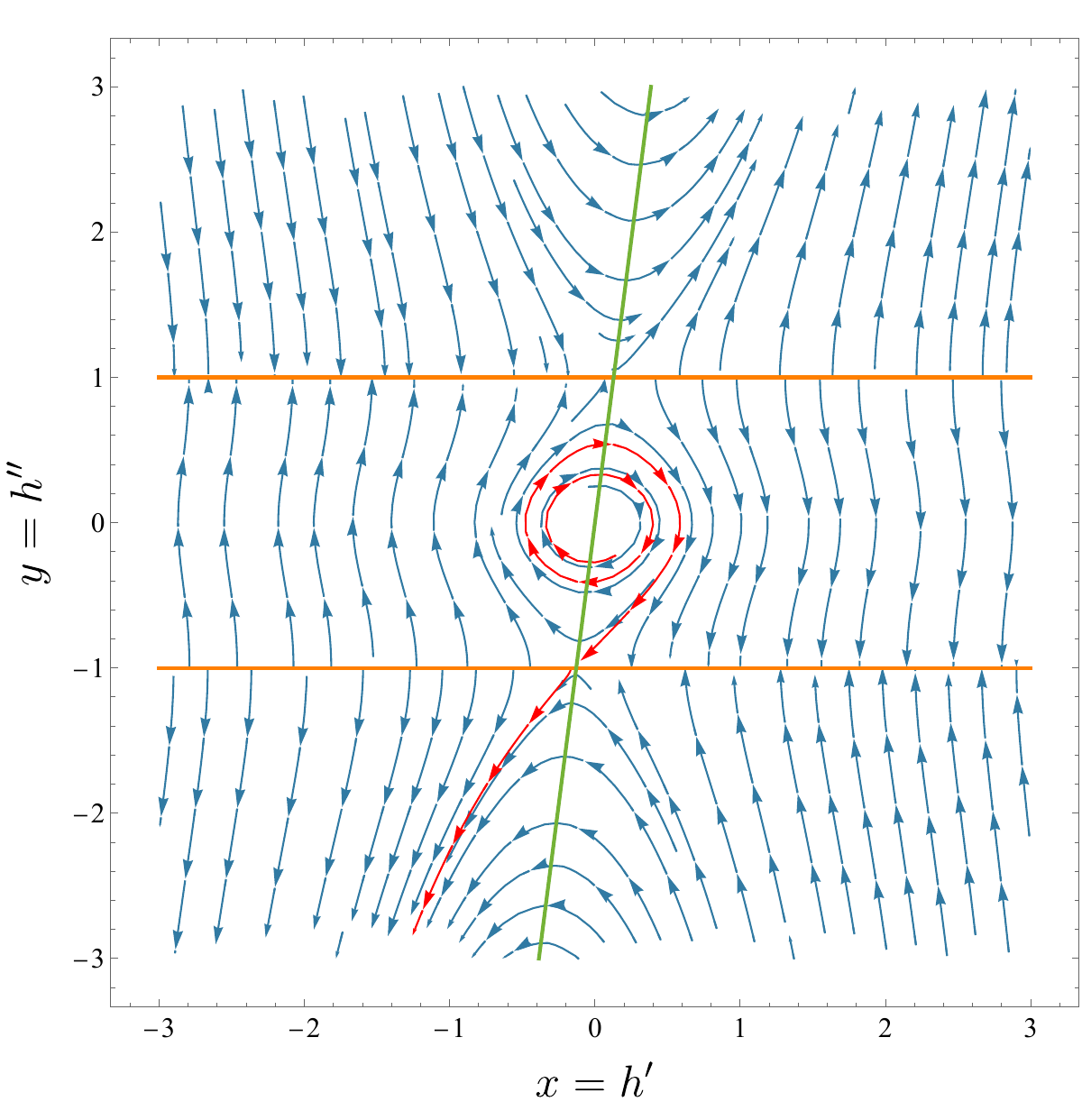}\quad
\includegraphics[scale=.27,valign=t]{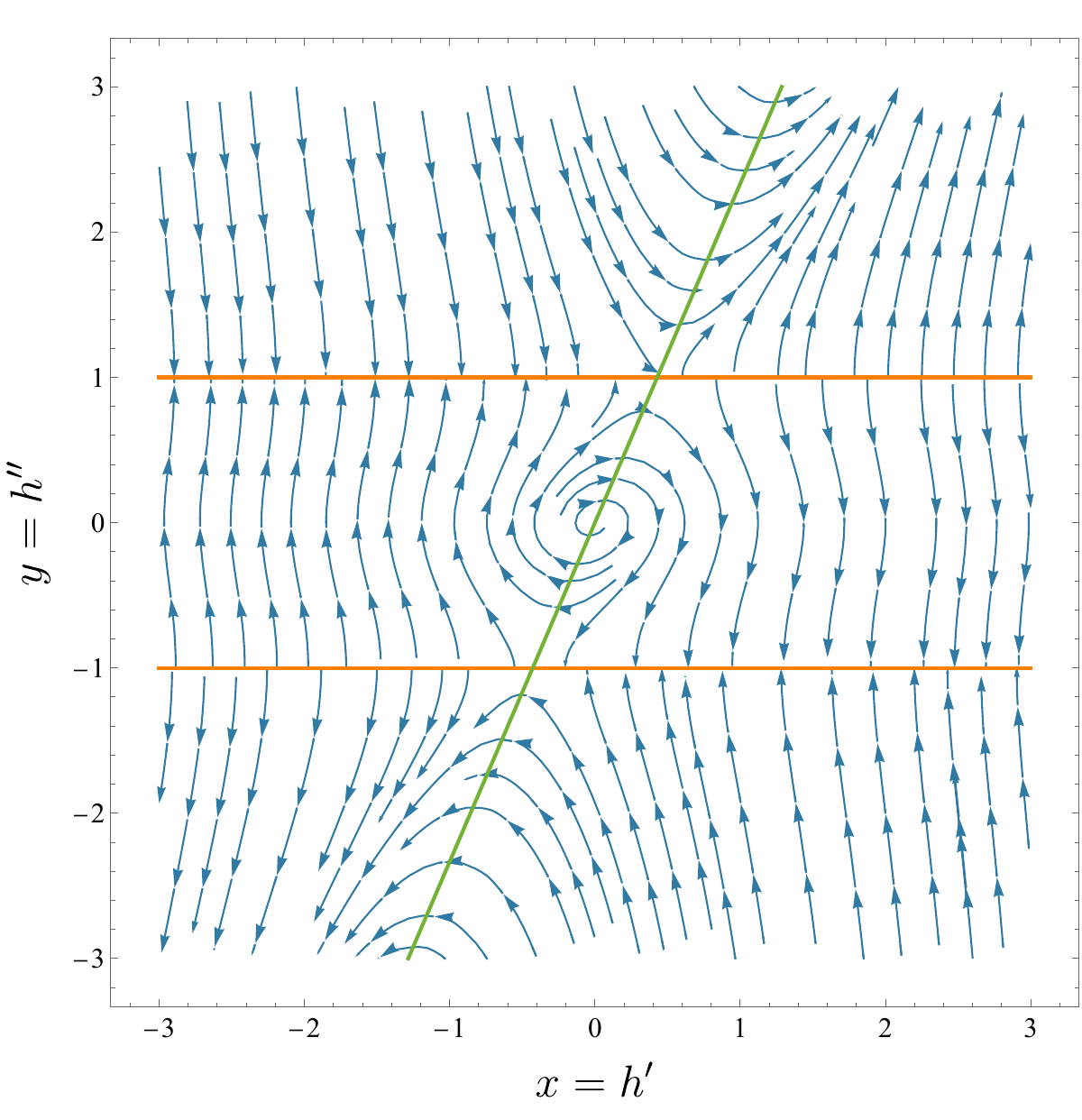}\quad
\includegraphics[scale=.27,valign=t]{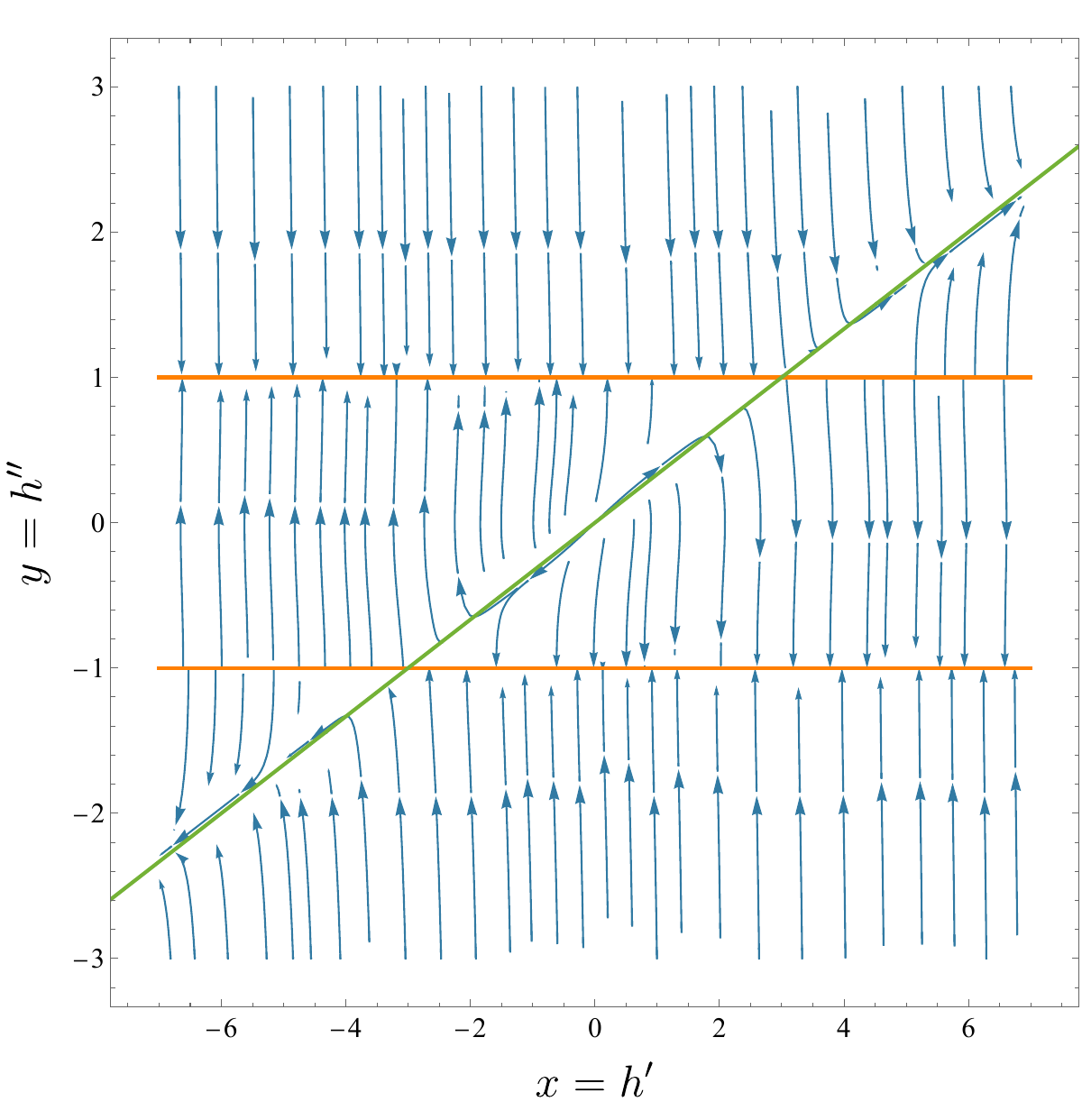}
\caption{Snapshots of the dynamical system \eqref{eq:dyn-syst} with $\gamma=0.3$, for (from left to right) $ {\vphi}=0.3$, $ {\vphi}=1$, and $ {\vphi}=7$.
The horizontal orange lines represent the locus of singularities at $ {h}''=\pm1$.
The green line represents the nullcline \eqref{eq:nullcline}.
In the first panel we have plotted in red a separatrix that we claim corresponds to a solution passing smoothly across the locus of singularities.
}
\label{fig:snapshots-1}
\end{figure}

\begin{figure}[H]
\centering
\includegraphics[scale=.27,valign=t]{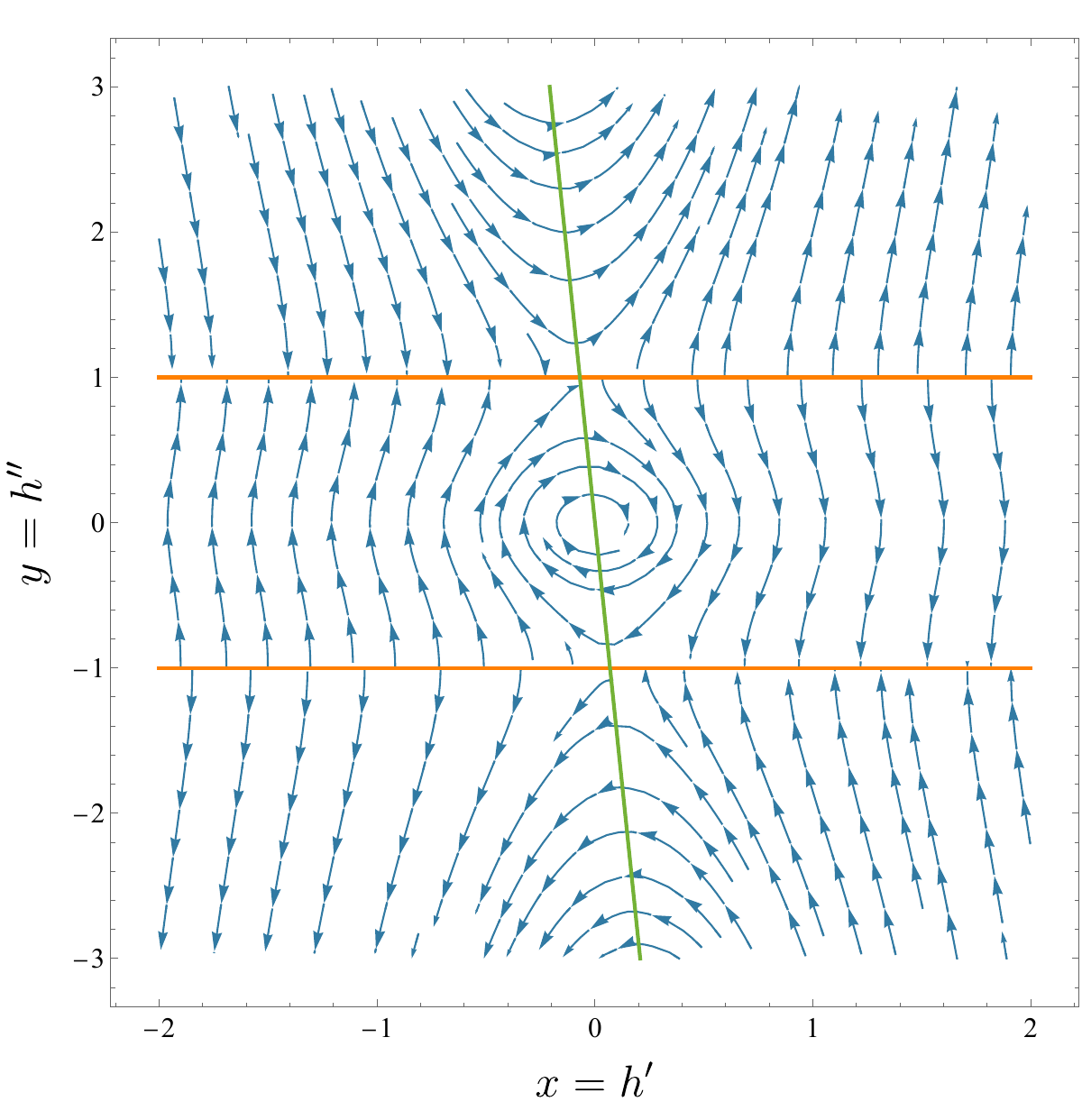}\quad
\includegraphics[scale=.27,valign=t]{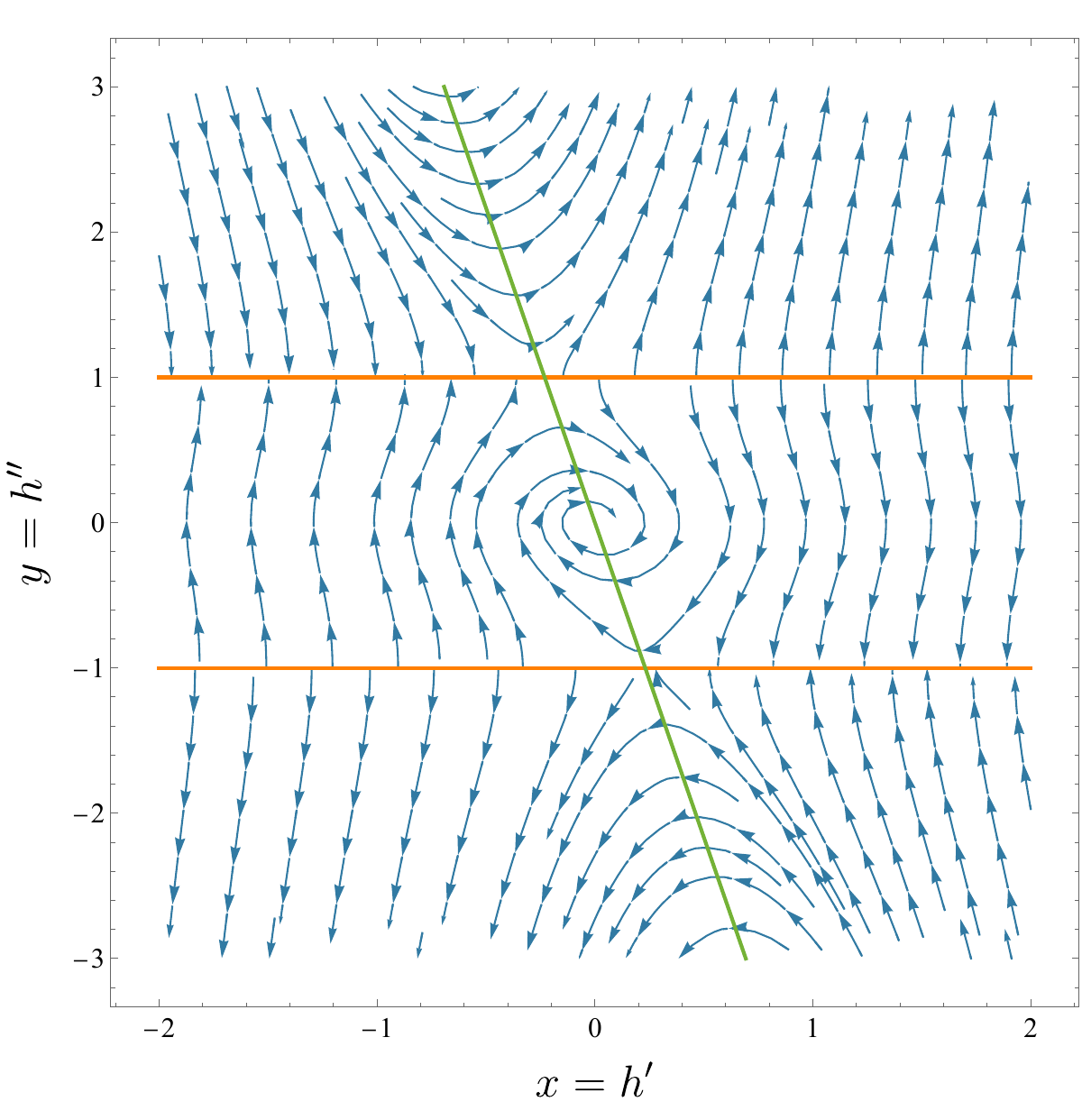}\quad
\includegraphics[scale=.27,valign=t]{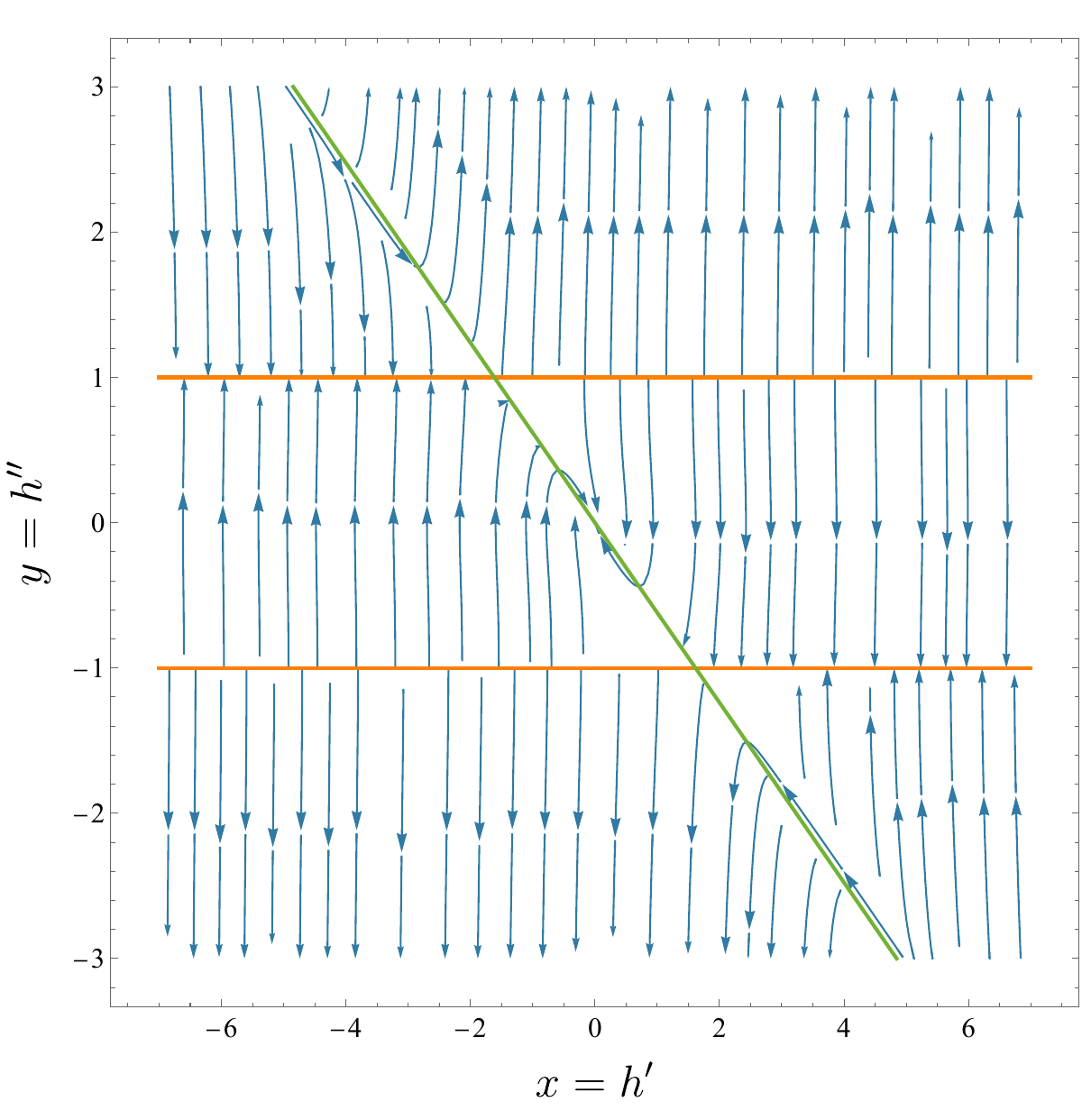}
\caption{Snapshots of the dynamical system \eqref{eq:dyn-syst} with $\gamma=-0.3$, for (from left to right) $ {\vphi}=0.3$, $ {\vphi}=1$, and $ {\vphi}=7$.
The horizontal orange lines represent the locus of singularities at $ {h}''=\pm1$.
The green line represents the nullcline \eqref{eq:nullcline}.}
\label{fig:snapshots-2}
\end{figure}

We have seen that writing the fixed-point ODE as in \eqref{eq:normal_ode} requires a choice of branch.
We can instead avoid choosing a branch if we take one derivative of \eqref{eq:FP-pureIm}, and solve for $  {h}'''$, which appears linearly. This leads to \eqref{eq:h3}, which can also be recast as a non-autonomous dynamical system, if we define $ {h}'( {\vphi})=x( {\vphi})$, $ {h}''( {\vphi})=y( {\vphi})$, thus obtaining:
\ba \label{eq:dyn-syst}
& x' = y \\ 
& y' = \frac{ ((1- \gamma) x -\g\, {\vphi}\, y) (1+y^2)^2}{y^2-1} .
\ea
The system \eqref{eq:dyn-syst} is  non-autonomous because of the explicit dependence on $ {\vphi}$, which plays the role of time.
Some snapshot of the vector field at different values of  $\g$ and $ {\vphi}$ are shown in figures \ref{fig:snapshots-1} and \ref{fig:snapshots-2}.
Although this rewriting of our ODE does not help in solving it analytically, it allows us to deduce a number of useful properties.

\paragraph{Nullclines and trivial fixed point.}
First of all, we identify the nullclines, that is, the curves where either $x'=0$ or $y'=0$.
These are given by the equations $y=0$ and 
\be \label{eq:nullcline}
(1- \gamma) x -\g\, {\vphi}\, y=0,
\ee
respectively.
The former coincides with the locus of initial conditions $ {h}''(0)=0$, $ {h}'(0)=- {g}/(1-\gamma)$.
The latter is the locus of points where the derivative of the discriminant  in \eqref{eq:normal_ode} vanishes.

Fixed points of a dynamical system are obtained when all the nullclines intersect.
The dynamical system \eqref{eq:dyn-syst} has only one fixed point at the origin $x=y=0$, that we recognize as the Gaussian theory in the FRG interpretation of the system.

\paragraph{Stability of the origin.}
We notice from the  figures \ref{fig:snapshots-1} and \ref{fig:snapshots-2} that the stability (not to be confused with RG stability) of the origin depends on the sign of $\gamma$.  
More precisely, from the linearization of  \eqref{eq:dyn-syst}, we obtain the stability eigenvalues
\be
\vartheta_{\pm} = \f{\gamma  {\vphi} \pm \sqrt{ (\gamma  {\vphi})^2 - 4 (1-\gamma) }}{2} .
\ee
At large $ {\vphi}$ (or $(\gamma  {\vphi})^2>4 (1-\gamma)$), the eigenvalues are real, one of them grows in absolute value, $\vartheta_+\sim\gamma {\vphi}$, and the other approaches zero, $\vartheta_-\sim\f{1-\gamma}{\gamma {\vphi}}$. The eigenvector associated to the latter approaches the nullcline \eqref{eq:nullcline}.
Taking into account the nonautonomous nature of the system, one finds that these eigenvalues, rather than leading to usual exponential behaviour of eigenperturbations in autonomous systems,  lead to the asymptotic behaviours $\exp(\f{\g}{2} {\vphi}^2)$ and $ {\vphi}^{\f{1}{\g}-1}$, consistently with \eqref{eq:asymp-negDelta} and the discussion around \eqref{eq:lin-test}.
For $\gamma<0$ these behaviours are consistent with the linearization of the system, and thus the origin is stable, with a spiralling behaviour that disappears for sufficiently large $ {\vphi}$.

The large-time stability of the origin at $\gamma<0$ means that in this case there exist a continuum of global solutions to the LPA equation that asymptote to zero for large field values, in agreement with \eqref{eq:asymp-negDelta} and the counting argument below it. Starting at small $ {\vphi}$, in a neighbourhood of the origin, orbits in the $\{x,y\}$ plane wind around the origin a number of times, and every time they cross the $x\equiv  {h}'=0$ axis the potential $ {h}$ has a minimumum or maximum. As $ {\vphi}$ increases, the spiralling behaviour disappears and $ {h}$ goes to zero monotonically.
This will be confirmed from the numerical integrations of section \ref{sec:numerical}.

Notice that even at $\g<0$ the origin stops being an attractor if we start the flow with initial condition sufficiently far in the $x$ variable: beyond some critical value for the initial condition of $x$, the flow ends on the $y=\pm 1$ singularity.


\paragraph{Singularities.}
The dynamical system \eqref{eq:dyn-syst} has obvious singularities at $y=\pm 1$.
In general these are the singularities for $ {h}'''$ discussed around \eqref{eq:sing-h3}: at positive $\gamma$, for any initial conditions inside the strip $y\in(-1,1)$, except of course the origin, they are reached at a finite value of $ {\vphi}$. For negative $\gamma$, the same happens only for initial conditions outside the basin of attraction of the origin, that at $y=0$ corresponds to a finite interval in $x$.

\paragraph{Separatrices and global solutions.}
In order to further understand the properties of the dynamical system \eqref{eq:dyn-syst}, and its possible global solutions, it is important to consider also its separatrices. Since the system is non-autonomous, separatrices in the vector field at fixed time are not necessarily true separatrices of the full time-dependent system, in particular they are not invariant curves. Nevertheless, they help in qualitatively understanding some of the behaviour that we observe in numerical integrations. 

Consider first the case $\g>0$. It turns out that as we change the initial condition $x(0)=x_0$, $y(0)=0$, we can jump from solutions that hit the singularity at $y=-1$ to  solutions that hit the singularity at $y=+1$, with a corresponding jump in the value of $ {\vphi}_s$. The solution at the jump is what we call a separatrix, and at frozen time it might look like the one pictured in red in the first diagram of figure \ref{fig:snapshots-1}.

Going back to the behaviour in equation \eqref{eq:discrim-condition}, we claim that this happens for solutions that at $\vphi=\vphi_s$ lie precisely on a separatrix of this sort.
As can be seen in figure \ref{fig:snapshots-1}, the separatrix in red is one of the only two curves (at fixed $ {\vphi}$) on which the arrows preserve their orientation on the two sides of the $y=-1$ singularity (the other one, not shown in  figure \ref{fig:snapshots-1}, has arrows in the opposite direction, and it intersects the red separatrix on the $y=-1$ line). Since $y^2-1$ in the denominator in the $y$-component of \eqref{eq:dyn-syst} changes sign across the singularity, the sign of $y'$ can only remain the same if also the numerator changes sign, meaning that it has a first-order zero at $ {\vphi}_s$. Since the numerator of the $y$-component of \eqref{eq:dyn-syst} is proportional to the derivative of the discriminant $1-4\, ( {h}- \gamma\,   {\vphi}\,  {h}')^2$, and the discriminant itself vanishes at the singularity, it follows that \eqref{eq:discrim-condition} must hold when the separatrix reaches the singularity.
In other words, the separatrix intersects the locus of singularities at the same point as the nullcline \eqref{eq:nullcline} does, as indeed can be seen in the plots in figures \ref{fig:snapshots-1} and \ref{fig:snapshots-2}.

As we mentioned above, the important next question is whether the continuation of such solution to the $|y|>1$ region leads to a global or singular solution. 
From the asymptotic analysis, we know that a global solution at $0<\gamma<1/2$ must behave like in \eqref{eq:asymp-posDelta}, and thus it will satisfy 
\be
(1-\gamma)  {h}'( {\vphi}) - \gamma  {\vphi}  {h}''( {\vphi}) \sim \bigO{ {\vphi}^{1-\frac{1}{\gamma}}} ,
\ee
that is, it must approach the nullcline.
Stream plots suggest that this is happening, but we do not have an analytic proof, nor a satisfactory understanding of how this behaviour is associated to one or multiple initial conditions.

For $\g<0$, we can similarly identify a separatrix with a curve that smoothly hits the singularity at $|y|=1$, and then escapes to $|y|>1$. The main difference is that in this case rather than separating different singular behaviours, it separates global from singular solutions: see the discussion above, about the stability of the origin in the $\g<0$ case.

\subsection{Vanishing \texorpdfstring{$\Delta$}{Delta}}
\label{sec:Delta0}

At $\gamma=0$, i.e.\ $\Delta=0$, the equations \eqref{eq:coupled-rescaled} change drastically, because the first derivative term drops out.

While the solution \eqref{eq:unbound-sol} is still valid (and still unphysical), it is easy to see that all other solutions are affected drastically. In particular, the RHS cannot go to zero for $ {\vphi}\to\infty$, because this requires either $ {u}''$ or $ {h}''$ to blow up, which leads to a contradiction with the LHS, which would instead require both $ {u}$ and $ {h}$ going to zero.
Also the GFP solution is still valid, but its linear perturbations have periodic solutions.
These cannot really be considered small with respect to the $ {v}=1$, but we also have no reason to discard them, as they are not in obvious contradiction with the equations.
In fact, periodic solutions are present in this case, as we will now show.

\begin{figure}[H]
\centering
\includegraphics{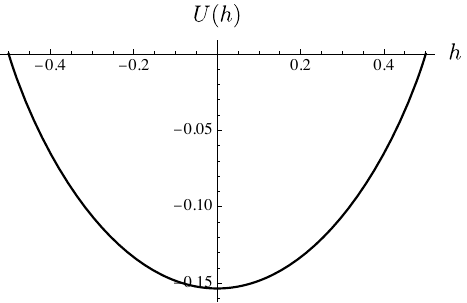}
\caption{The Newtonian potential of equation \eqref{eq:NewtonU}. Outside of the plotted range, $U( {h})$ becomes complex.}
\label{fig:NewtonU}
\end{figure}

\paragraph{Purely imaginary potential.}
In order to simplify the analysis, let us at first discard the real part of the potential. 
The equation \eqref{eq:FP-pureIm} reduces to
\be \label{eq:FP-pureIm-0}
 {h} = - \, \frac{ {h}''}{1+\( {h}''\)^2} ,
\ee
which can be algebraically solved for $ {h}''$. Of the two roots, we choose the one that admits the initial condition $ {h}(0)=0$:
\be \label{eq:NewtonEq}
 {h}'' = \frac{-1+\sqrt{1-4\,  {h}^2}}{2\,  {h}} =-\frac{d}{d\,  {h}} U( {h}).
\ee
This equation has the form of a Newton equation, with Newtonian potential
\be \label{eq:NewtonU}
U( {h}) = \f12 \( \ln\( 1+\sqrt{1-4\,  {h}^2} \) - \sqrt{1-4\,  {h}^2} \) ,
\ee
which is real, bounded and convex for $| {h}|\leq 1/2$ (see figure \ref{fig:NewtonU}).
As in the general case, we choose initial condition $ {h}(0)=0$, $ {h}'(0)=- {g}$ (notice that by equation \eqref{eq:NewtonEq} we have $ {h}'''(0)=- {h}'(0)$).
Introducing the conserved energy
\be \label{eq:energy}
E = \frac{ {h}'( {\vphi})^2}{2} + U( {h}( {\vphi})) = \frac{ {g}^2}{2}+U(0)=\frac{ {g}^2+\ln 2 -1}{2} \geq \frac{\ln 2 -1}{2} ,
\ee
the equation is solved in implicit form as
\be \label{eq:Newt-sol}
 {\vphi} = \f{1}{\sqrt{2}} \int_0^{ {h}} \frac{d\, z }{ \sqrt{E-U(z)} } .
\ee

The minimal energy $E=(\ln 2-1)/2$ corresponds to the stationary point at the bottom of the Newton potential. 
This has $ {h}(0)= {h}'(0)=0$, and thus we recognize it to be the GFP at $d=2$ (because $ {g}=0$ and $\Delta=0$ imply $d=2$).

For $(\ln 2-1)/2 < E < 0$, i.e.\ for $0< {g}^2<1-\ln 2$, the solutions are periodic with period
\be
T = \sqrt{2} \int_{ {h}_1}^{ {h}_2} \frac{d\, z }{ \sqrt{E-U(z)} } ,
\ee
where $ {h}_1$ and $ {h}_2$ are the roots of the equation $U( {h})=E$. Periodic solutions exist and are regular for all $ {\vphi}\in\mathbb{R}$, hence they are all acceptable fixed point solutions. 
Therefore, in the LPA at $d=2$, where $\Delta=0$ for any $ {g}$, the GFP is not an isolated fixed point.
In the LPA', the $\gamma=0$ analysis is not sufficient, because for fixed $d$, we have $\Delta=0$ only at one value of $ {g}^2$.

For $E>0$, $ {h}( {\vphi})$ reaches the value $\pm 1/2$ with finite velocity (i.e.\ finite $ {h}'( {\vphi})$), hence it is pushed towards the complex domain. Moreover, $ {h}'''( {\vphi})$ diverges when $ {\vphi}$ is such that $ {h}( {\vphi})^2=1/4$ and $ {h}'( {\vphi})\neq 0$, as it is seen by deriving once equation \eqref{eq:NewtonEq}.
This behaviour is precisely that of a moveable singularity with finite $ {h}''$, discussed around equation \eqref{eq:sing-h3}.
We conclude that none of the solutions with $E> 0$ is a global solution.
Notice also that the singularity is always reached in finite ``time": in fact, since $-U(z)>0$ for $z\in[-1/2,1/2]$, for $E>0$ we have
\be
 {\vphi}_s =  \f{1}{\sqrt{2}} \int_{0}^{1/2} \frac{d\, z }{ \sqrt{E-U(z)} } <  \f{1}{\sqrt{2}} \int_{0}^{1/2} \frac{d\, z }{ \sqrt{-U(z)} } \simeq 1.3132 .
\ee

The situation at $E=0$ requires some extra care. The points $ {h}( {\vphi})=\pm 1/2$, are reached with vanishing $ {h}'$ and thus finite $ {h}'''$.
Expanding \eqref{eq:Newt-sol} at $E=0$ near $ {h}=1/2$, and plugging it into \eqref{eq:NewtonEq}, we find
\be
 {h}''( {\vphi}) \simeq -1 + \sqrt{2} | {\vphi}_s- {\vphi}| .
\ee
When continued to $ {\vphi}> {\vphi}_s$, this leads to a periodic global solution of class $C^2(\mathbb{R})$, with discontinuous third derivative $ {h}'''$.

\begin{figure}[H]
\centering
\includegraphics[scale=.4,valign=t]{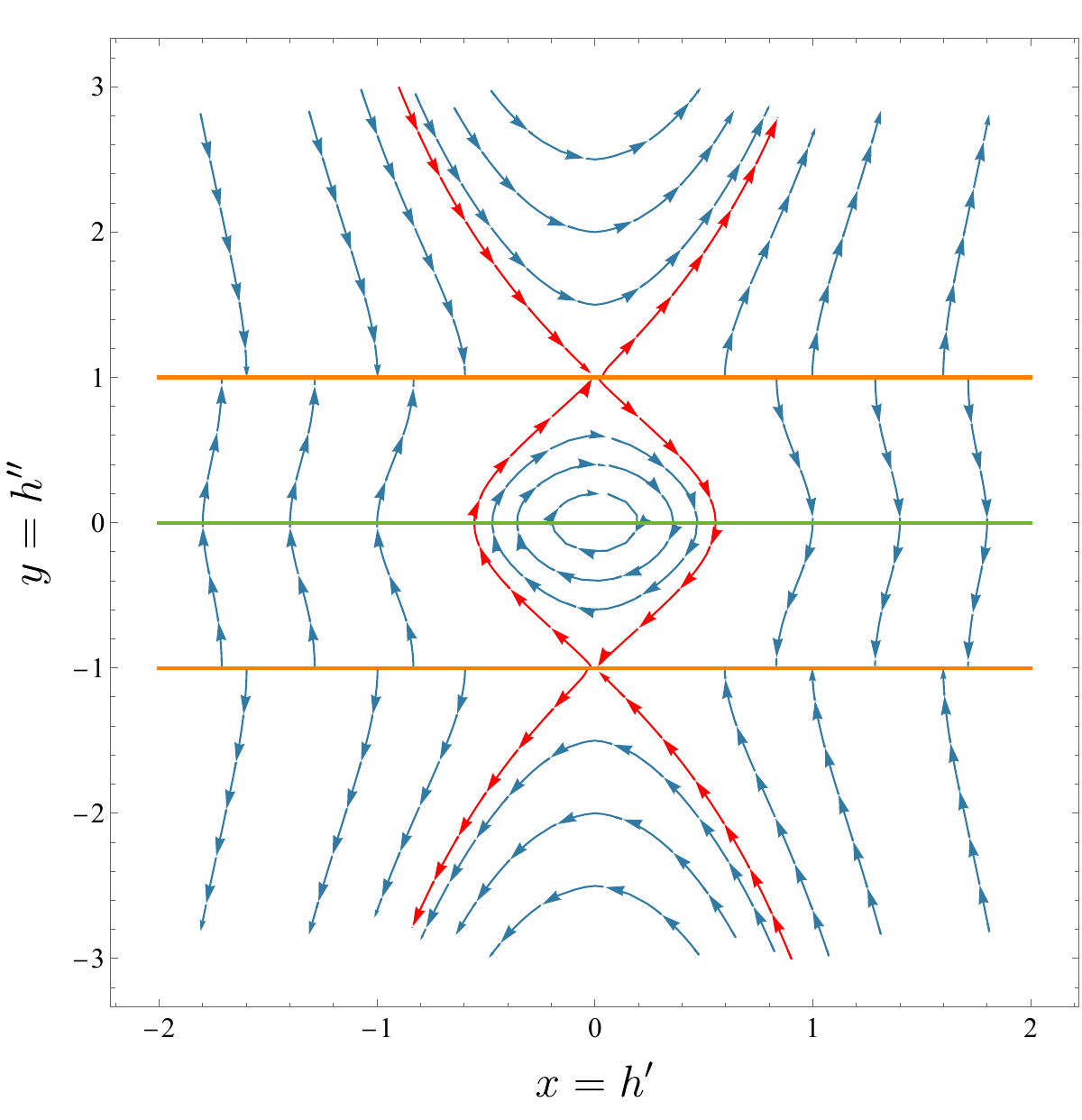}
\caption{Stream plot of the dynamical system \eqref{eq:dyn-syst} with $\gamma=0$.
The green line is the nullcline $y=0$, on which we choose our initial conditions $x(0)=x_0$.
The horizontal orange lines represent the locus of singularities at $ {h}''=\pm1$.
The red lines are the separatrices: starting from a point on the green line, the separatrix corresponds to the $E=0$ solution.
}
\label{fig:snapshot-0}
\end{figure}

Alternatively, we could do the continuation by means of the other branch:
\be \label{eq:h2-branches}
 {h}'' = \begin{cases} \frac{-1+\sqrt{1-4\,  {h}^2}}{2\,  {h}}  &\; \text{for }\,  {\vphi}< {\vphi}_s \\
 \frac{-1-\sqrt{1-4\,  {h}^2}}{2\,  {h}}  &\; \text{for }\,  {\vphi}> {\vphi}_s ,
\end{cases}
\ee
leading to
\be
 {h}''( {\vphi}) \simeq -1 + \sqrt{2} ( {\vphi}_s- {\vphi}) , \quad \text{for }  {\vphi}\simeq  {\vphi}_s. ,
\ee
and continuous derivatives.
This can be understood also by using the dynamical system representation \eqref{eq:dyn-syst}, which bypasses the need to choose a branch, and whose stream plot is shown in figure \ref{fig:snapshot-0}.
The red curves in the figure denote the separatrices, and from their plot it is clear that the periodic orbit obtained by joining the two red trajectories in the $y\in[-1/2, 1/2]$ domain is not smooth, while extending one of them to $|y|>1/2$ we can obtain a smooth curve.

However, such smooth trajectory is not a global solution. Indeed, when $| {h}''|>1$, we replace the Newton potential by
\be
U( {h}) = \f12 \( \ln\( 1-\sqrt{1-4\,  {h}^2} \) + \sqrt{1-4\,  {h}^2} \) ,
\ee
but we can still apply the formula \eqref{eq:Newt-sol} in order to compute the ``time" it takes to go from $ {h}=1/2$, where $ {h}'=0$ and $ {h}''=-1$, to $ {h}=0^+$, where $ {h}''\to-\infty$.
This gives a finite result, meaning that the $ {h}''\sim -\infty$ is reached at finite $\vphi_s$.
This can also be understood from \eqref{eq:h2-branches}, as $ {h}''\to-\infty$ requires $ {h}\to 0^+$, but this combined behaviour cannot be obtained as an asymptotic limit for $\vphi\to\infty$.

In conclusion, at $\gamma=0$ we have:
\begin{itemize}
\item global smooth periodic solutions for $0< {g}^2<1-\ln 2$;
\item singular solutions for $ {g}^2>1-\ln 2$;
\item a global class-$C^2$ periodic solution for $ {g}^2=1-\ln 2$.
\end{itemize}

\paragraph{Dimension at which $\Delta=0$.}

With some simple assumption, we can straightforwardly give an upper bound for the dimension $d_0$ at which $\Delta=0$ for  the fixed point of the $i {\vphi}^3$ theory, without using the FRG. 
We expect that the anomalous dimension of such theory is negative for all $d<6$, and in particular we know that at $d=2$ it coincides with the scaling dimension predicted by the $M(2,5)$ minimal model \cite{Cardy:1985yy}, namely $\eta\vert_{d=2}= \Delta\vert_{d=2}=-4/5$.
Assuming that $\eta$ (and thus $\Delta$) depends monotonically in $d$, as observed from numerical results, and thus that the 2d result provides a lower bound for the anomalous dimension of the Lee-Yang model for $2\leq d<6$, then $d_0>2$ and we must have
\be\label{eq:asympto-vanishingbound}
-\frac{4}{5} < \eta\vert_{d=d_0} = 2-d_0 \quad \Leftrightarrow \quad d_0 <2.8.
\ee

A more precise prediction for $d_0$, within the (purely-imaginary) LPA' approximation of the FRG, can be obtained from the following reasoning, for which we anticipate some of the ideas behind the methods of section \ref{sec:numerical}.

We know that when integrating the ODE \eqref{eq:FP-pureIm} starting from the origin, most of the solutions end up at a singularity of the type discussed in section \ref{sec:sing}. Only at some isolated point $ {g}=g_*(d)$ we expect to find a global solution.
At $d= 6-\epsilon$, with $\epsilon\ll 1$, one finds only one nontrivial global solution, that we can interpret as the fixed point corresponding to the Lee-Yang universality class of the $i {\vphi}^3$ theory, so let us denote the corresponding initial condition by  $ {g}_{\rm LY}(d)$. 
At small $\epsilon$, we find that at such fixed point the scaling dimension is in the range  $0<\Delta<\frac{d}{2}$, i.e.\ we are in the case with asymptotic behaviour \eqref{eq:asymp-posDelta}.
Following such solution to lower $d$, $\D$ decreases and at some value $d_0>2$ it reaches $\Delta=0$, i.e.\ we end in the $\gamma=0$ case discussed above.
Further decreasing the dimension to $2\leq d<d_0$, we find $\Delta<0$, corresponding to the asymptotic behaviour \eqref{eq:asymp-negDelta}.

On the other hand, given that $\Delta=(d-2+\eta(d,g))/2$, for any $d$ there exists a value $ {g}= {g}_0(d)$ at which $\Delta=0$. By definition, $ {g}_{\rm LY}(d)= {g}_0(d)$ only at $d=d_0$.
Therefore, if at $d>d_0$ we only find a global solution at $ {g}= {g}_{\rm LY}(d)$, and not also at $ {g}= {g}_0(d)$, then the solution at  $ {g}= {g}_0(d)$, with $d>d_0$, must correspond to $E>0$ (see \eqref{eq:energy} and discussion below that).
We thus conclude that if $E$ at  $ {g}= {g}_0(d)$ varies continuously with $d$, at $d_0$ we must have $E=0$.
By the definition \eqref{eq:energy}, this is equivalent to $ {g}_0(d_0)=\pm\sqrt{1-\ln 2}$. Using \eqref{eq:eta-init}, which in the rescaled variables that we use in this section becomes $\eta=-\frac{d(d+2)}{2(d-\Delta)} {g}^2$, we find that solving $\Delta=0$ for $d_0$ gives
\be \label{eq:predict-d0}
d_0 = 2 \frac{3-\ln 2}{1+\ln 2} \simeq 2.7249 .
\ee
As we will see in section \ref{sec:numerical}, this prediction matches beautifully with the results from the numerical integration.

\section{Numerical solutions}
\label{sec:numerical}

A functional fixed-point equation such as \eqref{eq:realode} is standardly solved using one of the following two methods:
\begin{itemize}
    \item \textbf{Polynomial truncation}: the functional potential is further expanded in a polynomial $v(\vphi) = \sum_n \lambda_n \vphi^n$ truncated at some finite order $n\le N$. Then, the equation \eqref{eq:realode} translates into an algebraic linear system equivalent to the system of beta functions $\kdk \lambda_n = \beta(\lambda_n)=0$ for each coupling. While the system is now easily solvable analytically or numerically, as the order of truncation rises, the order of the polynomial equations for the beta functions increases alongside, and a number of spurious solutions appeares. It thus becomes harder to identify the physical solution or vary the dimension.
    \item \textbf{Numerical shooting}: the potential is kept fully functional and the differential equation is solved by integrating numerically starting from various initial conditions until either  it reaches a movable singularity or it becomes clear that we have reached a large-$\vphi$ asymptotic regime and we thus have a global solution.
As explained before, global solutions are in general unstable, in the sense that they should appear as isolated points, or occasionally they exist for some interval of the initial condition parameters. 
Isolated global solutions in general require an infinite fine tuning of the initial conditions, therefore their search relies on a graphical method. The location of the singularity is plotted against the initial conditions, and the global fixed point solutions are identified by ``spikes" or discontinuities in the plot, which later we refer to as \textit{spike plot} \cite{Morris:1994ki,Codello:2012sc,Hellwig:2015woa}.  
\end{itemize}

For the Lee-Yang universality class, the method of polynomial truncations has been already used in \cite{Zambelli:2016cbw}, where it was found that it becomes increasingly difficult to use below $d\simeq 4$.

On the other hand, numerical shooting is complicated by the coupled nature of the system \eqref{eq:coupled}.
Indeed in the case of the complex potential, the coupled differential equations \eqref{eq:coupled} are characterized by a two dimensional space of parameters, namely the $\lambda_2$ and $g$ of the initial conditions \eqref{eq:init-cond}.
Numerical shooting is thus more involved, and identifying spikes in a continuous 2$d$ space requires fine-tuning\footnote{See \cite{Defenu:2017dec} for a 2$d$ spike plot in the case of real potential but to second order of the derivative expansion.} and hints from other methods to reduce the searching range.
We have attempted such search, but were unable to identify a clear pattern singling out possible global solutions.

We can partially avoid the limitations of the two methods by considering a combination of them, namely truncating the real part to a polynomial (in practice of quadratic order at most, as we explain below) and solve functionally the imaginary part.
Also this approach has been employed in \cite{Zambelli:2016cbw}, however again with some difficulty in reaching low dimensions.
Below, we will study the LPA and LPA' for the imaginary part of the potential, with the aim to extend the results from \cite{Zambelli:2016cbw} to lower dimensions, compare the numerical results for Lee-Yang critical exponents to the known one in 2$d$, and explore the fate of the multicritical Lee-Yang universality classes as $d\tend2$.

\paragraph{Truncation of the real part.}

Starting from \eqref{eq:realode}, we first assume that the potential is purely imaginary.
In practice, the real part of the coupled flow equation \eqref{eq:coupled} is discarded and only the imaginary part is kept, with $u(\vphi)=0$.
Deviating from the standard approach to FRG approximations after rescaling of both the field and potential, the differential equation simplifies to 
\eqref{eq:FP-pureIm}, with the associated anomalous dimension taken from \eqref{eq:eta-rescaled}:
\be
\eta = \begin{cases}
0 & \quad \text{(LPA)}\\
\frac{d+2}{2}\(1-\sqrt{1+\frac{4d}{d+2}g^2}\) & \quad \text{(LPA')},
\end{cases}
\ee
For stability, the third derivative differential equation \eqref{eq:h3} is used in the numerics, with initial conditions
\be \label{eq:init-cond-im}
h(0)=0, \quad h'''(0)=g\in \R , \qquad \Rightarrow \qquad  h'(0) = -\frac{g}{1-\gamma}, \quad h''(0) = 0 .
\ee
Note that $g\in\R$ implies that the anomalous dimension $\eta$ is always negative, a manifestation of the nonunitarity of theories with a complex potential, while the full scaling dimension is negative for\footnote{As a reminder, the coefficient $c_d$ in \eqref{eq:cd} is always positive for $2\le d \le 6$.}
\be\label{eq:trunc-im_bound}
\Delta \le 0\quad \Leftrightarrow\quad g \ge \sqrt{2-\frac{8}{d+2}}, \quad d\ge 2.
\ee

With a purely imaginary potential, the real part of the coupled flow equation \eqref{eq:coupled} is violated even for small values of $\phi$.
We can try to improve numerical treatment of the flow equation by including a non-trivial real even part to the potential truncated up to quadratic order. However, the resulting fixed point equation for the potential $h$ can be mapped to \eqref{eq:FP-pureIm} by a rescaling of both the field and the potential, such that it is fully equivalent to the study of the purely imaginary potential.
Details on this are given in appendix \ref{app:higher-truncation}.

It turns out that higher-order truncations lead to convergence problems, thus we do not consider them. 
We suspect that the reason could be that the asymptotic behaviour for the multicritical Lee-Yang models at $\Delta>0$ is given by \eqref{eq:asympto-nontrivial} with $A=0$, that is, with a real part that tends rapidly to zero at large $\vphi$. 
This is supported also by the scaling argument at the end of Sect.\ \ref{sec:perturbation}, where, translating to current notation, we saw that at $\vphi\sim (\f{1}{\eps})^{\f{1}{4n-2}}$, the imaginary part of the potential scales like $h(\vphi)\sim (\f{1}{\eps})^{\f{1}{2n-1}}$, while the real part scales like $u(\vphi)\sim 1$:
comparing to \eqref{eq:asympto-nontrivial}, we find precisely the same bahaviour if we set $A=0$ and $B=\eps^{1/2}$, and $\g$ at the upper critical dimension.
In this case, including in the equation for $h$ a polynomial truncation of $u$, that has a wrong large-$\vphi$ behaviour, strongly destabilizes the search for the fixed point.

Therefore, in the following we will only consider the fixed point equation  \eqref{eq:FP-pureIm} for the imaginary part of the potential $h(\vphi)$.

\paragraph{Eigenperturbations.}
As in the perturbative case, the IR dimensions of composite operators $:\phi^n:$ are estimated from the linear stability analysis around a given fixed point solution $h_*(\vphi)$: 
we write $h(\vphi,k)= h_*(\vphi)+\delta h(\vphi)\,\(\frac{k}{k_0}\)^{-\theta}$, with $\theta$ the critical exponent, and $k_0$ a reference scale, and expand both the flow equation and the anomalous dimension \eqref{eq:realode} to linear order in the fluctuation $\delta h(\vphi)$. 
The resulting linear equation is:
\ba\label{eq:opcompo}
&\mathcal{L}\, \delta h (\vph)  + \frac{\delta\eta}{2} \,\vphi\,h_*'(\vph)
= \Delta_n\, \delta h (\vph), \\
&\mathcal{L} = 
- d\frac{1-(h_*'')^2}{\(1+(h_*'')^2\)^2}\, \partial_{\vphi}^2
+ \frac{\Delta_*}{2}\,\vphi\,\partial_{\vphi}
\ea
In the LPA case, with $\delta\eta=0$, this reduces to a standard Sturm-Liouville problem,\footnote{By a change of variables one can also remove the first derivative and thus recast the eigenvalue problem as a standard one-dimensional time-independent Schrödinger equation \cite{Benedetti:2013jk,Hellwig:2015woa,Mandric:2023hlb}.}
where the eigenvalues $\Delta_n = d-\theta_n$ correspond to the scaling dimension for the composite operators.
In the LPA' instead, the variation of the anomalous dimension around the fluctuation, $\delta\eta$, depends linearly on the initial conditions of $\delta h$ through its second and third derivative, hence it is a non-standard eigenvalue equation.
 
By linearity, the fluctuation $\delta h$ is normalized and split into a $\Z_2$-even and $\Z_2$-odd contribution:
\be
\delta h''_{\text{odd}}(0)= 0,\quad \delta h'''_{\text{odd}}(0)= 1,\quad \text{and} \quad \delta h''_{\text{even}}(0)= 1,\quad \delta h'''_{\text{even}}(0)= 0.
\ee

For the above eigenvalue problem one finds easily two exact solutions \cite{Hellwig:2015woa}: $\delta h(\vphi) = \vphi$, with eigenvalue $\Delta$, and $\delta h(\vphi) = h'_*(\vphi)$, with  eigenvalue
\be\label{eq:shadow}
\Delta_{h'} = d - \Delta.
\ee
In CFT language,  such a relation between scaling dimensions is known as \emph{shadow relation} \cite{Ferrara:1972kab}.
This result might look strange, because in general CFTs do not have pairs of operators satisfying a shadow relation.\footnote{A notable exception are long-range models \cite{Paulos:2015jfa,Behan:2017emf,Behan:2025ydd}.}
The reason one finds them in the LPA is simple: in the FRG we do not automatically exclude redundant operators, and $h'_*(\vphi)$ is precisely a redundant operator \cite{Wegner:1974sla}. In fact, it is the equations of motion operator,\footnote{The full equations of motion should at least have also a $\p^2\vphi$ term, but in the LPA $\vphi$ is constant. Moreover, in the FRG the composite operators are obtained as perturbations of the action, meaning that they are introduced as integrated operators, hence total derivative operators (i.e.\ descendants) are not included in the spectrum.}
which is the prototype of a redundant operator, as it can be removed by a field redefinition. Moreover, in terms of the bare action $S[\phi]$, this operator should correspond to $\delta S/\delta\phi$, which by the Schwinger-Dyson equations satisfies $\la \f{\delta S}{\delta\phi}(x) \phi(y) \ra = \d(x-y)$, and it is thus recognized to be the shadow of $\phi$.

Other eigenperturbations need to be found numerically, and we will discuss them below.

\subsection{Fixed-points structure}\label{sec:fp-struct}

The differential equation \eqref{eq:FP-pureIm} is solved numerically by shooting from different values of the initial condition parameter $g$, and a global solution is searched via a spike plot, as briefly described at the beginning of this section. As noted in section \ref{sec:flow}, the  equation \eqref{eq:FP-pureIm}, together with the initial conditions \eqref{eq:init-cond-im}, is invariant under the transformation $g\to -g$, $h\to -h$, hence we will only consider $g\geq 0$.

\begin{figure}[H]
\centering
\includegraphics[scale=1,valign=t]{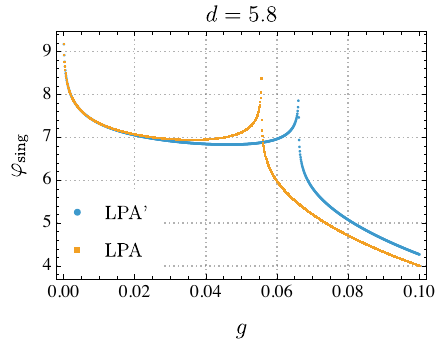}\qquad
\includegraphics[scale=1,valign=t]{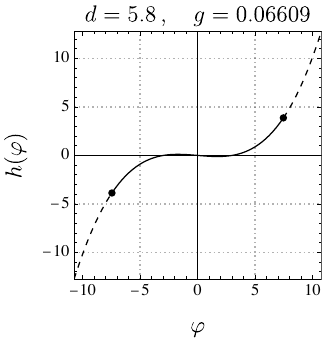}
\caption{Left: Integration endpoint for \eqref{eq:FP-pureIm} with different initial conditions $h'''(0)=g$ at $d=5.8$, in the LPA and the LPA'. Right: Potential solution $h(\vphi)$ at the "spike" in the LPA' in $d=5.8$, approached from the "left" (dashed) and from the "right" (up to dots) of the spike.
}
\label{fig:trunc-im_phi3spike}
\end{figure}

In dimensions $d\geq 6$, one finds only one spike at $g=0$, corresponding to the Gaussian fixed point $h(\vphi)=0$.
As soon as we move to $d=6-\epsilon$, for arbitrarily small $\epsilon$, a new spike is detected, besides the Gaussian one, as illustrated in figure \ref{fig:trunc-im_phi3spike}.
The new spike at $g=g_*>0$ indicates the presence of a nontrivial global solution, that is, a nontrivial fixed point.
Based on the fact that $d=6$ is the upper critical dimension for a $\vphi^3$ interaction, and on the perturbative analysis of section \ref{sec:perturbation}, we identify the nontrivial fixed point with the Lee-Yang universality class.
Such identification is also supported by the shape of the fixed-point potential  in figure \ref{fig:trunc-im_phi3spike}.

Next, we explore the landscape of spikes by gradually decreasing the space dimension until it reaches $d=2$, where we wish to compare the critical exponents derived from the resulting fixed point solutions to the known exact CFT results.
Following the analysis in section \ref{sec:analytic}, at any given $d$, the spike plots can be divided into three regions, in which the global solutions have widely different asymptotic behaviours: the region with $\Delta>0$, the one with  $\Delta<0$, and the interface with $\Delta=0$.

\begin{figure}[H]
\centering
\includegraphics[scale=1,valign=t]{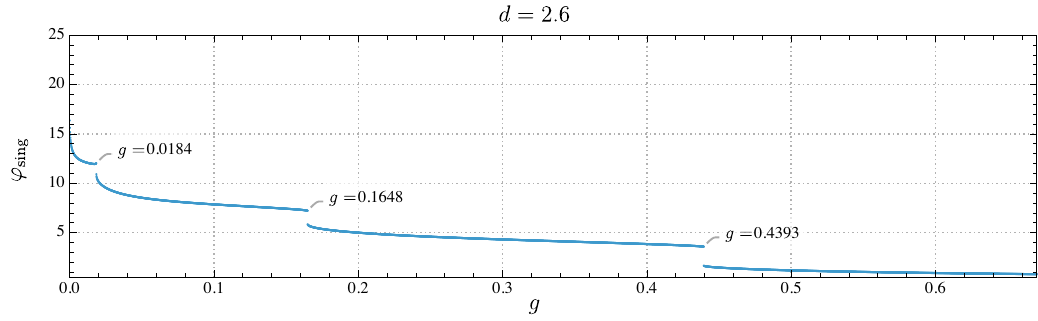}\\\hspace{1em}
\includegraphics[scale=1,valign=t]{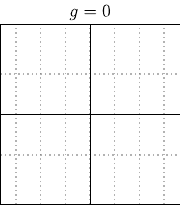}\hspace{3em}%
\includegraphics[scale=1,valign=t]{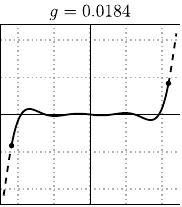}\hspace{3em}%
\includegraphics[scale=1,valign=t]{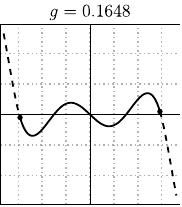}\hspace{3em}%
\includegraphics[scale=1,valign=t]{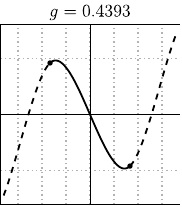}
\caption{Above: Integration endpoint for \eqref{eq:FP-pureIm} with different initial conditions $h'''(0)=g$ in $d=2.6$, in the LPA. The precision is given by the parameter steps in $g$, here of $10^{-4}$. Below: Solutions $h(\vphi)$ on both sides of a discontinuity are drawn in solid and dashed line.
The potential profiles let us clearly identify (e.g.\ by the number of zeros) fixed points as corresponding to, from left to right,   $h(\vph)=0$ (the Gaussian), $h(\vph)=\vph^7$, $h(\vph)=\vph^5$, and $h(\vph)=\vph^3$.
}
\label{fig:trunc-im_phi26spikeLPA}
\end{figure}

\begin{figure}[H]
\centering
\includegraphics[scale=1,valign=t]{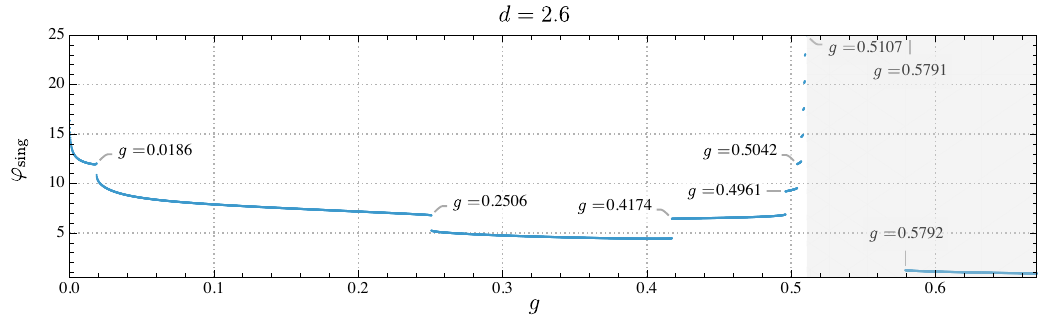}\\\hspace{1em}
\includegraphics[scale=1,valign=t]{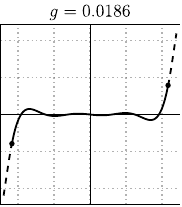}\hspace{3em}%
\includegraphics[scale=1,valign=t]{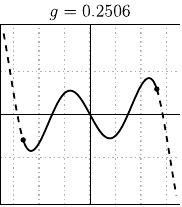}\hspace{3em}%
\includegraphics[scale=1,valign=t]{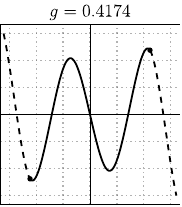}\hspace{3em}%
\includegraphics[scale=1,valign=t]{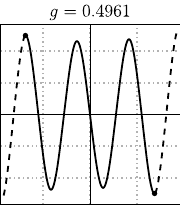}\hspace{3em}\\\medskip\hspace{1em}
\includegraphics[scale=1,valign=t]{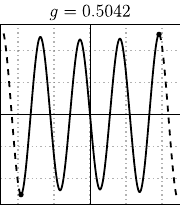}\hspace{3em}%
\includegraphics[scale=1,valign=t]{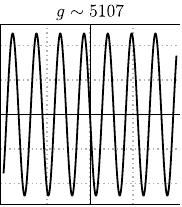}\hspace{3em}%
\includegraphics[scale=1,valign=t]{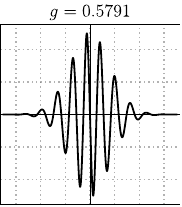}\hspace{3em}%
\includegraphics[scale=1,valign=t]{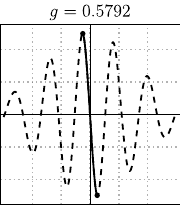}\hspace{3em}%
\caption{Above: Integration endpoint for \eqref{eq:FP-pureIm} with different initial conditions $h'''(0)=g$ in $d=2.6$, in the LPA'. The precision is given by the parameter steps in $g$, here of $10^{-4}$. In grey, we indicate the region where $\Delta\le 0$. For $0.5107 < g < 0.5791$, the LPA' integrate to numerical infinity. Below: Solutions $h(\vphi)$ on both sides of a discontinuity are drawn in solid and dashed line. The $g\sim 0.5107$ and $g=0.5791$ potential solutions integrate to infinity but were truncated to highlight their main features.
}
\label{fig:trunc-im_phi26spike}
\end{figure}

In the LPA, the scaling dimension appearing in the ODE \eqref{eq:FP-pureIm} is always positive for $d>2$, hence the non-trivial global solutions all diverge at large $\vphi$ and the corresponding spikes are easily identified, as in figure \ref{fig:trunc-im_phi3spike} and  \ref{fig:trunc-im_phi26spikeLPA}.
On the other hand, in the LPA', the scaling dimension in \eqref{eq:FP-pureIm} can vanish at any $d>2$, if $g$ is sufficiently large.
This occurrence is well illustrated by the spike plot in figure \ref{fig:trunc-im_phi26spike} (to be compared with figure \ref{fig:trunc-im_phi26spikeLPA} in the LPA), obtained at $d=2.6$, where the three regions are clearly distinguishable. We now proceed to describe the observed features of the solutions in such regions.

\begin{itemize}
    \item \underline{$\Delta > 0$}: As $d$ is reduced, new isolated scaling solutions branch off the Gaussian solution at specific critical dimensions $d_n=2\f{2n+1}{2n-1}$, corresponding to the upper critical dimensions of $\vphi^{2n+1}$ interactions (see table \ref{tab:dc}). As we keep decreasing $d$, such spikes move to the right, as predicted from the perturbative treatment. While in the large $\vphi$ limit the global solutions are dominated by the asymptotic behaviour \eqref{eq:asymp-posDelta}, near $\vphi=0$, they behave like an odd polynomial potential of order $2n+1$ (in particular they have $2n+1$ zeros) corresponding to the critical dimension at which they appeared. 
    For example, in figure \ref{fig:trunc-im_phi26spike}, the first two potentials correspond to the $i\vphi^{7}$ and $i\vphi^{5}$ theories respectively. 
    \begin{table}[H]\centering
    \begin{tabular}{|c||c|c|c|c|c|}
        \hline
          $ i\vphi^n$ & $i\vphi^3$ & $i\vphi^5$  & $i\vphi^7$ &  $i\vphi^9$&$i\vphi^{11}$  \\
          \hline
         $d_n$ & 6 & $\frac{10}{3}\sim3.33$& $\frac{14}{5}=2.8$ & $\frac{18}{7}\sim 2.57$ & $\frac{22}{9}\sim 2.44$\\
        \hline
    \end{tabular}
    \caption{Upper critical dimensions $d_n$, at which $n\Delta =d$, for $n=1,\ldots,5$..}
    \label{tab:dc}
    \end{table}
    On the right of the first two spikes, of perturbative origin, we see an infinite ladder of $i\vphi^{2n+1}$-type spikes branching out not from the Gaussian, but from the boundary $\Delta=0$, all at once below the dimension $d_0$ at which the Lee-Yang fixed point reaches $\Delta=0$. We will discuss these nonperturbative spikes in detail in section \ref{sec:im-phi5}.

    As $d=2.6 <d_0$ (see \eqref{eq:predict-d0} as well as the next subsection), the field $\vphi$ corresponding to the Lee-Yang scaling solution has already acquired a negative scaling dimension, hence we no longer have a Lee-Yang spike in the region $\Delta>0$.

    \item \underline{$\Delta = 0$}: As discussed in section \ref{sec:Delta0}, the solutions at exactly $\Delta=0$ can either be singular or be global periodic solutions, depending on the energy level $E$ in \eqref{eq:energy}, fixed at a given $d$ by the value $g=g_0(d)$ at which $\Delta=0$. 
As argued at the end of that section, at  $d=2.6 <d_0$ we expect to have a global periodic solution with $E<0$, and this is corroborated by the numerical solution.
    
    \item \underline{$\Delta < 0$}: The region of negative scaling dimension is further split in two subregions with widely different asymptotic behaviour.
    At sufficiently large coupling, the differential equation always presents a movable singularity, and the integration is stopped ($g\geq 0.5792$ in figure \ref{fig:trunc-im_phi26spike}). In contrast, closer to the $\Delta = 0$ boundary, the differential equation gets integrated up to numerical infinity. The associated solutions for the  potential $h(\vphi)$ take the shape of a wavepacket with periodic oscillations enveloped by an exponential decay, as can be seen in the plot at $g=0.5791$ in figure \ref{fig:trunc-im_phi26spike}. These two behaviours were predicted in section \ref{sec:dyn-syst} on the basis of the stability properties of the Gaussian theory, and the snapshots of the vector fields in figure \ref{fig:snapshots-2}, as well as on the asymptotic behaviour \eqref{eq:asymp-negDelta}.

\end{itemize}

\begin{figure}[H]
\centering
\includegraphics[scale=1,valign=t]{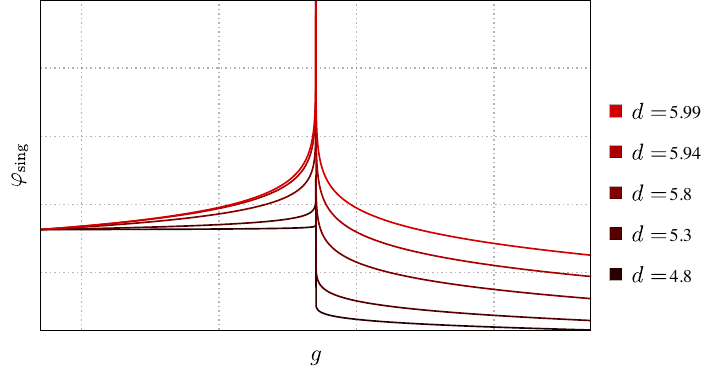}
\caption{Left: Interpolation of the integration endpoints at various dimensions over ranges in $g$ of identical length, shifted in both axis such that the Lee-Yang spikes overlap. A clear shrinking of the spike's width is visible.}
\label{fig:spikewidth}
\end{figure}

As a last comment, one should note that the width of the spike as shown for $d=5.8$ in figure \ref{fig:trunc-im_phi3spike} shrinks as we lower the dimension, up to the point where the spike cannot be resolved numerically.\footnote{We tried for the Lee-Yang fixed point up to 25 digits precision in $d=2.6$.}. Only a  step discontinuity is left, as can be seen in figure \ref{fig:spikewidth}. 
This suggests that the global solution at $g=g_*$ is not approached gradually: the singularities at $g=g_*\pm \delta$ converge to a finite value $\vphi_*^\pm<\infty$ as $\delta\tend 0$, with a behaviour like $\vphi_*^\pm-\vphi_{\rm sing}\sim \delta^{\a_\pm}$, with exponents $\a_\pm>0$ that become smaller and smaller as we lower $d<6$.
This is consistent with the understanding gained in the previous section. Indeed, we know that a global solution with $\Delta>0$ must have the asymptotic behaviour \eqref{eq:asymp-posDelta}, which leads to $\lim_{\vphi\to\infty} |h''(\vphi)|=\infty$.
Therefore, a global solution is not reached by pushing farther and farther the value of $\vphi_{\rm sing}$ at which we hit the singularity with $|h''(\vphi_{\rm sing})|=1$, but rather by tuning to a separatrix that manages to go smoothly through such a singularity (see figure \ref{fig:snapshots-1}). In such a situation, the approach to the global solution is completely discontinuous, with solutions hitting an early singularity as soon as they deviate infinitesimally from the separatrix.

\subsection{The Lee-Yang universality class} \label{sec:im-phi3}

\begin{figure}[H]
\centering
\includegraphics[scale=1,valign=t]{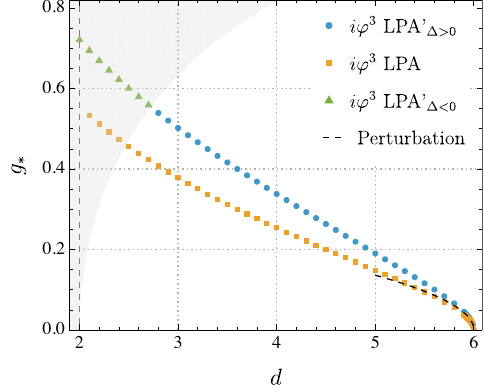} \quad
\includegraphics[scale=1,valign=t]{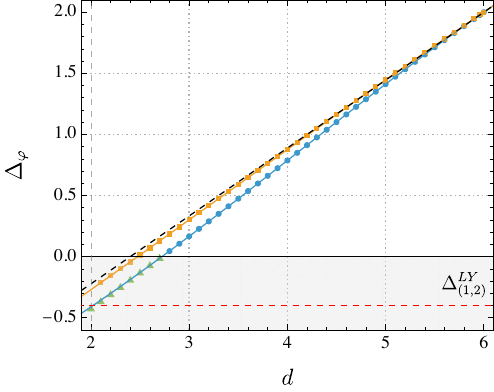}
\caption{Left: Critical values $g_*$ corresponding to a spike (see figure \ref{fig:trunc-im_phi3spike}) for the LPA, LPA' and the perturbative solution \eqref{eq:perturbation}. In gray, we indicate the region for which $\Delta<0$. Right: Comparison between the scaling dimensions $\Delta_\vphi$ in the LPA values, the LPA' and the perturbative result (at order $\epsilon$), as well as their extrapolation by means of a fit with a cubic polynomial in $d$ to $d=2$. The conformal dimension $\Delta^{LY}_{(1,2)}=-2/5$ for the primary $\phi_{(1,2)}$ of the Lee-Yang minimal model $\mathcal{M}_{(2,5)}$ is indicated in red for reference.
}
\label{fig:trunc-im_phi3}
\end{figure}

\begin{figure}[H]
\centering
\includegraphics[scale=1,valign=t]{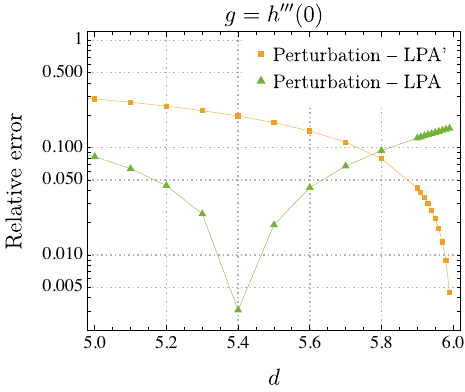}\qquad
\includegraphics[scale=1,valign=t]{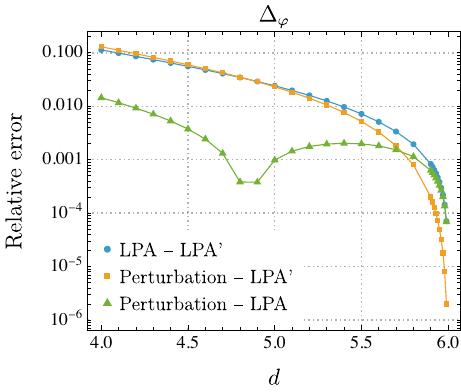} 
\caption{
Left: Relative error of the perturbative potential \eqref{eq:perturbation} with respect to both the LPA and LPA' solution at $\varphi=0$.
Right: Relative error in the scaling dimension between the LPA, the LPA' and the perturbation results. 
The local drop in the relative error of LPA vs perturbation is due to the fact that the LPA curve for $\D_\vph$ crosses the one for perturbation theory, see figure \ref{fig:trunc-im_phi3}.
}
\label{fig:lpavspert}
\end{figure}

Let us concentrate on the non-trivial fixed point solution branching off the Gaussian solution below $d=6$.
As discussed above, based on perturbation theory, we can interpret it as corresponding to the $i\phi^3$ theory, i.e.\ the Lee-Yang universality class.
The dependence of the critical point $g_*$ and scaling dimension $\Delta_\vphi=\Delta|_{g=g_*}$ 
on $d$ is shown in figure \ref{fig:trunc-im_phi3}. A comparison to the perturbative results, confirming the interpretation of the spike, is provided in figure \ref{fig:lpavspert}.

The spike reaches the boundary $\Delta=0$ at
\be 
d_0\lesssim2.7249(1),
\ee 
matching perfectly with the predicted dimension \eqref{eq:predict-d0}.
At lower $d$, when we expect spike to enter the region of negative scaling dimension, we actually find a whole range of values of $g\in[g_0, g_{\rm up}]$ where no singularity is encountered, as explained before, and as illustrated in figure \ref{fig:trunc-im_phi26spike} (with $g_0=4.336$ and $g_{\rm up}=4.846$).
We suspect that solutions in this range of $g$ should be excluded, except for the one at $g=g_{\rm up}$, but we do not have a good theoretical argument to support this.
Nevertheless, the point at $g=g_{\rm up}$, describing the sharp transition between the global wavepacket solutions and the singular solutions,\footnote{We also notice that based on figure \ref{fig:snapshots-2} and the discussion around it, we believe that the solution at $g=g_{\rm up}$ has an asymptotic expansion as in \eqref{eq:asymp-negDelta} with $b_2=0$, i.e.\ without exponential part.} is observed to follow closely the extrapolation of the location of $g_*$ from $d>d_0$ to $d<d_0$, see figure \ref{fig:trunc-im_phi3}.
Therefore, at $d<d_0$ we identify the scaling solution at $g=g_{\rm up}$ as the one associated to the Lee-Yang universality, i.e.\ we identify $g_{\rm up}$ as the continuation of $g_*$.

We report in table \ref{tab:trunc-im_LY2d} the scaling dimensions at the fixed point in various approximations and compare to the most relevant conformal dimension of the Lee-Yang model $\mathcal{M}_{(2,5)}$. The LPA' approximates surprisingly well the Lee-Yang CFT value, with a relative error between $2.6\%$ and $7\%$ compared to the exact value. 
As a comparison, in \cite{Hellwig:2015woa}, using a similar setup as in this paper, the scaling dimension of the Ising model in $d=2$ was obtained with a relative error of $74.52\%$ with respect to the exact value (see table VI in \cite{Hellwig:2015woa}). We believe that the main difference is that the scaling dimension for the complex potential depends only on the initial conditions, and not on the full profile of the potential. Indeed, when the potential is even and the third derivative of the potential vanishes at $\vphi=0$, the LPA' is done iteratively by computing the anomalous dimension at the minimum of the critical potential and feeding it back into the equation until convergence. In the complex case instead, as we have seen, the scaling dimension only depends on the initial conditions and no iteration is needed, making the numerical scheme more stable.
We also note that in the Lee-Yang case, the scaling dimension computed at the potential's minimum instead of the origin moves away from the expected values in low dimensions. Results closer to the LPA' should be obtainable by iterating, but the iteration scheme becomes increasingly difficult in the negative-$\Delta$ region.
Lastly, we observe that the value in $d=2$ obtaining by extrapolating the $d>d_0$ data is closer to the exact value than the  critical coupling computed directly at $d=2$ using the scaling solution at $g=g_{\rm up}$. 

\begin{table}[H]
    \centering
    \begin{tabular}{|l|c|c|c|c|}
        \hline
        Approximation &$\Delta_\vphi^{d=5}$ & $\Delta_\vphi^{d=4}$ & $\Delta_\vphi^{d=3}$ & $\Delta_\vphi^{d=2}$\\
        \hline\hline
        Perturbation (leading order) \eqref{eq:perturbation} &$1.4444(1)$&$0.8888(1)$ & $0.3333(1)$ & $-0.2222(1)$ \\
        \hline
        LPA at $\vphi=\text{min}\, h$ &$1.43995$&$0.81572$ & $-0.01731$ & $-1.71197$ \\
        \hline
        LPA at $\vphi=0$ &$1.44587$ &$0.87621$ &$0.30122$ &$-0.27004$ \\
        \hline
        $\text{LPA'}_{\Delta>0}$ & $1.41178$ &$0.78675$ &$0.16659$& $-0.41035$ \\
        \hline
        $\text{LPA'}_{\Delta<0}$ &  &  &  & $-0.42809$ \\
        \hline\hline
        Truncated bootstrap \cite{Hikami:2017hwv,Gliozzi:2014jsa}  & $1.43807$ & $0.823283$ & $0.174343$ &  $-0.39777(2)$ \\
        \hline
        5-loops resummation \cite{Borinsky:2021jdb} & $1.4245(10)$ & $0.827(6)$ & $0.215(10)$ & $-0.390(15)$ \\
        \hline
        6-loops resummation \cite{Gracey:2025rnz} & $1.4240(5)$ & $0.823(3)$ & $0.211(2)$ &  \\
        \hline
        Regularized sphere method \cite{ArguelloCruz:2025zuq} &  & $0.874$ & $0.214(2)$ & $-0.406277$ \\
        \hline
        CFT $\mathcal{M}_{(2,5)}$& & & & $-0.4$ \\
        \hline
    \end{tabular}
    \caption[Caption for LOF]{Numerical values of the scaling dimension $\Delta_\vphi$ using various approximations, based on the $i\vphi^3$ spikes' values given in figure \ref{fig:trunc-im_phi3}. The values in $d=2$ are extrapolated from a cubic polynomial fit in $d$, except for the value $\text{LPA'}_{\Delta<0}$ which is taken from the solution at $g=g_{\rm up}$ in the $\Delta<0$ region, directly at $d=2$, and the perturbation result. 
    Numerical errors of our data are taken at machine precision, but do not include the larger undetermined systematic error\footnotemark{} from the LPA and LPA', and corresponding values are arbitrarily rounded at the fifth digit.}
    \label{tab:trunc-im_LY2d}
\end{table}%
\footnotetext[18]{\label{foot:errors}Systematic errors arise from the truncation of the derivative derivative expansion and from discarding the real part of the potential, as well as from cutoff dependence. A rough estimate for them is usually obtained by either choosing half the range of values obtained by different cutoffs \cite{Balog:2019rrg} or by comparing to the previous order of the truncation \cite{DePolsi:2020pjk}, but unfortunately these methods cannot be applied in our limited setup.}%

\paragraph{Scaling operators.}
The scaling dimensions of composite operators are estimated by solving numerically \eqref{eq:opcompo} with $h_*(\vphi)$ given by the numerical fixed-point solutions corresponding to the points in figure \ref{fig:trunc-im_phi3}. 
However, as explained at the end of section \ref{sec:fp-struct}, the global solution is not approached gradually, hence any numerically obtained $h_*(\vphi)$ is only valid on a finite interval $(-\vphi_{\text{sing}},\vphi_{\text{sing}})$, which cannot be enlarged at will by increasing the numerical precision of $g_*$. 
As a consequence, the solution terminates at a singularity before shaping a full potential with the appropriate number of extrema, see figure \ref{fig:trunc-im_phi26spike} for some examples.
Nevertheless, a workaround to this caveat is to project the full potential $h(\vphi)\tend h_*(\vphi)+\delta h(\vphi)\,\(\frac{k}{k_0}\)^{-\theta}$ on a polynomial basis truncated up to some finite order.
The $\cP\cT$-symmetry condition imposes that the $h_*(\vphi)$ has an expansion in odd powers of $\vphi$,
 $n \le N$:
\be \label{eq:poly-fp}
h_*(\vphi) = \sum_{n=0}^N \frac{g_{2n+1}}{(2n+1)!}\,\vphi^{2n+1}, 
\ee
while the fluctuations can be separated in $\Z_2$-even and $\Z_2$-odd, 
\be \label{eq:poly-fluct}
\delta h_{\text{odd}}(\vphi) = \sum_{n=0}^N \delta_{2n+1}\, \vphi^{2n+1}, \quad 
\delta h_{\text{even}}(\vphi) = \sum_{n=0}^N \delta_{2n}\, \vphi^{2n}.
\ee

Substituting \eqref{eq:poly-fp} into \eqref{eq:FP-pureIm} and expanding in $\vphi$, we obtain the fixed point equations $\b_{2n+1}=0$, where the beta functions\footnote{Substituting \eqref{eq:poly-fp} in the flow equation we would of course obtain $\b_{2n+1}=k\p_k g_{2n+1}$.} $\b_{2n+1}$ are polynomial in the couplings $\{g_{2m+1}\}_{m\leq n+1}$, with the coupling $g_{2n+3}$ appearing as a linear monomial. For $n=0$, also $g_1$ appears linearly.
Therefore, we can easily solve iteratively the fixed-point equations $\{\b_{2n+1}=0\}_{n=1,\ldots, N-1}$ for the couplings $\{g_{2n+1}\}_{n\neq 1}$ as functions of the third derivative coupling $g_3=g$. The last equation, $\b_{2N+1}=0$, would fix $g_3$ because there is no $g_{2N+3}$ in the truncation, but we choose to leave it unsolved, and use instead the numerical values for $g$ from the spikes plots, leading to an approximate solution. This has the advantage of avoiding any spurious solutions resulting from the high-degree polynomial equation $\b_{2N+1}(g_3)=0$, and it amounts to Taylor expanding the numerical solution $h_*(\vphi)$.

For the perturbations, the functional eigenvalue problem \eqref{eq:opcompo} is replaced by a finite-dimensional eigenvalue problem for a matrix of size $(2N+2)\times (2N+2)$, by injecting in \eqref{eq:poly-fp} and \eqref{eq:poly-fluct}, as we have done for the fixed point equation. 
Alternatively, we can start from the truncated system of beta functions, with the anomalous dimension replaced by its implicit definition \eqref{eq:eta-rescaled-eq}, and do the usual finite-dimensional stability analysis, that is, compute the stability matrix  $B_{mn}=\p\b_{2n+1}/\p g_{2m+1}|_{g=g_*}$, its eigenvalues $-\theta_i$, and thus obtain the scaling dimensions $\Delta_i=d-\theta_i$. 
The second method only works for the odd perturbations, reducing to a matrix of size $(N+1)\times (N+1)$, unless we start from the original equation \eqref{eq:realode}, and compute the beta functions for both even and odd order couplings, as in \cite{Zambelli:2016cbw}.

\begin{figure}[H]
\centering
\includegraphics[scale=1,valign=t]{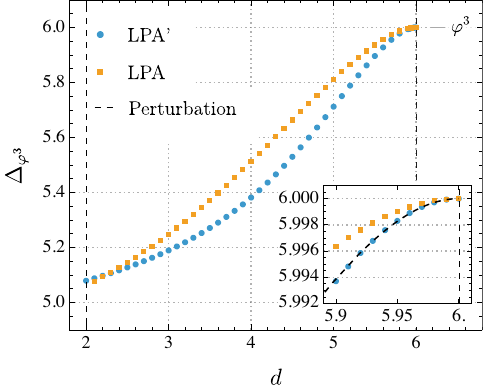}\qquad
\includegraphics[scale=1,valign=t]{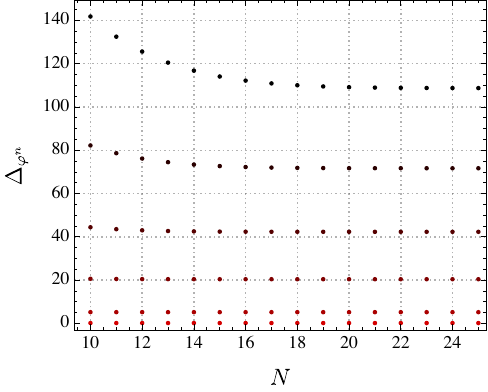}
\caption{
Left: Plot of $\Delta_{\vphi^3}$ in the LPA' and LPA, with a truncation at $N=19$ computed from the stability matrix, and benchmarked against standard perturbation theory at order $\eps^3$ \cite{deAlcantaraBonfim:1980pe,deAlcantaraBonfim:1981sy}.
Right: LPA' eigenvalues in $d=3$, in terms of the truncation order $N$ computed from the stability matrix. 
Only $\Z_2$ odd fluctuations are shown.
}
\label{fig:LY-dimPhi3}
\end{figure}

The first method with a truncation up to $N=19$  allowed us to verify the relation \eqref{eq:shadow} between the $\vphi$ and $h_*'(\phi) $ operators in the LPA' with an absolute error of order $10^{-17}$ near $d=6$, to $10^{-4}$ near $d=2$. The error is controlled by both the truncation order and the precision of the numerical fixed point $g_*$, taken here up to machine precision. Increasing either of them decrease the numerical error in the relation. The higher errors in lower dimensions are due to the increased sensibility of the critical solution on initial conditions.
Despite the lack of information on the even perturbations, the second method shows better convergence of the eigenvalue problem at the same truncation order, as shown on the right of figure \ref{fig:LY-dimPhi3}. However, the dimension of irrelevant operators such as $\vphi^3$ appears to be badly approximate at lower dimension in the LPA'.
At $d=5$, with $N=19$ we have $\Delta_{\vphi^3}=5.712$ (LPA'), to be compared with $5.701-5.702$ of the two-sided Padé estimate from \cite{ArguelloCruz:2025zuq}, and at $d=4$ we have $\Delta_{\vphi^3}=5.382$ (LPA'), to be compared with $5.206-5.212$ \cite{ArguelloCruz:2025zuq}.
However, both the LPA and LPA' estimates stay above $\Delta_{\vphi^3}>5$ in all dimensions, while in the Landau-Ginzburg description \cite{ArguelloCruz:2025zuq}, the operator $:i\vphi^3:$ is identified with the $T\overline{T}$ operator, that has dimension 4.

\subsection{Higher multicritical models}\label{sec:im-phi5}

\begin{figure}[H]
\centering
\includegraphics[scale=1,valign=t]{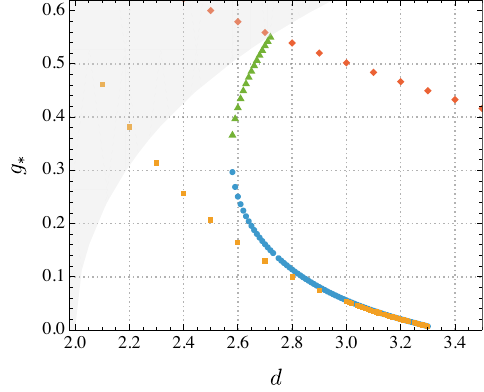} \quad
\includegraphics[scale=1,valign=t]{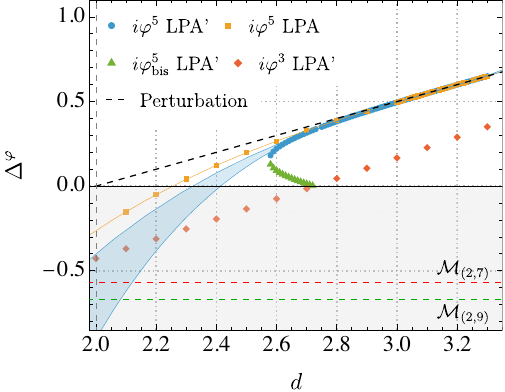}
\caption{Left: Critical values $g_*$ corresponding to a spike for the the LPA, LPA’. In gray, we indicate the region for which $\Delta<0$. Data legends are on the right panel. Right: Comparison between the scaling dimensions in the LPA, the LPA', and perturbation theory \eqref{eq:perturbation5} as well as their extrapolation at cubic order to $d=2$. In blue, we indicate the range of the fits for the LPA' as we vary the set of points taken into account, from all the points being fitted (lowest curve) to neglecting the points below $d_0$ \eqref{eq:predict-d0} (upper curve).
The conformal dimensions $\Delta_{\vphi}^{(2,7)}=-\frac{4}{7}$ and $\Delta_{\vphi}^{(2,9)}=-\frac{2}{3}$ for the primary $\phi_{(1,2)}$ in $\mathcal{M}_{(2,7)}$ and $\mathcal{M}_{(2,9)}$ respectively are indicated in red and green for reference.
}
\label{fig:trunc-im_phi5}
\end{figure}

The perturbative spikes for higher multicritical $i\vphi^{2n+1}$ theories are followed numerically in lower dimensions, starting from their upper critical dimensions given in table \ref{tab:dc}. 
The trajectory in the $d$-dimensional RG space for the $i\vphi^5$ theory is shown in figure \ref{fig:trunc-im_phi5}, for both LPA and LPA'.
While in the LPA case the $i\vphi^5$ solution can be continued from the upper critical dimension $d\simeq 3.33$ to arbitrarily close to $d= 2$, in the LPA' it disappears below $d\simeq 2.57$, annihilating with another fixed point that branches off from the $i\vphi^3$ solution, when the latter reaches $\Delta_\vph=0$.
Indeed, as illustrated in the spike plot \ref{fig:trunc-im_phi26spike}, when $d<d_0$, in the LPA' we observe an infinite ladder with spikes/steps corresponding to nonperturbative $i\vphi^{2n+1}$-like theories, emerging from the periodic solution at $\Delta=0$. For simplicity, we denote as ``$i\vphi^{2n+1}_{\text{bis}}$" such nonperturbative fixed points.
We will discuss their reliability and interpretation in the section \ref{sec:conclusions}.

While in $d=3>d_0$ a unique $i\vphi^5$ spike is observed, we were not able to resolve numerically the scaling dimensions of composite operators, as no convergence was observed in the eigenvalues with increasing precision in the fixed point solution not increasing truncation order in \eqref{eq:poly-fp}, using both methods.
Nevertheless, the values in $d=3$ for the scaling dimension $\Delta_{\vphi}$ are reported in table \ref{tab:phi5-3d}, with a relative error of $3.63\%$ between the LPA' and the two-sided Padé value \cite{Katsevich:2025ojk}.

\begin{table}[H]
    \centering
    \begin{tabular}{|c|c|c|c||c|}
        \hline
        Approximation &Perturbation & LPA & LPA'& Two-sided Padé \cite{Katsevich:2025ojk}\\
        \hline
        $\Delta_\vphi^{d=3}$ & $0.49997$ 
        & $0.49579$ &$0.49534$ & 0.478 \\
        \hline
    \end{tabular}
    \caption[Caption for LOF]{Numerical values of the scaling dimension $\Delta_\vphi$ using various approximations, based on the $i\vphi^5$ spike's value in $d=3$. Numerical errors for our data are taken at machine precision, but do not include the larger undetermined systematic error (see footnote \ref{foot:errors}) from the LPA and LPA', and corresponding values are arbitrarily rounded at the fifth digit.}
    \label{tab:phi5-3d}
\end{table}

\begin{figure}[H]
    \centering
    \includegraphics[scale=1,valign=t]{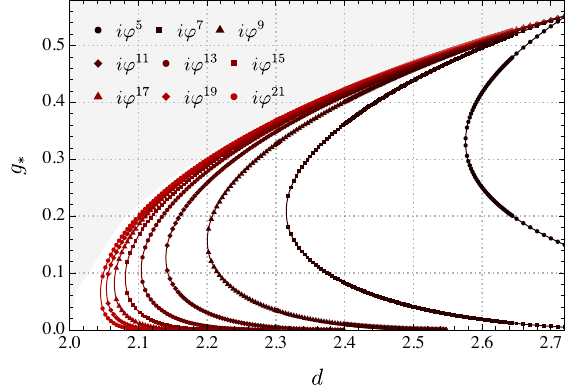}\hspace{3em}
    \begin{tabular}[t]{|c|c|}
            \hline
          $ i\vphi^n$ & $d_c$\\
          \hline\hline
          $i\vphi^3$ &  /\\\hline
          $i\vphi^5$  & 2.5758\\\hline
          $i\vphi^7$ &   2.315\\\hline
          $i\vphi^9$ &  2.200\\\hline
          $i\vphi^{11}$ & 2.1399 \\\hline
          $i\vphi^{13}$ & 2.104 \\\hline
          $i\vphi^{15}$ & 2.080\\
        \hline
    \end{tabular}
    \caption{On the left, we interpolate the critical values $g_*$ corresponding to multicritical spikes in the LPA' up to $d=2.06$. In gray, we indicate the region for which $\Delta<0$.
    The associated numerical upper bound on the dimension at which the spikes collide in the LPA' are given on the right.}
    \label{tab:db}
\end{figure}

More in general, as we decrease $d$, the spike $i\vphi^{2n+1}$, $n>1$ collides with its non-perturbative image $i\vphi^{2n+1}_{\text{bis}}$ before it reaches the boundary $\Delta_{\vphi}=0$, as illustrated in figure  \ref{tab:db}.
We report in the table of figure \ref{tab:db} the upper bound for the collision dimensions $d_c$ for the first six multicritical models.

We conclude that the multicritical $i\vphi^{2n+1}$ fixed points cannot be extrapolated to $d=2$ in the LPA' due to the presence of their non-perturbative counter-parts, leaving open the Landau-Ginzburg interpretation of the $\cM(2,2n+3)$ minimal models with $n>1$.


\section{Summary and discussion}
\label{sec:conclusions}

We have presented a detailed study of the fixed-point equation for the local potential approximation of the Wetterich equation in the case of a $\cP\cT$-symmetric complex potential for a single scalar field.
Our main motivation was the development of nonperturbative methods for multicritical Lee-Yang models, especially in view of the conjectural status of their link to minimal models in two dimensions \cite{Becker:1991nr,Becker:1991nq,vonGehlen:1994rp,Zambelli:2016cbw,Lencses:2022ira,Katsevich:2025ojk,Amoruso,Lencses:2024wib}.
And on the technical level, the FRG approach to complex potentials had remained so far largely unexplored (with few exceptions, e.g.\ Refs.\ \cite{Zambelli:2016cbw,An:2016lni}),\footnote{For studies of the ordinary real LPA equation in the complex field plane, from the viewpoint of convergence of polynomial
expansions, see instead \cite{Litim:2016hlb,Juttner:2017cpr,Litim:2018pxe}.} thus providing in itself motivation for their further investigation.
The main lessons from our analysis are:
\begin{itemize}
\item The perturbative construction of solutions, given in section \ref{sec:perturbation}, has shown that weakly coupled fixed points appear at $d=d_{2n+1}-\eps  = \frac{4n+2}{2n-1}-\eps$, as for $i\vph^{2n+1}$ theories in standard perturbation theory \cite{Codello:2017epp}. The coefficients of the series expansion in $\eps$ of the effective potential $v(\vph)$ are expressed as linear combinations of Hermite polynomials in $\vph$, and, because of their growth, the expansion breaks down when $\vph \sim(\f{1}{\eps})^{\f{1}{4n-2}}$.
For larger values of $\vph$, the asymptotic analysis of section \ref{sec:analytic} shows that  the leading large-field behaviour is $v(\vph)\sim \vph^{d/\Delta_\vph}$, which is a noninteger power for $\eps\neq 0$, thus explaining why the polynomial expansion cannot hold at large $\vph$. Nevertheless, the perturbative results serve as a guidance in identifying universality classes in the numerical analysis, and thus even at finite $\eps$ we will refer to such universality classes as $i\vph^{2n+1}$ fixed points.

\item The detailed analysis of singularities and asymptotic expansion in section \ref{sec:analytic} has shown that the properties of the fixed-points equations change drastically when $\Delta$ changes sign: in particular, while at $\Delta>0$ we expect only isolated global solutions, at $\Delta\leq 0$ a continuum of global solutions is possible. This fact affects crucially the LPA' of the $i\vph^{2n+1}$ theories, that have $\Delta_\vph=\Delta|_{g=g_*}>0$ near the upper critical dimension $d_{2n+1}$, but are expected to have $\Delta_\vph<0$ at $d=2$, based on their conjectured CFT correspondence.

\item The numerical work presented in section \ref{sec:numerical} has allowed us to verify the appearance of the $i\vph^{2n+1}$ fixed points below their respective upper critical dimension $d_{2n+1}$, and to continue them to the nonperturbative domain of lower dimensions.
For the Lee-Yang universality class $n=1$, we found that $\Delta_\vph$ becomes zero at $d=d_0\simeq 2.7249$, in agreement with the analytic prediction from section \ref{sec:analytic}, and to be compared with truncated bootstrap result $d_0\simeq 2.6199$ \cite{Hikami:2017hwv} and the two-sided Pad\'e result $d_0\simeq 2.65$ from the five-loop perturbative expansion \cite{ArguelloCruz:2025zuq}.\footnote{Private communication by Igor Klebanov.} 
Estimates for $d_0$ of $2.649(25)$ and $2.655(35)$ have also been 
established using the six-loop perturbative approach of \cite{Gracey:2025rnz},
where the values were derived from the exponents $\sigma$ and 
$\Delta_\phi$, respectively.\footnote{Private communication by John A. Gracey.}

For $d<d_0$, we found a continuous interval of initial conditions leading to global solutions, again in agreement with the anaysis of section \ref{sec:analytic}. Identifying the Lee-Yang universality class with the end point of such interval, we were able to follow it all the way to $d=2$, finding a value of $\Delta_\vph$ with a relative error  between $2.6\%$ and $7\%$ compared to the exact value from $\cM(2,5)$.

\item Unexpectedly, at $d<d_0$, we discovered the appearance of an infinite sequence of new nonperturbative fixed points, that we denoted ``$i\vphi^{2n+1}_{\text{bis}}$", branching off from the $i\vph^3$ one, and annihilating one-by-one with the perturbative $i\vph^{2n+1}$ fixed points with $n>1$, at some dimension $2<d<\min(d_{2n+1},d_0)$.
As a consequence, we could not continue the multicritical Lee-Yang universality classes down to $d=2$ and we thus could not compare them to the minimal models $\cM(2,2n+3)$ or $\cM(2,4n+1)$.

\end{itemize}

\

There are many possible directions for improving our analysis, but they all require substantial effort, so we leave them to future work. However, we stress that there are at least two strong motivations for pursuing these studies: the first is to aim at a quantitative improvement of our predictions, in order to be able to provide a reliable benchmark for other approaches; the second is to clarify the status of the nonperturbative fixed points found within the LPA'. 

The easiest improvement of our analysis would be to change or generalize our choice of cutoff, in order to test the dependence of our results on such choices, and to exploit that for optimization purposes.
In fact, it is well known that while critical exponents (scaling dimensions) should be universal, i.e.\ cutoff independent, approximations such as the LPA lead to a cutoff dependence of the results.
Optimization then amounts to tuning the cutoff to minimize the dependence of critical exponents on its variations.
This approach is known as \emph{principle of minimal sensitivity} \cite{Stevenson:1981vj}, and it has been successfully implemented in the derivative expansion \cite{Canet:2002gs,Balog:2019rrg,Balog:2020fyt}.

The next obvious improvement would be to enlarge our truncation space, that is, to go beyond our LPA/LPA'.
First, it would be interesting to include also the real part of the potential. As we explained, truncating it to a polynomial destabilizes the fixed point equation for the imaginary part, due to the wrong large-field behaviour of the polynomials. Therefore, one should either employ a polynomial expansion for both real and imaginary parts, as done already in \cite{Zambelli:2016cbw}, or treat both functionally, thus looking for global solutions of the coupled system of ODE \eqref{eq:coupled}.
Second,  it would be important to move to the next order in the derivative expansion, thus promoting $Z_k\to Z_k(\vph)$ in \eqref{eq:lpa-ansatz} \cite{Morris:1994ie,Defenu:2017dec}.
Other possibilities to go beyond the LPA/LPA' include the BMW scheme \cite{Blaizot:2005xy}, and the LPA''  \cite{Rose:2018tpn}.

The improvements suggested above would certainly answer the question on whether the predictions for Lee-Yang in low dimensions, e.g.\ for  $\D_\vph$ and $\D_{\vph^3}$, could be improved.
In particular it would be nice to be able to get better results for the scaling dimensions of $\vph^m$, which is of course the ultimate goal of any method.
However, the most pressing question raised by our work is without doubt the one about the nonperturbative fixed points.

Are the ``$i\vphi^{2n+1}_{\text{bis}}$" fixed points, and their annihilation with the $i\vphi^{2n+1}$ ones, a spurious result of the LPA', or are they a genuine nonperturbative feature brought to light by the FRG?\footnote{We notice that a annihilations of perturbative fixed points with unexpected nonperturbative fixed points have been observed before within the FRG, for example in the $O(N)$ model at large $N$ \cite{Yabunaka:2017uox,Yabunaka:2018mju,Yabunaka:2021fow}.}
The second option would probably be the most exciting, as it would be a nice case of truly nonperturbative physics, unforeseeable with perturbative methods, and it would force us to completely rethink about the Ginzburg-Landau description of the minimal models  $\cM(2,2n+3)$.
However, at present we lack any interpretation for the fact that an infinite series of fixed points seems to emerge from the Lee-Yang CFT just below $d=d_0$, i.e.\ the dimension at which it has $\Delta_\vph=0$.
The mechanism could have similarities to how the $\phi_{1,3}$-perturbed unitary minimal models emerge from the restricted sine-Gordon model \cite{LeClair:1989wy,Smirnov:1990vm} (see also \cite{Mussardo:2020rxh} for a clear exposition), but in noninteger dimension, and without the power of 2d CFTs and integrability, it is hard to make sense of it.
About the annihilation, it should also be said that since we are dealing with a complex potential from the start, it is not obvious that complex solutions for $h(\vph)$ should be discarded: what they would really imply, in the full system including both $u(\vph)$ and $h(\vph)$ is that what we mean by real and imaginary part gets mixed up. However, given the transformation properties of $u(\vph)$ and $h(\vph)$ under  $\vph\to-\vph$, this would also imply an explicit breaking of $\cP\cT$ invariance when $d<d_0$, which currently seems at odds with the  properties of $\cM(2,2n+3)$.

Given the puzzles above, and in the absence of independent confirmation,\footnote{It would be interesting to use truncated conformal bootstrap methods in order to test this annihilation scenario. In fact our annihilation of fixed points is somewhat reminiscent of the one found in \cite{Nakayama:2021zcr} for the supersymmetric Lee-Yang model.} right now the most conservative answer is that the ``$i\vphi^{2n+1}_{\text{bis}}$" fixed points and the related annihilation are probably spurious.
There is actually a simple argument for why we might expect that in the pure-imaginary LPA' it is impossible to follow the $i\vphi^{2n+1}$ fixed points all the way down to $d=2$, and that this problem might be lifted by a larger truncation of the derivative expansion.
The argument is the following. Denoting $\Delta_{\vph,n}= (d-2+\eta_n)/2$ the scaling dimension of $\vph$ at the $i\vphi^{2n+1}$ fixed point, if the latter should correspond in $d=2$ to the nonunitary minimal model $\cM(2,2n+3)$ or $\cM(2,4n+1)$, with the identification $\vph \sim \phi_{1,2}$, then we must have $\Delta_{\vph,n}\vert_{d=2} < \Delta_{\vph,1}\vert_{d=2}$, for $n>1$.
On the other hand, in the LPA' the anomalous dimension decreases monotonically with $g$ (we have $\eta\propto -g^2$ in the unrescaled variables, see \eqref{eq:eta-init}).\footnote{Note that this statement is independent of the choice of cutoff, as can be seen from \eqref{eq:etak}: in the complex potential case, the $V_k^{(3)}(0)^2$ in front always produces a $-g^2$ factor, while at $\phi=0$ the integral over $q$ is  $g$-independent and negative, thus canceling the explicit minus in front.} Since  the $i\vphi^{2n+1}$ fixed points, with corresponding initial condition $g^*_n$, pop out of the Gaussian one at $d=d_{2n+1}-\eps$ with $g^*_n\sim\sqrt{\eps}$, then if $n>1$, at such dimension the $i\vph^3$ fixed point $g^*_1$ has already moved farther away at larger values, i.e.\  $g^*_1\gg g^*_n$. As a consequence, we have $\Delta_{\vph,n}\vert_{d_{2n+1}-\eps} > \Delta_{\vph,1}\vert_{d_{2n+1}-\eps}$, for $n>1$ and $\eps\ll 1$.
Therefore, at some dimension $d$ such that $2<d<d_{2n+1}$, we must have $g^*_n=g^*_1$, i.e.\ the $i\vphi^{2n+1}$ fixed points with $n>1$ should cross the $i\vphi^{3}$ fixed point (and in fact also cross all the other model with $n'<n$).
Such crossings are clearly problematic and cannot occur without harm.
Notice that in the LPA, where we do not see nonperturbative fixed points or annihilations, we can push all the multicritical fixed points to $d\to2$, but indeed with wrong scaling dimensions, as the fixed points are always ordered on the $g$ axis with decreasing $n$, as can be seen in Fig.\ \ref{fig:trunc-im_phi26spikeLPA}, and thus we have $\Delta_{\vph,n} > \Delta_{\vph,1}$ even at $d=2$.

The problem described above could be circumvented within an approximation in which $\eta$ depends on more than one parameter, e.g.\ including the real part of the potential or going to next-to-leading order in the derivative expansion. For example, keeping also the real part of the potential, we also have the $\l_2$ parameter, see  \eqref{eq:eta-init}. And in a two- or higher-dimensional space of parameters the $i\vphi^{2n+1}$ fixed points could avoid each other even when their $\Delta_{\vph,n}$ cross each other.
On the other hand, it is clear that staying within the pure-imaginary LPA', but choosing a different cutoff, would not affect this aspect of the problem, as the anomalous dimension would still depend on only one parameter. 

Nevertheless, although the argument above explains why we might expect problems for the  $i\vphi^{2n+1}$ fixed points in the LPA', it does not explain the origin of the nonperturbative $i\vphi^{2n+1}_{\text{bis}}$ fixed points, and why the annihilation occurs with these new fixed points, rather than for example pairwise among the $i\vphi^{2n+1}$ ones, when they should cross each other.
Only further study will clarify this issue.

\paragraph{Acknowledgments.}
We thank Alessandro Codello, Bertrand Delamotte, Nicolas Dupuis, John Gracey, Igor Klebanov, Daniel Litim, Gilles Tarjus, and Matthieu Tissier for useful discussions and comments.

\appendix

\section{$\epsilon$-expansion of local potential equations}
\label{sec:osborn}

A procedure by O'Dwyer and Osborn can be applied to find the $\epsilon$-expansion of multicritical model of local potential approximations \cite{ODwyer:2007brp}.

\subsection{Polchinski equation}
\label{sect:osborn-polchinski}

%
It is simpler to describe the procedure using Polchinski's equation,\footnote{%
Polchinski's equation is generally presented as the RG flow with respect to an ultraviolet scale $\Lambda$ as opposed to the infrared scale $k$ of FRG. However, the methods of this appendix translate to the FRG case replacing $\Lambda \partial_\Lambda v(\varphi)=0$ with $k\partial_k v(\varphi)=0$ straightforwardly. To the leading order of the derivative expansion any dependence on the cutoff function can be factored out by appropriate rescalings \cite{Ball:1994ji}.
} in which case the RG flow of the local potential is
\be\label{eq:polchinski}
 -\Lambda \partial_\Lambda v(\varphi) = -d v(\varphi)
 + \frac{d-2}{2} \varphi v'(\varphi) - v''(\varphi) + v'(\varphi)^2
 \,,
\ee
and fixed points are solutions of $\Lambda \partial_\Lambda v(\varphi)=0$ as usual.
In the derivation of eq.\  \eqref{eq:polchinski} the potential $v(\varphi)$ is rescaled in such a way that the cutoff dependence is factored out completely.
We limit the first analysis of this appendix to the local potential case in which $\eta=0$
for clarity of the presentation, but all the results can be extended to include the anomalous dimension (see also Ref.\ \cite{ODwyer:2007brp} and section \ref{sect:polchinski-odd-beyond} below), as we have done for the case of the odd models that are considered in this paper in section \ref{sec:perturbation}.

We perform the rescaling $x = \frac{1}{2}(d-2)^{1/2} \varphi$
and define $w(x)=v(\varphi)$, so that the fixed point equation is equivalent to
\be\label{eq:polchinski-rescaled}
 \frac{2d}{d-2} w(x) - x w'(x) + \frac{1}{2} w''(x) -\frac{1}{2} w'(x)^2 = 0
 \,.
\ee
Assuming a regime in which a solution $w(x)$ is small, i.e.\ a solution that is sufficiently close to the Gaussian one, the linearized equation becomes 
\be\label{eq:polchinski-linearized}
 \frac{2d}{d-2} w(x) - x w'(x) + \frac{1}{2} w''(x) \approx 0 \,,
\ee
which is identical to the FRG counterpart given in eq.\  \eqref{eq:wetterich-linearized} of the main text. The polynomially bounded solutions are Hermite polynomials, which we repeat here for convenience,
\be\label{eq:w-hermite}
 w(x) \approx  \lambda \, H_n(x) \,, \qquad \qquad n = \frac{2d}{d-2} \,,
\ee
for some undetermined constant $\lambda$ and $n$ a positive integer. This shows that the requirement is satisfied only at critical dimensions such that $d =d_n = \frac{n}{n-2}$. The parameter $\lambda$ could be though of as the coupling of some multicritical model $\varphi^n$, because at large $x$ we have $H_n(x)\sim x^n$, which is marginally relevant below $d_n$. However, rather than finding its beta function and fixed point as a function of $d$, the strategy that we adopt is to solve \eqref{eq:polchinski-rescaled} using an $\epsilon$-expansion below $d_n$ and solving the couplings iteratively.

\subsection{Even models}\label{sect:polchinski-even}

If we require that $w(x)$ in eq.\  \eqref{eq:w-hermite} is bounded from below, we must also enforce that $\lambda>0$ and that $n \to 2n $ is even ($n\geq 1$). The $\epsilon$-expansion can then be set up below the (upper) critical dimensions $d_{2n}=\frac{n}{n-1}$, i.e.\  for $d= d_{2n}-\epsilon$, and
the nonlinear equation for $w(x)$ is solved iteratively using the inner product
$
 \left(f,g\right) = \int {\rm d}x \, {\rm e}^{-x^2} f(x) g(x) \,,
$
which was introduced also in the main text in \eqref{eq:w-norm} and is such that the Hermite polynomials form an orthogonal basis $\left(H_p,H_q\right)=2^p \sqrt{\pi} p! \delta_{p,q}$.

We start with the leading-order ansatz
\be
 w(x) = \epsilon \lambda_n H_{2n}(x) + O(\epsilon^2)\,,
\ee
which by definition solves eq.\  \eqref{eq:polchinski-rescaled} at the leading order in $\epsilon$ when $d=d_{2n}-\epsilon$. In order to fix the coefficient $\lambda_n$ we consider the order $\epsilon^2$ of eq.\  \eqref{eq:polchinski-rescaled}, which includes a contribution coming from expanding the coefficient $2d/(d-2)$ at the leading order in $\epsilon$ and a contribution coming from the nonlinear term $w'(x)^2$. Together they are
\be\label{eq:eps2-even}
 2 \left(n^2-2 n+1\right) \lambda_n H_{2 n}(x)-16 n^2 \lambda_n^2 H_{2 n-1}(x){}^2 \,,
\ee
where we have used the fact that $H_{2n}'(x)=4n H_{2n-1}(x)$.
Now we make use of the inner product, i.e.\ we project the above equation on $H_{2n}(x)$
as
\be\label{eq:even-cn-eqn}
 \int {\rm d}x \, {\rm e}^{-x^2} H_{2n}(x) \left\{2 \left(n^2-2 n+1\right) \lambda_n H_{2 n}(x)-16 n^2 \lambda_n^2 H_{2 n-1}(x){}^2\right\}=0 \, .
\ee
After integration, the first term is proportional to $\lambda_n$ up to an $n$-dependent coefficient. The second term requires knowledge of the integral of three Hermite polynomials, which is known in general (the coefficients satisfy a triangular inequality). For the purpose of this paper we need integrals of up to four Hermite polynomials, so it may be convenient to make repeated use of
\be
 H_p(x) H_q(x) = \sum_{r=0}^{{\rm min}(p,q)} r! \binom{p}{r}\binom{q}{r} H_{p+q-2r}(x) \,.
\ee
Applying this formula to the product $H_{2 n-1}(x){}^2$ and integrating over $x$, only the coefficient such that $r=n-1$ gives a contribution. We find the unique nontrivial solution for \eqref{eq:even-cn-eqn}
\be\label{eq:cn-even}
 \lambda_n= \frac{ (n-1)^2 n!^3}{2^{n}n(2 n)!^2} \,.
\ee
Notice that the constant $\lambda_n$ is greater than zero for $n\geq 1$, so the coefficient of the leading Hermite polynomial is positive, i.e.\  $\epsilon\, \lambda_n\geq 0$, only if $\epsilon>0$, implying that the $d_{2n}$s are actually \emph{upper} critical dimensions.
Now we expand the ansatz to include subleading corrections and we do it in such a way that the solution remains symmetric
\be
 w(x) = \epsilon \lambda_n H_{2n}(x) + \epsilon^2 \lambda_n  \sum_{p \geq 0} \alpha_{2p,n} H_{2p}(x)\,.
\ee
The normalization involving one power of $\lambda_n$ in the subleading terms
is chosen for convenience because it gives equations that are homogeneous in $\lambda_n$, which then factors out in the equations determining the coefficients $\alpha_{2p,n}$.
The subleading corrections do not influence the determination of $\lambda_n$ as long as we consider the case $p\neq n$. We can thus insert the ansatz in eq.\  \eqref{eq:polchinski-rescaled}
and at order $\epsilon^2$ we find that in addition to the terms \eqref{eq:eps2-even} there are
also
\be\label{eq:eps2-even-part2}
 \lambda_n \sum_{p\neq n}\left[4 n  \alpha_{2p,n} H_{2 p}(x)
 +8 p (2 p-1 -x) \alpha_{2p,n} H_{2 p-2}(x)\right] \,.
\ee
where the instance of the variable $x$ can be replaced by the polynomial $H_1(x)=2x$.
Now we again use the inner product to project the sum of \eqref{eq:eps2-even} and \eqref{eq:eps2-even-part2} but this time on a general $H_{2p}(x)$ with $p\neq n$. The projection has contributions from both equations, but the requirement that $p\neq n$ ensures that the subleading terms do not affect the original determination of $\lambda_n$.
Due to the triangular inequality on the integral of three polynomials, the nonlinear term of the Polchinski equation contributes only if $p$ is bound to $0\leq p \leq 2n-1$.
The solution to the equation is
\be\label{eq:apn-even}
 \alpha_{2p,n} = \frac{(n-1)^2 (n!)^3 2^{n-p-1} (2 n-p)}{n (p!)^2 (n-p) (2 n-p)!} \,,
\ee
and analytic continuation of this formula verifies that $\alpha_{2p,n}=0$ if $p \geq 2n$.
Finally, the correction to the leading term, that is, the $\alpha_{2n,n}$ contribution, can be obtained by including the case $p=n$ and solving eq.\  \eqref{eq:polchinski-rescaled} to order $\epsilon^3$
\be\label{eq:ann-even}
 \alpha_{2n,n}= \frac{n-1}{2}- \frac{(2 n)! \lambda_n}{(n-1)^2} \sum_{p \neq n} \frac{2^{p+1} p  (2 p)! }{ (p!)^2 (2
   n-p)!}  \alpha_{2p,n} \,,
\ee
which, together with the previous solutions, confirms the result reported in Ref.\ \cite{ODwyer:2007brp}.
The final solution is 
\be
 w(x) = \epsilon \lambda_n H_{2n}(x) + \epsilon^2 \lambda_n  \sum_{p \geq 0}^{2n-1} \alpha_{2p,n} H_{2p}(x)+ O(\epsilon^3) \,,
\ee
with the coefficients given in Eqs.\ \eqref{eq:cn-even}, \eqref{eq:apn-even} and \eqref{eq:ann-even}. The potential $v(\varphi)$ can be obtained easily by substituting $x= \frac{1}{2}(d-2)^{1/2}\vphi = (\frac{1}{\sqrt{2} \sqrt{n-1}}-\frac{\sqrt{n-1} \epsilon }{4 \sqrt{2}})\vphi+O(\epsilon^2) $ and expanding once more in $\epsilon$.

We can estimate the convergence of the expansion as a function of $x$ by comparing the leading contribution versus the strongest subleading one.
The highest value of the number $p$ is $2n-1$, so the dominating subleading term is $H_{4n-2}(x)$, and the comparison gives
$ \epsilon H_{2n} \sim \epsilon^2 H_{4n-2}$. At large $x$ we have that $H_p(x) \propto x^p$, which means that the expansion fails when $\epsilon x^{2n-2} \sim 1$. By fixing $\epsilon$ to a small but finite value this conforms with the notion that the solution enters a nonperturbative large-field regime and the expansion in terms of Hermite polynomials is no longer valid.
Notice also that, because of our rescaling $\varphi \propto \Delta^{-1/2} x$, where $\Delta$ is the dimension of the operator $\varphi$, so in any limit in which $\Delta \to 0$ (e.g., when $d\to 2$ in the LPA) a finite $\varphi$ requires a large $x$, which inevitably brings the equation in its nonperturbative regime.

The analysis of fluctuations around the given fixed point $w(x)$ solution can be set up as a Hamiltonian problem. Let $v(\varphi) \to v(\varphi) + {\rm e}^{-\theta t} \delta v(\varphi)$ in eq.\  \eqref{eq:polchinski}, where $\delta v(\varphi)$ is a deformations of the fixed point solution with critical exponent $\theta$ ($t=-\log\Lambda$ instead of $t=\log k$ because $\Lambda$ is an ultraviolet scale). After the same rescaling of the field $\varphi$, the fluctuation obeys the linear equation
\be\label{eq:polchinski-fluctuations}
 {\cal L} \, \delta w \equiv \hat{\theta} \delta w(x)
 -  \left(x+w'(x)\right) \delta w'(x)
 +\frac{1}{2}\delta w''(x)
 =0
 \,,
\ee
where
$\hat{\theta}=\frac{2(d-\theta)}{d-2}$ (the notation in the same as Ref.\ \cite{ODwyer:2007brp}).
This is a Hamiltonian problem 
\be\label{eq:polchinski-fluctuations-2}
 {\cal H}\delta w 
 =\hat{\theta} \delta w
 \,,
 \qquad \qquad 
 {\cal H} \equiv -\frac{1}{2} \frac{{\rm d}^2}{{\rm d}x^2}+\left(x+w'(x)\right)\frac{{\rm d}}{{\rm d}x}  \,,
\ee
with eigenvalue $\hat{\theta}.$
In the limit $\epsilon\to 0$ the solution is such that $w(x)\to 0$, then the requirement that fluctuations are also polynomially bounded gives a spectrum bounded from below with leading contributions
$\delta w(x) \approx H_r(x)$ and
$\hat{\theta} \approx r \in \mathbb{N}$, which is the spectrum of fluctuations of the Gaussian fixed point. Notice that $r$ is a natural number (not necessarily even) such that both even and odd deformations are included.
At the leading nontrivial order in $\epsilon$, we parametrize
\be
 \delta w(x) = H_r(x) + O(\epsilon) \,, \qquad \qquad \hat{\theta}_r = r + \epsilon \, \hat{\theta}_r^{(1)} + O(\epsilon^2)\,.
\ee
Obviously the spectrum is not degenerate as long as $\epsilon$ is sufficiently small.
In order to determine $\hat{\theta}_r^{(1)}$, we project $\left(H_r, {\cal L}\delta w\right)=0$ and solve
the order $\epsilon$ equation.
The result is expressed in general as
\be
 \hat{\theta}_r^{(1)}=\frac{2 (n-1)^2 k! n!}{(2 n)! (r-n)!} \,,
\ee
and it is easy to verify that $\hat{\theta}_r^{(1)}=0$ when $r<n$ by analytic continuation.
Further corrections to the eigenfunctions would require standard quantum mechanical perturbation theory, but as long as we are interested in the leading order corrections to the critical exponents, then linear perturbation theory is sufficient. For the exponent $\theta_r$ we find, inverting the expression of $\hat{\theta}_r$ in terms of $\theta_r$
and expanding $d$,
\be
 \theta_r =
 \frac{r-2 n}{1-n} + \epsilon  \left(\frac{r}{2}-1-\frac{\hat{\theta}_r^{(1)}}{n-1}\right)+ O(\epsilon^2)
   =
   \frac{r-2 n}{1-n} + \epsilon  \left(\frac{r}{2}-1-\frac{2 (n-1) r! n!}{(2 n)! (r-n)!}\right)+ O(\epsilon^2)
 \,,
\ee
for $r<2n$.
Further results and a discussion on the universal character of the above results can be found in Ref.\ \cite{ODwyer:2007brp}.

\subsection{Odd models}\label{sect:polchinski-odd}

Now we build on the method of Ref.\ \cite{ODwyer:2007brp}
by relaxing the requirement that the solutions in eq.\  \eqref{eq:w-hermite} are bounded from below.
As such, we can include the cases in which $n$ is odd, i.e.\  $n\to 2n+1$. This reveals the critical dimensions of the odd models $d_{2n+1} = \frac{4n+2}{2n-1}$. The expansion $d=d_{2n+1}-\epsilon$
shares some similarities with the even case, but also some key differences in the form of the expansion and the order in which equations are solved that must be taken into account carefully.

To understand the structure of the expansion first, we first take
$w(x) \approx c H_{2n+1}(x)$ in eq.\  \eqref{eq:polchinski-rescaled}
and project the result onto $H_{2n+1}(x)$ itself using the Hermite norm \eqref{eq:w-norm}.
The term $w'(x)^2 \propto H_{2n}(x)^2$ is an even polynomial, so it does not contribute to the projected equation and we would assume that $c=0$ is the only solution. However, the fact that $w'(x)^2$ is even implies that a nonzero $c$ generates even subleading terms proportional to $c^2$, which, if taken into account, contribute to $w'(x)^2$ with odd polynomials proportional to $c^3$.
The net result is that $c$ satisfies a cubic equation which, schematically, is of the form $-\epsilon c + c^3=0$ with two nontrivial complex conjugate roots proportional to $\pm i \sqrt{\epsilon}$ as long as $\epsilon>0$.
This suggests that the proper parameter for the expansion of the fixed point solutions is actually $\sqrt{\epsilon}$, rather than $\epsilon$, in the case of the odd models. This also agrees with the expectations coming from perturbation theory and dimensional regularization \cite{Codello:2017epp}.

In view of the preliminary considerations, we take the ansatz
\be\label{eq:w-ansatz}
 w(x) = \sqrt{\epsilon} \lambda_n  H_{2n+1}(x) + \epsilon \lambda_n^2 \sum_{q=0}^\infty \alpha_{q,n} H_q(x) + O(\epsilon^{3/2}) \, ,
\ee
where the power $\lambda_n^2$ in front of the subleading term is chosen to make the equations for the subleading corrections homogeneous in analogy to the even case.
By construction, the ansatz solves eq.\  \eqref{eq:polchinski-linearized}
with $d = d_{2n-1} -\epsilon$
at the leading order in $\epsilon^{1/2}$, but $\lambda_n$ is still arbitrary.
In order to solve the equation at order $\epsilon$ we perform the projection
on $H_p$ for arbitrary $p$, but $p$ turns out to be even for the reasons discussed above. The computation requires the integral of three Hermite polynomials, thus $p$ is also bound to satisfy a triangular inequality.
Solving $\left(H_{2p}, \Lambda\partial_\Lambda v\right)=0$, we find
\be\label{eq:apn-sol-polchinski-even}
 \alpha_{2p, n} = \frac{2^{2 n-p+1} (2 n+1)^2 (2n)!^2}{ (2n-2p+1)p!^2 }\,,
\ee
and holds for $0\leq p \leq 2n$, which is also confirmed by analytic continuation, while $\alpha_{2p+1,n}=0$.\footnote{%
There is an important difference with the main text in which, for the Wetterich equation, the label $p$ satisfies $0\leq p \leq 2n-1$, see Eq.\ \eqref{eq:w-ansatz-main-general}. The difference is relevant when comparing odd and even parts of the solution.
}
Notice that $\alpha_{p=0,n}=0$ and that the coefficients $\alpha_{2p, n}$ become negative for $p>n$.
The solution to $\lambda_n$ can be found projecting $\left(H_{2n+1},\Lambda\partial_\Lambda v\right)=0$ at order $\epsilon^{3/2}$,
which gives
\be\label{eq:cn-odd-equation}
\sqrt{\pi } 2^{2n-1}(2 n+1)!\lambda_n\left( 
   (2 n-1)^2 
   -\lambda_n^2 \sum_{p=1}^{2n}\frac{ 2^{p+4} (2 n+1) p^2  
   (2 n)!   (2 p-1)!
   }{p!^2 (2 n-p+1)!} \alpha_{2p,n}
   \right)=0 \,,
\ee
having used the fact that $\alpha_{0,n}=0$.
To determine the nontrivial solutions, we need to insert the explicit form of $\alpha_{2p,n}$ and perform the summation over $p$.
Let 
\be
 F_n \equiv -\sum_{p=1}^{2n}\frac{ 2^{p+4} (2 n+1) p^2  
   (2 n)!   (2 p-1)!
   }{p!^2 (2 n-p+1)!} \alpha_{2p,n} 
   =
   \sum_{p=1}^{2n}\frac{ 2^{2 n+5} (2 n+1)^3 p^2  
   (2 n)!^3   (2 p-1)!
   }{(2p-2n-1) p!^4 (2 n-p+1)!}
   \,,
\ee
which is a \emph{positive} number for $n\geq 1$ that could be expressed as a hypergeometric function (notice the overall negative sign).
Then we find the expected nontrivial solutions
\be
 \lambda_n = \pm i \frac{2n-1}{\sqrt{F_n}} \,,
\ee
that appear as a complex conjugates pair thanks to the fact that $F_n>0$.

A solution for $w(x)$ is obtained by inserting any of the two $\lambda_n$ roots and the coefficients $\alpha_{2p,n}$ in eq.\  \eqref{eq:w-ansatz}. Differently than in the even case, the solution is not invariant under the parity transformation $x\to -x$, though there is another type of ``symmetry.'' To see it, first notice that
complex conjugation acts on any of the two solution by sending it to the other one, i.e.\  $ w(x) \to w^\star(x)$ that is equivalent to $\lambda_n \to -\lambda_n$. Even terms are multiplied by $\lambda_n^2$ and $\alpha_{2p,n}$ that are real numbers, so the even part of the solution is unaffected, while the odd part changes by an overall sign. This means that the solutions are invariant under the combination of parity and complex conjugation,
that is
\be
 w(x) = w^\star(-x) \qquad \Longrightarrow \qquad v(\varphi) = v^\star(-\varphi) \,,
\ee
as long as $\epsilon$ is sufficiently small.
The expansion of the complex conjugate pair of solutions can, in principle, be iterated to further orders by incorporating in the ansatz \eqref{eq:w-ansatz} more and more terms.
The general structure reveals that semi-odd powers of $\epsilon$
multiply odd Hermite polynomials and have purely imaginary coefficients,
while the integer powers of $\epsilon$ multiply even Hermite polynomials
with real coefficients in agreement with $v(\varphi) = v^\star(-\varphi)$.
To estimate the convergence, we need the dominating power of the $\epsilon^{3/2}$ term,
which is
\be
 w(x)|_{\epsilon^{3/2}} \propto H_{6n-1}(x)\,.
\ee
We can estimate the convergence of the expansion as a function of $x$ comparing
the leading and subleading terms, but in the spirit of the main text we shall do it independently
for the even and odd contributions.
Using the odd part, the comparison gives $\epsilon^{1/2} H_{2n-1} \sim \epsilon^{3/2} H_{6n-1}$, implying $\epsilon x^{4n-2} \sim 1$ at large $x$, so the solution must enter a nonperturbative regime at a finite value of $x$ as expected. Interestingly, the comparison of the even part reveals the same onset (it can be obtained by replacing $2n\to 4n$ in the pertinent equation for the even multicritical models of section \ref{sect:polchinski-even}).

Similarly to the even case, we can carry out 
the analysis of fluctuations around a given fixed point, which could be any of the two roots as the spectrum is the same (to fix ideas, take the one proportional to $+i$). The Hamiltonian equation is exactly the same as in the even case given in eq.\  \eqref{eq:polchinski-fluctuations-2} and we use the same definitions for the spectrum.
At the leading nontrivial order in $\epsilon$, we parametrize
\be
 \delta w(x) = H_r(x) + O(\epsilon^{1/2}) \,, \qquad \qquad \hat{\theta}_r = r + \epsilon \, \lambda_n^2\, \hat{\theta}_r^{(1)} + O(\epsilon^2)\,.
\ee
Notice that the critical exponents are analytic functions of $\epsilon$, which can be confirmed by a direct computation, so we have not included a term proportional to $\epsilon^{1/2}$. 
In order to determine $\hat{\theta}_r^{(1)}$, we could follow the same strategy as in section \ref{sect:polchinski-even}
of projecting $\left(H_r, {\cal L}\delta w\right)=0$ to find
a complicate expression, that we do not give for brevity, in which only $0\leq p\leq r$ contribute
to the $r$th critical exponent.
For the exponent $\theta_r$ we find, inverting the expression of $\hat{\theta}_r$ in terms of $\theta_r$
and expanding $d$,
\be
 \theta_r = \frac{4n+2-2r}{2n-1}+\frac{1}{2}\epsilon\left(r+2 
 -\frac{4\hat{\theta}_r^{(1)} \lambda_n^2}{2n-1} \right)
  + O(\epsilon^2)\,,
\ee
for $r\leq 2n+1$.
This spectrum is universal in the sense that it does not depend on the cutoff, but it does receive contributions from the inclusion of the anomalous dimension. We return on this in section \ref{sect:polchinski-odd-beyond}.

\subsection{The special cases $\vphi^3$ and $\vphi^5$}
\label{sect:polchinski-special}

The Lee-Yang $\varphi^3$ fixed point is obtained by setting $n=1$ in the solution $w(x)$.
Using $x = \frac{1}{2}(d-2)^{1/2} \varphi$ and re-expanding $d=6-\epsilon$, we find
\be
 v(\varphi) = \frac{i \vphi  \left(2 \vphi ^2-3\right) \epsilon^{1/2} }{24
   \sqrt{3}}+\frac{\left(12 \vphi ^4-60 \vphi ^2+19\right)
   \epsilon }{1152}+O\left(\epsilon ^{3/2}\right) \,.
\ee
The spectrum for the operators $H_r\sim \varphi^r$ is
\begin{eqnarray}
\theta_r
 &=& \left\{6-\epsilon, \, 4-\frac{1}{2} \epsilon, \, 2- \frac{1}{2} \epsilon,  \, -\epsilon \,, \cdots \right\}+O(\epsilon^2)\,.
\end{eqnarray}
The first multicritical model $\varphi^5$ is obtained by setting $n=2$ and has upper critical dimension $d_5=\frac{10}{3}$. Following the same steps we obtain
\be
 v(\vphi)=\frac{i \vphi  \left(4 \vphi ^4-60 \vphi ^2+135\right)
   \epsilon^{1/2} }{4800 \sqrt{6}}+\frac{\left(80 \vphi ^8+2400
   \vphi ^6-130680 \vphi ^4+793800 \vphi ^2-478467\right)
   \epsilon }{55296000}+O\left(\epsilon ^{3/2}\right) \,.
\ee
and
\begin{eqnarray}
\theta_r
 &=&
 \left\{
 \frac{10}{3}-\epsilon, \, \frac{8}{3}-\frac{1}{2}\epsilon, \, 2+\frac{3}{160}\epsilon, \, \frac{4}{3}+\frac{7}{32}\epsilon, \, \frac{2}{3}-\frac{1}{2}\epsilon,\, -3\epsilon,\, \cdots 
 \right\}+O\left(\epsilon ^2\right)\,.
\end{eqnarray}

\subsection{Odd models beyond the LPA}
\label{sect:polchinski-odd-beyond}

In order to go beyond the LPA we include to eq.\  \eqref{eq:polchinski} the effects of interactions with two derivatives and a corresponding potential $z(\vphi)$, so that the RG equations of $v(\varphi)$ and $z(\varphi)$ should be solved as a coupled system. The next-to-leading order of the LPA of Polchinski equation is generally presented in a form such that the limit $z\to 0$ corresponds to the LPA approximation because $z(\vphi)$ represents an interaction term \cite{Ball:1994ji}. To make the analogy with the Wetterich LPA' more transparent we have redefined $z(\vphi) \to z(\vphi)-1$ in the results of Ref.\ \cite{Ball:1994ji}\footnote{%
The LPA corresponds to $z=0$ and $\eta=0$ in Ref.\ \cite{Ball:1994ji}.
}
and this results in a coupled system of equations
\begin{eqnarray}\label{eq:polchinski-beyond}
 -\Lambda \partial_\Lambda v(\varphi) &=& -d v(\varphi)
 + \frac{d-2+\eta}{2} \varphi v'(\varphi) - v''(\varphi) + v'(\varphi)^2 - 2 K_1 \left(z(\vphi)-1\right)
 \,,
 \\
 -\Lambda \partial_\Lambda z(\varphi) &=& -\frac{\eta}{2}+\eta \, z(\varphi)
 + \frac{d-2+\eta}{2} \varphi z'(\varphi) - z''(\varphi) + 2 v'(\varphi) z'(\varphi) + 4 v''(\varphi) (z(\varphi)-1) -K_2 v''(\varphi)^2\,,
 \nonumber
\end{eqnarray}
where $K_1$ and $K_2$ are cutoff dependent constants that cannot be rescaled away \cite{Ball:1994ji}.
Eqs.\ \eqref{eq:polchinski-beyond} reduce to \eqref{eq:polchinski} in the limit $z(\vphi)\to 1$ and $\eta\to 0$, which is cutoff independent.
The important property of \eqref{eq:polchinski-beyond} is that, even in the limit
$z(\vphi) \to 1$, the second equation provides a way to determine $\eta$ as in the LPA' expansion of the Wetterich equation. As LPA' expansion of Polchinki equation we thus take the system
\begin{eqnarray}\label{eq:polchinski-beyond-2}
 \Lambda \partial_\Lambda v(\varphi)|_{z\to 1} &\propto& -d v(\varphi)
 + \frac{d-2+\eta}{2} \varphi v'(\varphi) - v''(\varphi) + v'(\varphi)^2 
 \,,
 \\
 \Lambda \partial_\Lambda z(\varphi)|_{z\to 1} &\propto& \eta -2 K_2 v''(\varphi)^2\,,
 \nonumber
\end{eqnarray}
and now only one cutoff dependent constant has survived the limit. We should stress that this is potentially an improper way to solve Polchinski's equation, with the correct procedure bein described in Ref.\ \cite{ODwyer:2007brp} for the full second order of the derivative expansion. For example the numerical analysis generally reveals that $z(0)\neq 1$, while here we are treating it as if it was a wavefunction renormalization that can be rescaled as an inessential coupling. This presentation however is useful for its analogy with section \ref{sec:perturbation} of the main text in which FRG is used.

The method outlined in section \ref{sect:osborn-polchinski} is almost unaltered
except that now it is necessary to consider the rescaling $x = \frac{1}{2}(d-2+\eta)^{1/2} \varphi$ and $w(x)=v(\vphi)$, which is almost the same that was considered in the main text for the Wetterich equation. The rescaling includes $\eta$, so the final formula for $v(\vphi)$ will require an additional expansion, but this will be possible order-by-order in $\epsilon$.
For the expansion we modify the ansatz \eqref{eq:w-ansatz} to
\begin{eqnarray}\label{eq:w-ansatz-beyond}
 w(x) &=& \sqrt{\epsilon} \lambda_n  H_{2n+1}(x) + \epsilon \lambda_n^2 \sum_{q=0}^\infty \alpha_{q,n} H_q(x) + O(\epsilon^{3/2}) \, ,
 \\
 \eta &=& \epsilon \lambda_n^2 \eta^{(1)} + O(\epsilon^{3/2}) \, ,
\end{eqnarray}
which is essentially the same as the one given in eq.\  \eqref{eq:w-ansatz-main} of the main text. The leading correction to the anomalous dimension is determined by solving
\be
 \left(1, \Lambda \partial_\Lambda z(\varphi)|_{z\to 0}\right)=0
 \qquad \Longrightarrow \qquad
 \eta^{(1)} = \tilde{K}_2 \frac{2^{2 n+5} n^2 (2 n+1)^3 (2 n+1)!}{2 n-1}\,.
\ee
where $\tilde{K}_2 =  \frac{2 K_2}{d - 2 + \eta}$ is the rescaled coefficient of the $\Lambda \partial_\Lambda z$ equation.
The anomalous dimension should be determined by re-expanding $\eta$ in its own formula
\begin{eqnarray}
 \eta &=& \frac{2 K_2}{d - 2 +\eta} \frac{2^{2 n+5} n^2 (2 n+1)^3 (2 n+1)!}{2 n-1} \lambda_n^2 +O(\epsilon^2)
 \\
 &=& K_2 \frac{2^{2 n+4} n^2 (2n-1)(2 n+1)^3 (2 n+1)!}{2 n-1} \lambda_n^2 +O(\epsilon^2) \,,
\end{eqnarray}
but we still have to compute $\lambda_n$.
The coefficients $\alpha_{q,n}$ are determined as in the LPA case and their final form is the same as eq.\  \eqref{eq:apn-sol-polchinski-even}. The anomalous dimension in the scaling term of \eqref{eq:polchinski-beyond-2} contributes nontrivially in the determination of $\lambda_n$, to eq.\  \eqref{eq:cn-odd-equation}
we must add
\be\label{eq:cn-odd-equation-beyond}
\sqrt{\pi } 2^{2n-1}(2 n+1)! \lambda_n^3 \left( -2^{2n+5 } \tilde{K}_2 n^2 (2n+1)^3 (2n-1)! \right)\,,
\ee
which changes slightly the complex pair of solutions for $\lambda_n$.
The anomalous dimension also affects the spectrum directly
\be
 \theta_r = \frac{4n+2-2r}{2n-1}+\frac{1}{2}\epsilon\left(r+2 -\left(r \eta^{(1)} 
 +\frac{4\hat{\theta}_r^{(1)} }{2n-1} \right)\lambda_n^2 \right)\,
  + O(\epsilon^2)\,.
\ee
where $\hat{\theta}_r^{(1)}$ is determined as discussed in section \ref{sect:polchinski-odd}.
Notice that, when computed in this way the spectrum is still correctly determined only for relevant operators.

\section{Large-field analysis of the fixed point equations at \texorpdfstring{$\Delta\neq 0$}{Delta not 0}}
\label{app:asymptotics}

The asymptotic behaviour of $\tilde{u}$ and $\tilde{h}$ at large $\tilde{\vphi}$ can be classified according to the possible behaviours of the non-linear RHS of \eqref{eq:coupled-rescaled}, namely:
\begin{enumerate}
    \item diverging RHS (unbounded potential);
    \item constant RHS (Gaussian family);
    \item vanishing RHS (non-trivial potential).
\end{enumerate}

We do not consider the possibility that the RHS (or the potential itself) has no limit for $\tilde{\vphi}\to +\infty$, or in other words that it settles into an oscillatory regime, because we have found no hints of such behaviour at $\Delta\neq 0$.
One practical explanation is that any periodic behaviour of terms involving $\tilde{v}(\tilde{\vphi})$ and $\tilde{v}''(\tilde{\vphi})$ will in general be spoiled by the $\tilde{\vphi}\, \tilde{v}'(\tilde{\vphi})$ term.

Note that the behaviour of the RHS is mostly dominated by the denominator common to both ODEs \eqref{eq:coupled-rescaled}.
Thus, in general the RHS of the two ODEs present the same behaviour at large $\tilde{\vphi}$, although the behavior of $\tilde{u}$ and $\tilde{h}$ might differ, as we will see.\medskip

\subsection{Diverging RHS}

Because of the different powers in the numerator and denominator, the RHS of \eqref{eq:coupled-rescaled} can only diverge in the large $\tilde{\vphi}$ limit when the square norm of the shifted potential vanishes $\abs{1+\tilde{v}''}^2=(1+\tilde{u}'')^2+(\tilde{h}'')^2\tend0$, or in other words, when $\tilde{u}''\tend A$ and $\tilde{h}''\tend B$, with $A$, $B\in\R$ such that $(1+A)^2+B^2=0$.
The unique solution is given by $A=-1$ and $B=0$.

The leading order in the large $\tilde{\vphi}$ limit of the potential corresponding to this behaviour is
\be
\tilde{u}(\tilde{\vphi}) \sim -\half \tilde{\vphi}^2, \quad \tilde{h}(\tilde{\vphi})\sim B_1 \tilde{\vphi},\ B_1\in\R.
\ee
The subleading behaviour can be obtained by inverting the LHS and RHS of \eqref{eq:coupled-rescaled} 
\be\begin{systeme}
\frac{\(\tilde{h}''\)^2+\(1+\tilde{u}''\)^2}{1+\tilde{u}''}= \, \frac{1}{u- \gamma\, \tilde{\vphi}\, \tilde{u}'}\\
\frac{\(\tilde{h}''\)^2+\(1+\tilde{u}''\)^2}{\tilde{h}''}= -\, \frac{1}{h- \gamma\,  \tilde{\vphi}\, \tilde{h}'}\\
\end{systeme}
\ee
and plugging back the leading behaviour. 
The general solution is then given by
\be \label{eq:unbound-sol}
\begin{systeme}
    \tilde{u}(\tilde{\vphi}) = -\half \tilde{\vphi}^2 + A_1\, \tilde{\vphi} + \frac{2  }{1-2\gamma} \ln\(\tilde{\vphi}\) + A_2 + \bigO{\frac{1}{\tilde{\vphi}}}\\
    \tilde{h}(\tilde{\vphi}) = B_1 \tilde{\vphi}  + B_2 + \bigO{\frac{1}{\tilde{\vphi}}}.
\end{systeme}
\ee

The asymptotic solution is determined by four free parameters $A_1,A_2, B_1, B_2 \in\R$ as expected for two second-order ODEs.

Since the real part of the potential is unbounded from below, these solutions are unphysical.

\subsection{Constant RHS}

We now consider the case of the RHS tending to a nonvanishing constant. Because of the different powers in the numerator and denominator, we must have 
$(1+\tilde{u}'')^2+(\tilde{h}'')^2\tend\,$constant, i.e.\ $\tilde{u}''\tend A$ and $\tilde{h}''\tend B$, with $A, B \in \R$ such that $(1+A)^2+B^2=\const\neq0$.
Then, expanding the RHS at large $\tilde{\vphi}$ and keeping the leading behaviour, we find
\be\begin{systeme}
 \tilde{u}- \gamma\, \tilde{\vphi}\, \tilde{u}' \simeq  \, \frac{1+A}{\(1+A\)^2+B^2}\\
 \tilde{h}- \gamma\, \tilde{\vphi}\, \tilde{h}'\simeq -  \, \frac{B}{\(1+A\)^2+B^2},
\end{systeme}
\ee
with general solutions
\be\label{eq:ansatzconst}
\tilde{u}(\tilde{\vphi})= a\, \tilde{\vphi}^\frac{1}{\gamma} +  \frac{1+A}{\(1+A\)^2+B^2}+o(1),\quad \tilde{h}(\tilde{\vphi})= b\, \tilde{\vphi}^\frac{1}{\gamma} - \frac{B}{\(1+A\)^2+B^2}+o(1),
\ee
for some parameters $a,b \in \R$.
However, these solutions are consistent with the assumed behaviour of the second derivatives if and only if one of the following situation is realized:
\begin{enumerate}
\item  $\gamma=1/2$, and $a=A$ and $b=B$;
\item  $0<\gamma<1/2$, and $a=b=A=B=0$;
\item  $\gamma>1/2$ or $\gamma<0$, and $A=B=0$, but $a$ and $b$ arbitrary.
\end{enumerate}

The first case corresponds to $\Delta=d/2$, and we will not consider it further because it does not occur in the models we are interested in.

In the other two cases we always have $A=B=0$, hence we can write 
\be \label{eq:lin-ansatz}
\tilde{u}(\tilde{\vphi}) = 1 + \delta \tilde{u}(\tilde{\vphi}),\quad \tilde{h}(\tilde{\vphi})=\delta \tilde{h}(\tilde{\vphi}), \qquad  |\delta \tilde{u}''|,\, |\delta \tilde{h}''|\ll1.
\ee

In fact, we notice immediately that $\delta \tilde{u}=\delta \tilde{h}=0$ is an exact solution: the Gaussian fixed point (GFP), with constant potential $V(\vphi)=1$.\footnote{Notice that a vanishing potential is not a solution because in the way the Wetterich equation is derived the measure is normalized with respect to the unmodified quadratic part (i.e.\ without $R_k$).}

We can check whether the GFP solution is an isolated global solution by substituting \eqref{eq:lin-ansatz} in \eqref{eq:coupled-rescaled} and expanding the RHS in $\delta \tilde{u}''$ and $\delta \tilde{h}''$.
At linear order, the resulting equations are
\be \label{eq:lin-test}
\begin{systeme}
 \delta \tilde{u}- \gamma\, \tilde{\vphi}\, \delta \tilde{u}' =  -\delta \tilde{u}''\\
 \delta \tilde{h}- \gamma\, \tilde{\vphi}\, \delta \tilde{h}'= - \delta \tilde{h}''.
\end{systeme}
\ee
These are solved by linear combinations of ${}_1F_1(-1/2\g,1/2,\tilde{\vphi}^2\g/2)$ and  $\tilde{\vphi} \, {}_1F_1((\g-1)/2\g,3/2,\tilde{\vphi}^2\g/2)$, whose asymptotic behaviour for $\tilde{\vphi}\to\infty$ is a linear combination of $\tilde{\vphi}^{-1-1/\g} \exp(\tilde{\vphi}^2\g/2)$ and $\tilde{\vphi}^{1/\g}$.
Their fate depends on the value of $\gamma$:
\begin{itemize}
\item For or $0<\gamma<1/2$, both solutions are in contradiction with the hypothesis that $|\delta \tilde{u}''|$ and $|\delta \tilde{h}''|$ are small at large $\tilde{\vphi}$, hence they must be discarded and thus we conclude that the GFP is an isolated global solution.
In this case, if we want to find nontrivial global solutions, we need to change working hypothesis, as we will do in the next subsection.
\item For $\g>1/2$, the solutions are either exponentially growing, and thus to be discarded, or they are of the type in \eqref{eq:ansatzconst}, with $A=B=0$.  We thus find a 2-parameter family of solutions, with parameters $a$ and $b$ in \eqref{eq:ansatzconst}.
\item For $\g<0$, the exponential solutions are decaying ones, hence they are allowed. 
We thus find a 4-parameter family of solutions, with parameters $a$ and $b$ in \eqref{eq:ansatzconst}, plus two parameters for the exponential corrections.
\end{itemize}

\subsection{Vanishing RHS}

The last case we consider is when both RHS of \eqref{eq:coupled-rescaled} vanish, which happens if and only if $(1+\tilde{u}'')^2+(\tilde{h}'')^2\tend\infty$. We can further decompose this case in three sub-cases:
\begin{itemize}
    \item both $\tilde{u}''$ and $\tilde{h}''$ diverge, leading to a complex asymptotic solution,
    \item only $\tilde{u}''$ diverges, leading to a purely real asymptotic solution, and
    \item only $\tilde{h}''$ diverges, leading to a purely imaginary leading behaviour but complex subleading contributions.
\end{itemize}

For all cases, given a vanishing RHS, the two ODEs \eqref{eq:coupled-rescaled} decouple at leading order in $\tilde{\vphi}$ and they reduce to the classical scaling equation (LHS equal zero), which is solved by
\be\label{eq:vanishinglhs}
\tilde{u}(\tilde{\vphi}) \sim A \tilde{\vphi}^\frac{1}{\gamma}, \quad \tilde{h}(\tilde{\vphi}) \sim B \tilde{\vphi}^\frac{1}{\gamma}.
\ee
The assumption that at least one of the second derivatives of the potential diverges implies that
\be \label{eq:etabound}
\frac{1}{\gamma} > 2\quad \Leftrightarrow\quad d>2\Delta>0 \quad \Leftrightarrow\quad  2-d < \eta < 2.
\ee
Otherwise, for $-\infty<\frac{1}{\gamma}\leq 2$, we end up again in the previous case of asymptotically constant RHS.

To compute the subleading contributions in the large $\tilde{\vphi}$ limit up to some order $n$, we consider the two parameters ansatz:
\be\label{eq:vanishing-asympto}
\tilde{u}(\tilde{\vphi}) = A \tilde{\vphi}^\frac{1}{\gamma} + \sum_{k=1}^{n}\sum_{j=1}^{k} a_{j,k}(A,B) \tilde{\vphi}^{2j-\frac{1}{\gamma} k},\quad 
\tilde{h}(\tilde{\vphi}) = B \tilde{\vphi}^\frac{1}{\gamma} + \sum_{k=1}^{n}\sum_{j=1}^{k} b_{j,k}(A,B) \tilde{\vphi}^{2j-\frac{1}{\gamma} k}.
\ee
The coupled ODEs can be solved order per order by feeding iteratively the RHS with the ansatz, keeping the leading behaviour in $\tilde{\vphi}$ and solving for the coefficients $a_{j,k}$ and $b_{j,k}$.

It can be checked via a linearization test  \cite{Morris:1994ki}, that we are not missing other free parameters in the solutions.\footnote{For the linear perturbations we can consider the exponential ansatz
$\delta \tilde{u}(\tilde{\vphi})=A e^{S_u(\tilde{\vphi})}$, $\delta \tilde{h}(\tilde{\vphi}) = B e^{S_h(\tilde{\vphi})}$,
and using the method of dominant balance, it can then be shown the linearized equations lead to the contradiction $\delta \tilde{u} \gg \tilde{u}$ and $\delta \tilde{h} \gg \tilde{h}$.}
\medskip

\begin{itemize}\item \underline{$\tilde{u}'' \tend \infty$, $\tilde{h}''\tend \infty$}\end{itemize}\medskip

In this case, the potential is fully complex, with a non-trivial contribution for both the real and imaginary parts:
\ba\label{eq:asympto-nontrivial}
\tilde{u}(\tilde{\vphi}) = A \tilde{\vphi}^\frac{1}{\gamma} +  \frac{A \gamma^2}{2(A^2+B^2)(1-\gamma)^2}\tilde{\vphi}^{2-\frac{1}{\gamma}}-\frac{\(A^2-B^2\)\gamma^4}{(A^2+B^2)^2(1-\gamma)^2(3-4\gamma)}\tilde{\vphi}^{4-2\frac{1}{\gamma}}+\dots\\
\tilde{h}(\tilde{\vphi}) = B \tilde{\vphi}^\frac{1}{\gamma} - \frac{B \gamma^2}{2(A^2+B^2)(1-\gamma)^2}\tilde{\vphi}^{2-\frac{1}{\gamma}}+ \frac{2 AB \gamma^4}{(A^2+B^2)^2(1-\gamma)^2(3-4\gamma)}\tilde{\vphi}^{4-2\frac{1}{\gamma}}+\dots\\
\ea

Notice that the real part is even under the $\Z_2$ transformation $B\leftrightarrow-B$, while the imaginary part is odd. There is no symmetry in $A$ instead, because the equations \eqref{eq:coupled-rescaled} depend on the combination $1+u''$.
\medskip

\begin{itemize}\item \underline{$\tilde{u}'' \tend \infty$, $\tilde{h}''\tend 0$}\end{itemize}\medskip

The condition $\tilde{u}'' \tend \infty$, $\tilde{h}''\tend  c<\infty$, is inconsistent with the second equation in \eqref{eq:coupled-rescaled}, unless $c=0$. 
At leading order, we can then ignore $\tilde{h}''$ in the first equation of \eqref{eq:coupled-rescaled}, and find the leading solution in \eqref{eq:vanishinglhs} for $\tilde{u}$.
Plugging this back in the second equation in \eqref{eq:coupled-rescaled}, and looking for a solution for $\tilde{h}$ that tends asymptotically to zero, leads to an equation similar to those from the linearization test, and whose solution grows exponentially, thus leading again to a contradiction.
We conclude that in this case the imaginary part must vanish identically, and the solution corresponds to a purely real potential at all order, with first few orders given again by \eqref{eq:asympto-nontrivial} with $B=0$.

\medskip

\begin{itemize}\item \underline{$\tilde{u}'' \tend 0$, $\tilde{h}''\tend \infty$}\end{itemize}\medskip

The condition $\tilde{u}''\tend  c<\infty$, leads again to $c=0$, by consistency with the first equation of \eqref{eq:coupled-rescaled}. 
Proceeding similarly to the previous case, we now find a solution that corresponds to setting $A=0$ in  \eqref{eq:asympto-nontrivial}. 
Notice that in this case the asymptotic solution is not purely imaginary, but the real part is negligible at sufficiently large $\tilde{\vphi}$.

\section{Quadratic truncation of the real part}
\label{app:higher-truncation}

With a purely imaginary potential, the real part of the coupled flow equation \eqref{eq:coupled} is violated even for small values of $\phi$.
We can try to improve approximation of the coupled system by including a non-trivial real even part to the potential.
As a first improvement, we can introduce a quadratic polynomial for  the real part:
\be
u(\phi) = \lambda_0 + \half \lambda_2\, \vphi^2.
\ee
Following the usual approximation scheme of polynomial truncations of the FRG, we replace this ansatz in the flow equation, expand it in powers of $\vphi$, and truncate such expansion to quadratic order, effectively discarding the generation of higher-order terms.

Then, the real part of \eqref{eq:coupled} is cast into fixed-point equations for the beta functions of the real couplings, and after imposing the initial conditions \eqref{eq:init-cond}
\ba\label{eq:trunc2-eq}
\beta_{\lambda_0} &= -d\lambda_0 + \frac{c_d}{1+\lambda_2}\(1-\frac{\eta}{d+2}\) =0,\\
\beta_{\lambda_2} &= \(2\Delta-d\)\lambda_2 - 2c_d\frac{g^2}{(1+\lambda_2)^3}\(1-\frac{\eta}{d+2}\) =0.
\ea
For simplicity, the above system of equations is then solved for $g$, keeping the coupling $\lambda_2\in \R$ as our free parameter.
In the LPA, $\eta=0$, and the solutions reduce to:
\be
\lambda_0 = \frac{c_d}{d(1+\lambda_2)}, \quad g=\sqrt{-\frac{\lambda_2(1+\lambda_2)^3}{c_d}}, \quad -1<\lambda_2\le0,
\ee
while in the LPA',
\ba
\lambda_0 &= c_d\,\frac{2+\lambda_2+\sqrt{\frac{2(2+\lambda_2)^2+d(2+3\lambda_2)^2}{2+d}}}{4d(1+\lambda_2)^2}, \quad 
\eta = - c_d \frac{g^2}{(1+\lambda_2)^4},\quad -1<\lambda_2\le0\\
g&=\sqrt{-\frac{(1+\lambda_2)^3}{c_d}\((2+d)(2+3\lambda_2)-\sqrt{2+d}\sqrt{2(2+\lambda_2)^2+d(2+3\lambda_2)^3}\)}.
\ea
Note that \eqref{eq:trunc2-eq} is a second order polynomial equation in $g^2$, resulting in spurious solutions.
We lift the ambiguity by choosing the solution consistent with the GFP in $d=6$.
The condition on the parameter $\lambda_2$ ensues from the reality of $g$. \medskip

In the LPA', the full scaling dimension is negative for
\be
\Delta \le 0\quad \Leftrightarrow\quad -1<\lambda_2 \le -\frac{8-4d}{6-5d},\ d\ge2.
\ee
Noteworthily, the imaginary part of the differential equation \eqref{eq:coupled}
\be\label{eq:trunc-im2}
d\, h- \Delta\, \vphi\, h' = - c_d \(1-\frac{\eta}{d+2}\)\frac{h''}{(1+\lambda_2)^2+\(h''\)^2}, \quad \eta = -c_d \frac{g^2}{(1+\lambda^2)^4},
\ee
is only modified by a constant term in the denominator on the right-hand side, which can be eliminated via the rescaling:
\be
\vphi \tend \tilde{\vphi} = (1+\lambda_2)\, \vphi, \quad h(\vphi) \tend \tilde{h}(\tilde{\vphi}) = (1+\lambda_2)\, h(\vphi).
\ee
Therefore, we conclude that fixed-point equation for $h(\vphi)$ with a quadratic truncation of the real part is equivalent to the one with real part set to zero.

Higher-order truncations of the real part lead to a similar mixed system of beta functions for the $\lambda_{2n}$ couplings and beta functional for $h(\vphi)$, and they cannot be reduced to the pure-imaginary case by a rescaling.
However, as explained in section \ref{sec:numerical}, they lead to difficulties in the search of global solutions for $h(\vphi)$, hence we do not consider them.

\bibliographystyle{JHEP}
\bibliography{refs}
\addcontentsline{toc}{section}{References}


\end{document}